%% file: Wreg24_arXiv.tex
\documentclass[twoside,english,11pt]{article}

\input{macros_Wreg24_arXiv}

\usepackage{authblk}

\def\references{\bibliography{merged}}

\newcommand{\blind}{0}

\begin{document}

\single
\if0\blind
{
	\title{Wasserstein Regression\thanks{Research supported by NIH Echo and NSF DMS1712862.}}
	\date{}
	
	\author[1]{Yaqing Chen}	
	\author[2]{Zhenhua Lin}
	\author[1]{Hans-Georg M\"uller}
	
	\affil[1]{Department of Statistics, University of California, Davis}
	\affil[2]{Department of Statistics and Applied Probability, National University of Singapore}
	
	\maketitle
	\vspace{-2em}
} \fi

\if1\blind
{
	\title{\vspace{-2cm}Wasserstein Regression}
	\author{}
	\maketitle
	\vspace{-4em}
} \fi

\begin{abstract}
	The analysis of samples of random objects that do not lie in a vector space is gaining increasing attention in statistics. An important class of such object data is univariate probability measures defined on the real line. Adopting the Wasserstein metric, we develop a class of regression models for such data, where random distributions serve as predictors and the responses are either also distributions or scalars. To define this regression model, we utilize the geometry of tangent bundles of the space of random measures endowed with the Wasserstein metric for mapping distributions to tangent spaces. The proposed distribution-to-distribution regression model provides an extension of multivariate linear regression for Euclidean data and function-to-function regression for Hilbert space valued data in functional data analysis. In simulations, it performs better than an alternative transformation approach where one maps distributions to a Hilbert space through the log quantile density transformation and then applies traditional functional regression. We derive asymptotic rates of convergence for the estimator of the regression operator and for predicted distributions and also study an extension to autoregressive models for distribution-valued time series. The proposed methods are illustrated with data on human mortality and distributional time series of house prices. \\
	\noindent%
\end{abstract}


\section{Introduction}

Regression analysis is one of the foundational tools of statistics to quantify  the relationship between a response variable and predictors and  there have been many extensions of  simple models such as the  multiple linear regression model to more complex data scenarios. These include linear models for function-to-function regression,  where predictors and responses are both considered random elements in Hilbert space, with a variant where responses are scalars \citep{gren:50,rams:91}. Such linear functional regression models and their properties have been well studied  \citep{card:99,card:03:1,yao:05:2,cai:06,hall:07:1} 
and reviewed \citep{morr:15,wang:16}. 

Samples that include random objects, which are random elements in general metric spaces that by default do not have a vector space structure, are increasingly common. Such data 
cannot be analyzed with methods devised for Euclidean or functional data, which are usually viewed as random elements of  a Hilbert space \citep{marr:14,huck:15}. We focus here on the case where the random objects are random probability measures on the real line that satisfy certain regularity conditions. Specifically, at this time there are no in-depth studies with detailed statistical analysis of regression models that feature such random measures as predictors, in contrast to the situation where vector predictors are coupled with random distributions as responses \citep{pete:19}. 

Related work also includes a variety of methods that specifically target the case where Euclidean predictors are paired with responses that reside on a finite-dimensional Riemannian manifold \citep{davi:07,shi:09, hink:12,yuan:12,corn:17,lin:17:2}. Kernel and  spline type methods have been proposed for the case where both predictors and responses are elements of finite-dimensional Riemannian manifolds  
\citep{stei:09,stei:10,bane:16}. 
However, these methods do not cover spaces of probability measures under the Wasserstein metric, where the tangent spaces are subspaces of infinite-dimensional Hilbert spaces.
Additionally, no comprehensive investigation of the statistical properties and asymptotic behavior of distribution-to-distribution regression models seems to exist. To develop the proposed model, we utilize tangent bundles in the space of probability distributions with the Wasserstein metric and parallel transport to obtain asymptotic results for regression operators and predicted measures. 

A recent approach to including random distributions as predictors in complex regression models is to transform the densities of these distributions to unconstrained functions in the Hilbert space $\hilbert$, e.g., by the log quantile density (LQD) transformation \citep{pete:16:1} and then to employ functional regression models where the transformed functions serve as predictors and the responses are either also the transformed functions or scalars \citep{chen:19,koko:19,pete:19:8}, whence  established methods for functional regression become applicable. 
\revtwo{However, the LQD transformation does not take into account the geometry of the space of probability distributions and therefore the corresponding transformation map is not isometric and leads to deformations that change distances between pairs of objects. In contrast, the transformation method we develop here is closely adapted to the underlying geometry, leads to an isometric map and  fully  utilizes the geometric properties of the metric space of random measures equipped with the Wasserstein distance. 
	We also found in implementations and simulations that the proposed  geometric method that we refer to as Wasserstein regression works very well, especially when comparing it to a regression approach that is based on the  LQD transformation.} 
Other alternatives have been considered for regressing scalar responses on distribution-valued predictors \citep{pocz:13, oliv:14, szab:16, bach:17, thi:19}, \update{but these are either Nadaraya--Watson type estimators that suffer from a severe curse of dimensionality, or kernel-based methods that rely on tuning parameters whose choice could be sensitive in real applications}. \citet{bonn:16} approximate input histograms by the closest weighted barycenters of a database of reference histograms with respect to Wasserstein distance, which work when input histograms are not far from the references, aiming at applications in image processing. 


\update{Our goal is to develop a regression model where the predictors and responses are both distributions in $\manifold$. A good starting point is linear regression in Euclidean spaces, where for a pair of random elements $(X,Y)\in (\real^p,\real)$, $\expect(Y\mid X) = \linRegOp(X) = \expect Y + \linRegCoef^\top(X - \expect X)$. 
	The regression function $\linRegOp$ can be characterized by the following two properties: First, it maps the expectation of $X$ to the expectation of $Y$; second, conditioning on $X$, it transports the line segment between $\expect X$ and $X$ to that between $\expect Y$ and $\expect(Y\mid X)$. Specifically, }
\bal\label{eq:mvlr}
\expect Y = \linRegOp(\expect X) \qaq
\expect[\expect Y + t(Y - \expect Y)\mid X] = \linRegOp[\expect X + t(X - \expect X)],\text{ for all } t\in[0,1].\eal

\update{However, expectations and line segments are not well-defined for the space of distributions, since it is not a vector space. 
	In this paper, we develop a distribution-to-distribution regression model that is analogous to traditional linear regression models for Euclidean and functional data, with the decisive difference that both predictors and responses are univariate probability measures. 
	An example which we investigate later is to study the relationship of the age-at-death distributions of different countries in 2013 to the distributions 30 years before.}
We also discuss an extension of our approach to an autoregressive model for distribution-valued time series. 
In our estimation procedures and  theoretical analysis we cover the commonly encountered but more complex situation where neither predictor nor response  distributions are  directly observed and instead the available data consist of  i.i.d. samples that are generated by each of  these distributions.  
\update{After we submitted this paper, a preprint reporting independently conducted but related work on autoregressive modeling of distributional time series was posted by \citet{zhan:20}, where a simplified version of the distributional autoregressive model in \eqref{eq:autoreg} was studied.}

The remainder of the paper is organized as follows. We first propose a distribution-to-distribution regression model based on the tangent bundle of the Wasserstein space of probability distributions in Section~\ref{sec:method}, with estimation and asymptotic theory  in Section~\ref{sec:d2dEst}, and  then describe an  extension of  the model to an autoregressive model for time series of distributions in Section~\ref{sec:dts}. 
Simulation studies are illustrated in Section~\ref{sec:simu} to assess the  finite-sample performance of the proposed estimators and a competing approach. The wide applicability of the proposed methods is demonstrated  with applications to human mortality data and US house price data in Section~\ref{sec:app}. 

\section{Methodology}\label{sec:method}

\subsection{Tangent Bundle of the Wasserstein Space}\label{sec:geometry}

Let $\dom$ be $\real$ or a closed interval in $\real$, and $\borel(\dom)$ be the Borel $\sigma$-algebra on $\dom$. 
We focus on the Wasserstein space $\manifold = \manifold(\dom)$  of probability distributions on $(\dom, \borel(\dom))$ with finite second moments, endowed with the $\hilbert$-Wasserstein distance
\be\label{eq:wdist} \wdist(\distnOne,\distnTwo) = \left\{\int_0^1[\quantileDistnOne(p) - \quantileDistnTwo(p)]^2 \diffop p\right\}\half,\ee
for $\distnOne,\distnTwo\in\manifold$, where $\quantileDistnOne$ and $\quantileDistnTwo$ denote the quantile functions of $\distnOne$ and $\distnTwo$, respectively; 
specifically, for any distribution $\arbiDistn = \arbiDistn(\cdfArbiDistn)\in\manifold$ with cumulative distribution function (cdf) $\cdfArbiDistn$, we consider the quantile function $\quantileArbiDistn$  to be the left continuous inverse of  $\cdfArbiDistn$, i.e., \bgt\label{eq:quantileFromCDF}
\quantileArbiDistn(p) = \inf\{\argu\in\dom: \cdfArbiDistn(\argu)\ge p\},\quad \text{for }p\in(0,1).\egt 

As demonstrated for example in \citet{ambr:08,bigo:17:2,zeme:19}, basic concepts of Riemannian manifolds can be generalized to the Wasserstein space $\manifold$. 
\update{We assume in the following that $\refDistn\in\manifold$ is an atomless reference probability measure, i.e., it possesses a continuous cdf $\cdfRefDistn$. 
	For any $\arbiDistn\in\manifold$, the geodesic from $\refDistn$ to $\arbiDistn$, $\geodesic_{\refDistn,\arbiDistn}\colon [0,1]\ra\manifold$, is given by 
	\bal\label{eq:geodesic}
	\geodesic_{\refDistn,\arbiDistn}(t) = [t (\quantileArbiDistn \circ\cdfRefDistn-\id) + \id] \#\refDistn, \quad\text{for } t\in[0,1],\eal
	where for a measurable function $\arbiFctnTwo\colon \dom\ra\dom$, $\arbiFctnTwo\#\refDistn$ is a push-forward measure such that $\arbiFctnTwo\#\refDistn(A) = \refDistn(\{\argu\in\dom: \arbiFctnTwo(\argu)\in A\})$ for any set $A\in\borel(\dom)$. 
	The tangent space at $\refDistn$ is defined as 
	\bgt\nn 
	\tangentspace{\refDistn} = {\overline{\{t(\quantileArbiDistn \circ\cdfRefDistn - \id): \arbiDistn=\arbiDistn(\cdfArbiDistn)\in\manifold,\, t>0\}}}^{\hilbert_{\refDistn}}, \egt
	where $\hilbert_{\refDistn} = \hilbert_{\refDistn}(\dom)$ is the Hilbert space of $\refDistn$-square-integrable functions on $\dom \subset \real$, with inner product $\innerprod{\cdot}{\cdot}_{\refDistn}$ and norm $\|\cdot\|_{\refDistn}$. 
	The tangent space $\tangentspace{\refDistn}$ is a subspace of $\hilbert_{\refDistn}$ equipped with the same inner product and induced norm \citep[Theorem 8.5.1,][]{ambr:08}.} 

The exponential map $\Exp_{\refDistn}$ is then defined by the push-forward measures, which maps functions of the form $\arbiFctn = t(\quantileArbiDistn \circ\cdfRefDistn - \id)$ onto $\manifold$, with $\quantileArbiDistn$ being the quantile function of an arbitrary distribution $\arbiDistn\in\manifold$,  
\bal\label{eq:exp}
\Exp_{\refDistn}\arbiFctn = (\arbiFctn+\id)\#\refDistn. \eal 
\update{While this exponential map is not a local homeomorphism \citep{ambr:04},  
	any $\arbiDistn\in\manifold$ 
	can be recovered by $\Exp_{\refDistn}(\quantileArbiDistn\circ \cdfRefDistn-\id)$ 
}in the sense that $\wdist(\Exp_{\refDistn}(\quantileArbiDistn\circ \cdfRefDistn-\id), \arbiDistn)= 0$\note{\ due to the continuity of $\cdfRefDistn$}, 
and the logarithmic map $\Log_{\refDistn}\colon \manifold\ra \tangentspace{\refDistn}$, as the right inverse of the exponential map, is given by
\bal \label{eq:log}
\Log_{\refDistn}\arbiDistn = \quantileArbiDistn\circ \cdfRefDistn -\id, \quad\text{for }\arbiDistn\in\manifold.\eal 
Furthermore, restricted to the log image,  $\Exp_{\refDistn}|_{\Log_{\refDistn}(\manifold)}$ is an isometric homeomorphism \citep[e.g., Lemma 2.1,][]{bigo:17:2}. 



\subsection{Distribution-to-Distribution Regression}\label{sec:d2dreg}
Let $(\X,\Y)$ be a pair of random elements with a joint distribution $\jointdistn$ on $\manifold\times \manifold$, 
assumed to be square integrable in the sense that $\expect\wdist^2(\arbiDistn,\X)<\infty$ and $\expect\wdist^2(\arbiDistn,\Y)<\infty$ for some (and thus for all) $\arbiDistn\in\manifold$. 
Any element in $\manifold$ that minimizes $\expect\wdist^2(\cdot,\X)$ is called a Fr\'echet mean of $\X$ \citep{frec:48}. 
Since the Wasserstein space $\manifold$ is a Hadamard space \citep{kloe:10}, such minimizers uniquely exist  \citep{stur:03} and are given by 
\bgt\label{eq:Fmean}
\FmeanX = \argmin_{\arbiDistn\in\manifold} \expect\wdist^2(\arbiDistn,\X) \qaq
\FmeanY= \argmin_{\arbiDistn\in\manifold} \expect\wdist^2(\arbiDistn,\Y). \egt 
It is well-known that for univariate distributions as we consider here, the quantile functions of the Fr\'echet means are simply
\be
\quantileFmeanX(\cdot) = \expect \quantileX(\cdot)\quad\text{and}\quad \quantileFmeanY(\cdot) = \expect \quantileY(\cdot), \nn
\ee
where $\quantileFmeanX, \quantileFmeanY, \quantileX$ and $\quantileY$ are the quantile functions of $\FmeanX,\FmeanY, \X$ and $\Y$, respectively.


\update{As suggested by the multiple linear regression as per \eqref{eq:mvlr}, we replace expectations and line segments, which are not well-defined for the Wasserstein space, by Fr\'echet means and geodesics, respectively. Hence, }
a regression operator $\geoRegOp\colon \manifold\ra\manifold$ for the Wasserstein space would be expected to satisfy:
\bal\label{eq:geoReg}
\wdist(\FmeanY, \geoRegOp(\FmeanX)) = 0 \qaq
\wdist(\expect_\oplus\{\geodesic_{\FmeanY, \Y}(t) | \X\}, \geoRegOp\{\geodesic_{\FmeanX,\X}(t)\}) = 0,\text{ for all } t\in[0,1],\eal
where the conditional Fr\'echet mean $\expect_\oplus\{\geodesic_{\FmeanY, \Y}(t) \mid \X\} \coloneqq \argmin_{\arbiDistn\in\manifold}\expect[\wdist^2(\arbiDistn,\geodesic_{\FmeanY,\Y}(t))\mid\X]$. 

\update{We assume that the Fr\'echet means $\FmeanX$ and $\FmeanY$ are atomless so that they can be used as the reference probability measures as in Section~\ref{sec:geometry}. 
	Note that $\Log_{\FmeanX}\FmeanX=0$, $\FmeanX$-a.e., and $\Log_{\FmeanY}\FmeanY=0$, $\FmeanY$-a.e., and that $\Exp_{\arbiDistn}(0) = \arbiDistn$ for any $\arbiDistn\in\manifold$. 
	Furthermore, it follows from \eqref{eq:geodesic}--\eqref{eq:log} and the isometry property of $\Exp_{\FmeanY}|_{\Log_{\FmeanY}\manifold}$ that $\geodesic_{\FmeanX,\X}(t) = \Exp_{\FmeanX}(t\Log_{\FmeanX}\X)$ and that $\expect_\oplus\{\geodesic_{\FmeanY, \Y}(t) \mid \X\}=\argmin_{\arbiDistn\in\manifold}\expect(\|\Log_{\FmeanY}\arbiDistn-t\Log_{\FmeanY}\Y\|_{\FmeanY}^2\mid \Log_{\FmeanX}\X) =\Exp_{\FmeanY}[\expect(t\Log_{\FmeanY}\Y\mid \Log_{\FmeanX}\X)]$. Hence, \eqref{eq:geoReg} can be rewritten as
	\bal\label{eq:tanReg}
	\|\regOp(0)\|_{\FmeanY} = 0\qaq
	\|\expect(t\Log_{\FmeanY}\Y \mid \Log_{\FmeanX}\X) - \regOp(t\Log_{\FmeanX}\X)\|_{\FmeanY} = 0,\text{ for all } t\in[0,1], \eal
	where $\regOp\colon \tangentspace{\FmeanX} \ra \tangentspace{\FmeanY}$, $\regOp = \Log_{\FmeanY}\circ\geoRegOp\circ\Exp_{\FmeanX}$, is a regression operator between tangent spaces $\tangentspace{\FmeanX}$ and $\tangentspace{\FmeanY}$. 
}

As discussed in Section~\ref{sec:geometry}\note{, with the atomlessness of $\FmeanX$ and $\FmeanY$}, $\tangentspace{\FmeanX}$ and $\tangentspace{\FmeanY}$ are subspaces of $\hilbert_{\FmeanX}$ and $\hilbert_{\FmeanY}$, respectively. 
Distribution-to-distribution regression can then be viewed as function-to-function regression, which has been well-studied in functional data analysis \citep[see, e.g.,][]{ferr:03:2,yao:05:2,he:10,wang:16}. 
Specifically, we assume that the random pair of distributions $(\X,\Y)$ satisfy the model 
\bgt\label{eq:d2dreg}
\expect(\Log_{\FmeanY}\Y \mid \Log_{\FmeanX}\X) =\regOp(\Log_{\FmeanX}\X),\egt
where $\regOp\colon \tangentspace{\FmeanX} \ra \tangentspace{\FmeanY}$
is a linear operator defined as
\bgt\label{eq:linopt_coef}
\regOp \arbiFctn(t) = \innerprod{\regKernel(\cdot, t)} {\arbiFctn}_{\FmeanX}, \quad\text{for }t\in\dom \text{ and }\arbiFctn\in\tangentspace{\FmeanX}.\egt
Here, $\regKernel:\dom^2 \ra \real$ is a coefficient function (i.e., the kernel of $\regOp$) lying in $\hilbert_{\FmeanX\times \FmeanY}$, and $\FmeanX\times \FmeanY$ is a product probability measure on the product measurable space $(\dom^2, \borel(\dom^2))$ generated by $\FmeanX$ and $\FmeanY$. 
\update{We note that our model satisfies \eqref{eq:tanReg}. Furthermore, we assume 
	\ben[label = (A\arabic*), series = ass]
	\item\label{ass:convex} With probability 1, $\regOp(\Log_{\FmeanX}\X)+\id$ is non-decreasing. \een
	Assumption~\ref{ass:convex} guarantees that $\regOp(\Log_{\FmeanX}\X)\in\Log_{\FmeanY}\manifold$ with probability 1. We demonstrate the feasibility of the proposed model in \eqref{eq:d2dreg} by providing a framework in Section~\ref{sec:simu} to construct explicit examples that satisfy the model requirements and \ref{ass:convex}.} 


\bco
Since the model does not require quantile functions, this model also applies to the Wasserstein space $\manifold(\real^p)$ of probability measures on $\real^p$. Assume that the Fr\'echet means $\FmeanX$ and $\FmeanY$ of $\X$ and $\Y\in\manifold(\real^p)$ uniquely exist, and are absolutely continuous with respect to Lebesgue measure on $\real^p$ (referred to as regular measures in the following). 
Optimal transport maps from a regular probability measure $\refDistn$ to any probability measure $\arbiDistn\in\manifold(\real^p)$ uniquely exist \citep{mcca:95}, denoted by $\optTrans{\refDistn}{\arbiDistn}$.  
The tangent space and log map at $\refDistn$ are defined as $\tangentspace{\refDistn} = {\overline{\{t(\optTrans{\refDistn}{\arbiDistn} - \id): \arbiDistn\in\manifold, t >0 \}}}^{\hilbert_{\refDistn}}$, 
and $\Log_{\refDistn}\arbiDistn = \optTrans{\refDistn}{\arbiDistn} - \id$ for $\arbiDistn\in\manifold(\real^p)$, respectively \citep{ambr:08,zeme:19}.  
However, we only consider $p=1$ due to implementation difficulty for $p>1$.
\fi

\subsection[Covariance Structure and Regression Coefficient Function]{Covariance Structure, Regression Coefficient Function and Scalar Responses} \label{sec:cov_indep}
\update{Noting that $\expect (\Log_{\FmeanX}\X) = 0$, $\FmeanX$-a.e., and $\expect (\Log_{\FmeanY}\Y) = 0$, $\FmeanY$-a.e.,} we denote the covariance operators of $\Log_{\FmeanX}\X$ and $\Log_{\FmeanY}\Y$ by $\covopX = \expect(\Log_{\FmeanX}\X\otimes\Log_{\FmeanX}\X)$ and 
$\covopY = \expect(\Log_{\FmeanY}\Y\otimes\Log_{\FmeanY}\Y)$, respectively, and the cross-covariance operator by $\crosscovopXY =\expect (\Log_{\FmeanY}\Y \otimes \Log_{\FmeanX}\X)$. 
Since  the two covariance operators $\covopX$ and $\covopY$ are trace-class, they  have eigendecompositions \citep[Theorem 7.2.6,][]{hsin:15} as given below, \update{which can be viewed as an analog to multivariate principal component analysis \citep{daux:82,cast:86}}, yielding a corresponding decomposition for the cross-covariance operator $\crosscovopXY$, 
\bal\label{eq:cov_pop}
\covopX = \sum_{j=1}^\infty \egnvalX_j &\egnfctnX_j \otimes \egnfctnX_j,\quad 
\covopY = \sum_{k=1}^\infty \egnvalY_k \egnfctnY_k \otimes \egnfctnY_k,\quad
\crosscovopXY &= \sum_{k=1}^\infty \sum_{j=1}^\infty  \croval_{jk} \egnfctnY_k \otimes \egnfctnX_j.
\eal
Here  $\egnvalX_j = \expect[\innerprod{\Log_{\FmeanX}\X}{\egnfctnX_j}_{\FmeanX}^2]$ and 
$\egnvalY_k =  \expect[\innerprod{\Log_{\FmeanY}\Y}{\egnfctnY_k}_{\FmeanY}^2]$ are eigenvalues 
such that $\egnvalX_1\ge \egnvalX_2\ge\cdots\ge 0$ and  $\egnvalY_1\ge\egnvalY_2\ge \cdots \ge 0$, 
$\{\egnfctnX_j\}_{j=1}^\infty$ and  $\{\egnfctnY_k\}_{k=1}^\infty$ are eigenfunctions that are orthonormal in $\tangentspace{\FmeanX}$ and $\tangentspace{\FmeanY}$, respectively, 
and $\croval_{jk} = \expect[\innerprod{\Log_{\FmeanX}\X}{\egnfctnX_j}_{\FmeanX} \innerprod{\Log_{\FmeanY}\Y}{\egnfctnY_k}_{\FmeanY}]$. 
With probability 1, the log transformations $\Log_{\FmeanX}\X$ and $\Log_{\FmeanY}\Y$ admit the Karhunen--Lo\`eve expansions
\bal\nn
\Log_{\FmeanX}\X = \sum_{j=1}^\infty \innerprod{\Log_{\FmeanX}\X}{\egnfctnX_j}_{\FmeanX}\egnfctnX_j \quad\text{and}\quad
\Log_{\FmeanY}\Y = \sum_{k=1}^\infty \innerprod{\Log_{\FmeanY}\Y}{\egnfctnY_k}_{\FmeanY} \egnfctnY_k.\eal

Then as in the classical functional regression \update{\citep[e.g.,][]{bosq:91,card:99,yao:05:2}}, the regression coefficient function $\regKernel$ can be expressed as
\bgt\label{eq:coef_expansion} \regKernel = \sum_{k=1}^\infty \sum_{j=1}^\infty \regCoef_{jk} \egnfctnY_k\otimes \egnfctnX_j, \egt
with $\regCoef_{jk} = \egnvalX_j\inv \croval_{jk}$. 
In order to guarantee that the right hand side of \eqref{eq:coef_expansion} converges in the sense that 
\bal\nn \lim_{\nFPCX,\nFPCY\ra\infty}\int_{\dom} \int_{\dom} \left[\sum_{k=1}^{\nFPCY} \sum_{j=1}^{\nFPCX} \regCoef_{jk}  \egnfctnX_j(s) \egnfctnY_k(t) - \regKernel(s,t)\right]^2  \diffop \FmeanX(s) \diffop \FmeanY(t)= 0, \eal
we assume \citep[Lemma A.2,][]{yao:05:2} \be\label{eq:coef_cvg}\sum_{k=1}^\infty\sum_{j=1}^\infty\egnvalX_j^{-2}\croval_{jk}^2 <\infty.\ee

To keep notations simple,  we use the same notation $\arbiFctn_1\otimes \arbiFctn_2$ for the operator and its kernel throughout this paper. Namely, for $\arbiFctn_1\in \hilbert_{\distnOne}$ and $\arbiFctn_2\in \hilbert_{\distnTwo}$, 
$\arbiFctn_1\otimes \arbiFctn_2$ can represent either an operator on $\hilbert_{\distnTwo}$ such that $(\arbiFctn_1\otimes \arbiFctn_2)(\arbiFctn) = \innerprod{\arbiFctn_2}{\arbiFctn}_{\distnTwo} \arbiFctn_1$ for $\arbiFctn\in \hilbert_{\distnTwo}$ or its kernel, i.e., a bivariate function such that $(\arbiFctn_1\otimes \arbiFctn_2)(s,t) = \arbiFctn_1(t)\arbiFctn_2(s)$ for all $s,t\in\dom$.

A variant of the proposed distribution-to-distribution regression in \eqref{eq:d2dreg} is the pairing of distributions as predictors with scalar responses.  For a pair of random elements $(\X,\sY)$ with a joint distribution on $\manifold\times \real$, a distribution-to-scalar regression model is
\bgt\label{eq:d2sreg}
\expect(\sY\mid \Log_{\FmeanX}\X) = \expect(\sY) + \innerprod{\regKernelDtoS}{\Log_{\FmeanX}\X}_{\FmeanX}.\egt
Here, $\FmeanX$ is the Fr\'echet mean of $\X$ and $\regKernelDtoS\colon \dom \ra \real$ is a regression coefficient function in $\hilbert_{\FmeanX}$ which can be expressed as $\regKernelDtoS = \sum_{j=1}^\infty \egnvalX_j\inv \innerprod{\expect(\sY\Log_{\FmeanX}\X)}{\egnfctnX_j}_{\FmeanX} \egnfctnX_j$, 
where $\egnvalX_j$ and $\egnfctnX_j$ are the eigenvalues and eigenfunctions of the covariance operator $\covopX$ of $\Log_{\FmeanX}\X$ as in \eqref{eq:cov_pop}, and we assume that  $\sum_{j=1}^\infty  \egnvalX_j^{-2}\innerprod{\expect(\sY\Log_{\FmeanX}\X)}{\egnfctnX_j}_{\FmeanX}^2 <\infty$. This model can also be viewed as function-to-scalar regression, which has been well studied in functional data analysis \citep{card:99,card:03:1,cai:06,hall:07:1,yuan:10}.

\section{Estimation}\label{sec:d2dEst}
\subsection{Distribution Estimation}\label{sec:denEst}

While \citet{bigo:17:2} assume distributions are fully observed, in reality this is usually not the case, and this creates an additional challenge for the implementation of the proposed distribution-to-distribution regression model. 
Options to address this include estimating cdfs  \citep[e.g.,][]{agga:55,read:72,falk:83,lebl:12}, or estimating quantile functions \citep[e.g.,][]{parz:79,falk:84,yang:85,chen:97:2} of the underlying distributions. 
Given an estimated quantile function $\quantileArbiDistnEst$ (resp. cdf $\cdfArbiDistnEst$), we convert it to a cdf (resp. a quantile function) by right (resp. left) continuous inversion, 
\be\cdfArbiDistnEst(\argu) = \sup\{p\in[0,1]: \quantileArbiDistnEst(p)\le \argu\},\quad\text{for }\argu\in\real \label{eq:cdfFromQuantile}\ee
(resp. \eqref{eq:quantileFromCDF}). 
Alternatively, one can start with a density estimator to estimate densities \citep[][]{pana:16,pete:16:1} and then compute the cdfs and quantile functions by integration and inversion.

Suppose $\{(\Xi,\Yi)\}_{i=1}^n$ are $n$ independent realizations of $(\X,\Y)$. 
What we observe are collections of independent measurements $\{\dpX_{il}\}_{l=1}^{\nDpXi}$ and $\{\dpY_{il}\}_{l=1}^{\nDpYi}$, sampled from $\Xi$ and $\Yi$, respectively, 
where $\nDpXi$ and $\nDpYi$ are the sample sizes which may vary across distributions. 
Note that there are two independent layers of randomness in the data: The first generates independent pairs of distributions $(\Xi,\Yi)$; the second generates independent observations according to each distribution, $\dpX_{il}\sim \Xi$ and $\dpY_{il}\sim \Yi$. 

For a distribution $\arbiDistn\in\manifold$, denote by $\arbiDistnEst = \arbiDistn(\cdfArbiDistnEst)$ the distribution associated with some cdf estimate $\cdfArbiDistnEst$, based on a sample of measurements drawn according to $\arbiDistn$. 
Using $\XiEst$ and $\YiEst$ as surrogates of $\Xi$ and $\Yi$,
the theoretical analysis of the estimation of the distribution-to-distribution regression operator requires the following assumptions that  quantify the discrepancy of the estimated and true probability measures. 
\ben[label = (A\arabic*), resume = ass]
\item \label{ass:distn_est_rate}
For any distribution $\arbiDistn\in\manifold$, with some nonnegative decreasing sequences $\denRate_{\nDp} = o(1)$ as $\nDp\ra\infty$, the corresponding estimate $\arbiDistnEst$ based on a sample of size $\nDp$ drawn according to $\arbiDistn$ satisfies
\bal\nn 
\sup_{\arbiDistn\in\manifold} \expect[\wdist^2 (\arbiDistnEst, \arbiDistn)]= O(\denRate_{\nDp})\qaq 
\update{\sup_{\arbiDistn\in\manifold} \expect[\wdist^4 (\arbiDistnEst, \arbiDistn)]= O(\denRate_{\nDp}^2)}.\eal
\een

For example, for compactly supported distributions, the distribution estimator proposed by \citet{pana:16} satisfies \ref{ass:distn_est_rate} with $\denRate_{\nDp} = \nDp^{-1/2}$, 
while \citet{pete:16:1} consider a subset $\manifoldLQD$ of $\manifold$ containing distributions that are absolutely continuous with respect to Lebesgue measure on a compact domain  $\dom$ such that
\be\label{eq:denBound} \sup_{\arbiDistn\in\manifoldLQD}\sup_{\argu\in\dom_{\arbiDistn}} \max\{\den_{\arbiDistn}(\argu), 1/\den_{\arbiDistn}(\argu), |\den'_{\arbiDistn}(\argu)|\} \le \denBound, \ee
where $\den_{\arbiDistn}$ is the density function of a distribution $\arbiDistn\in\manifoldLQD$, $\dom_{\arbiDistn}$ is the support of distribution $\arbiDistn$ and $\denBound>0$ is constant,  and 
then obtain the rates $\sup_{\arbiDistn\in\manifoldLQD} \expect\wdist^2 (\arbiDistnEst, \arbiDistn)= O(\nDp^{-2/3})$ and $\sup_{\arbiDistn\in\manifoldLQD} \expect[\wdist^4 (\arbiDistnEst, \arbiDistn)]= O(\nDp^{-4/3})$ in \ref{ass:distn_est_rate} \citep[Proposition~1,][]{pete:19:2}.

The following assumption on the numbers of measurements per distribution $\nDpXi$ and $\nDpYi$ facilitates our analysis:
\ben[label = (A\arabic*), resume = ass]
\item \label{ass:nObsPerDens}
There exists a sequence $\nDp = \nDp(n)$ such that $\min\{\nDpXi,\nDpYi: i=1,\dots,n\} \ge \nDp$ and $\nDp\ra\infty$ as $n\ra \infty$.
\een

\subsection{Regression \update{Operator} Estimation}\label{sec:coef_est_indep}

\update{We note that  notations with `` $\wt{}$ '' refer to  estimators based on fully observed distributions, while those with `` $\wh{}$ '' refer to estimators for which the distributions, $\Xi$ and $\Yi$, are not fully observed and only samples of measurements drawn from the distributions are available.} 

Given independent realizations $\{(\Xi,\Yi)\}_{i=1}^n$ of $(\X,\Y)$, we first consider an oracle estimator for the regression \update{operator $\regOp$}, where we initially assume that $\{(\Xi,\Yi)\}_{i=1}^n$ are fully observed. 
First of all, the empirical Fr\'echet means are well-defined and unique due to the fact that we work in Hadamard spaces. 
Specifically, \update{replacing the expectation in \eqref{eq:Fmean} by that with respect to the empirical measure based on $\{(\Xi,\Yi)\}_{i=1}^n$ gives} 
\bgt\label{eq:oracleFrechetMean}
\FmeanXoracle=\underset{\mu\in\manifold}{\arg\min} \sum_{i=1}^{n}\wdist^2(\Xi,\mu)\quad\text{and}\quad \FmeanYoracle=\underset{\mu\in\manifold}{\arg\min} \sum_{i=1}^{\infty}\wdist^2(\Yi,\mu),\egt
where the corresponding quantile functions are the empirical means of quantile functions across the sample, 
\bgt\label{eq:quantile_oracleFrechetMean}
\quantileFmeanXoracle (\cdot) = \frac 1 n\s1n \quantileXi (\cdot) \quad\text{and}\quad \quantileFmeanYoracle (\cdot) = \frac 1 n\s1n \quantileYi (\cdot),\egt
and the corresponding distribution functions are given by right continuous inverses of the quantile functions as in \eqref{eq:cdfFromQuantile}. 
Then the log transforms $\Log_{\FmeanX}\Xi$ and $\Log_{\FmeanY} \Yi$ admit estimates $\Log_{\FmeanXoracle}\Xi$ and $\Log_{\FmeanYoracle}\Yi$. 
\update{The covariance operators $\covopX$ and $\covopY$ can be estimated by $\covopXoracle = n\inv\sum_{i=1}^n  \Log_{\FmeanXoracle} \Xi \otimes \Log_{\FmeanXoracle} \Xi$ and $\covopYoracle= n\inv\sum_{i=1}^n \Log_{\FmeanYoracle} \Yi \otimes \Log_{\FmeanYoracle} \Yi$. 
	We denote the eigenvalues and eigenfunctions of $\covopXoracle$ and $\covopYoracle$ by $\egnvalXoracle_j$ and $\egnfctnXoracle_j$, respectively by $\egnvalYoracle_k$ and $\egnfctnYoracle_k$, where the eigenvalues are in non-ascending order.  
	The cross-covariance operator $\crosscovopXY$ can be estimated by $\crosscovopXYoracle = n\inv\sum_{i=1}^n  \Log_{\FmeanYoracle} \Yi \otimes \Log_{\FmeanXoracle} \Xi$.}

Due to the compactness of $\covopX$, its inverse is not bounded, leading to an ill-posed problem \citep[e.g.,][]{he:03,wang:16}. 
Regularization is thus needed and can be achieved  through truncation. 
\update{Oracle estimators for the regression coefficient function $\regKernel$ and regression operator $\regOp$ are 
	\bgt\label{eq:regOpOracle}
	\regKernelOracle = \sum_{k=1}^{\nFPCY}\sum_{j=1}^{\nFPCX} \regCoefOracle_{jk} \egnfctnYoracle_k \otimes \egnfctnXoracle_j\quad\text{and}\quad 
	\regOpOracle \arbiFctn(t) = \innerprod{\arbiFctn}{\regKernelOracle(\cdot,t)}_{\FmeanXoracle},\ \text{for }\arbiFctn\in \Log_{\FmeanXoracle}\manifold,\ t\in\dom, \egt}
where $\regCoefOracle_{jk} = \egnvalXoracle_j\inv \crovalOracle_{jk}$, 
with $\crovalOracle_{jk} = n\inv\sum_{i=1}^n \innerprod{\Log_{\FmeanXoracle}\Xi} {\egnfctnXoracle_j} _{\FmeanXoracle} \innerprod{\Log_{\FmeanYoracle}\Yi} {\egnfctnYoracle_k} _{\FmeanYoracle}$, and $\nFPCX$ and $\nFPCY$ are the truncation bounds, i.e., the numbers of included eigenfunctions. 

Furthermore, we can construct an estimator based on the distribution estimation in Section~\ref{sec:denEst} which will be applicable in practical situations,  where typically $\Xi$ and $\Yi$ are observed in the form of samples generated from $\Xi$ and $\Yi$. 
Denote the estimated quantile functions by $\quantileXiEst$ and $\quantileYiEst$, respectively. Then the quantile functions of the empirical Fr\'echet means $\FmeanXest$ and $\FmeanYest$ of $\XiEst$ and $\YiEst$ for $i=1,\ldots,n$ are given by 
\bgt\label{eq:quantile_practicalFrechetMean}
\quantileFmeanXest(\cdot) = \frac 1 n\s1n \quantileXiEst(\cdot)  \quad\text{and}\quad \quantileFmeanYest(\cdot) = \frac 1 n\s1n \quantileYiEst(\cdot),\egt
and the corresponding distribution functions $\cdfFmeanXest$ and $\cdfFmeanYest$ can be obtained by right continuous inversion as per \eqref{eq:cdfFromQuantile}. 
Replacing $\Xi$ and $\Yi$ by the corresponding estimates $\XiEst$ and $\YiEst$, we can analogously obtain the estimates for the covariance operators, 
$\covopXest = n\inv\sum_{i=1}^n \Log_{\FmeanXest} \XiEst\otimes \Log_{\FmeanXest} \XiEst$ and 
$\covopYest = n\inv\sum_{i=1}^n \Log_{\FmeanYest} \YiEst \otimes \Log_{\FmeanYest} \YiEst$, \update{as well as the estimate for the cross-covariance operator, $\crosscovopXYest = n\inv\sum_{i=1}^n  \Log_{\FmeanYest} \YiEst \otimes \Log_{\FmeanXest} \XiEst$. 
	We denote the eigenvalues and eigenfunctions of $\covopXest$ and $\covopYest$ by $\egnvalXest_j$ and $\egnfctnXest_j$, respectively by $\egnvalYest_k$ and $\egnfctnYest_k$, where the eigenvalues are in non-ascending order. 
	Data-based estimators of the regression coefficient function $\regKernel$ and regression operator $\regOp$ in \eqref{eq:linopt_coef} are then 
	\be\label{eq:regOpEst}
	\regKernelEst = \sum_{k=1}^{\nFPCY}\sum_{j=1}^{\nFPCX} \regCoefEst_{jk}  \egnfctnYest_k \otimes \egnfctnXest_j,\quad\text{and} \quad \regOpEst \arbiFctn(t) = \innerprod{\arbiFctn}{\regKernelEst(\cdot,t)}_{\FmeanXest},\ \text{for }\arbiFctn\in\Log_{\FmeanXest}\manifold,\, t\in\dom,\ee}
where $\regCoefEst_{jk} = \egnvalXest_j\inv \crovalEst_{jk}$, 
and $\crovalEst_{jk} = n\inv\sum_{i=1}^n \innerprod{\Log_{\FmeanXest}\XiEst} {\egnfctnXest_j} _{\FmeanXest} \innerprod{\Log_{\FmeanYest}\YiEst} {\egnfctnYest_k} _{\FmeanYest}$. 

\update{Regarding the numbers of eigenfunctions included, $\nFPCX$ and $\nFPCY$, we note that larger values of $\nFPCX$ and $\nFPCY$ lead to smaller bias but larger variance and potential overfitting.} 
We discuss the selection of $\nFPCX$ and $\nFPCY$ further in  Section~\ref{sec:tuning} in the Supplementary Material. 

\revtwo{While this paper focuses on univariate distributions, we note that the proposed method in principle can be extended to the multivariate setting, where however the optimal maps and hence the log maps in general do not have closed-form expressions and the estimation is completely different from the univariate setting. 
	In addition, the required determination of the optimal transport maps is fraught with numerical difficulties \citep{cutu:13}. This is in contrast to the univariate case, where optimal transports just require the computation of quantile functions.  Furthermore, the corresponding asymptotic analysis is also different from the univariate setting; in particular, the expression of the parallel transport does not hold in the multivariate case.} See Section~\ref{sec:disc} in the Supplementary Material for further discussion. 

\subsection{Parallel Transport}\label{sec:paraTrans}
\update{Note that the true regression operator, $\regOp\colon \tangentspace{\FmeanX} \ra \tangentspace{\FmeanY}$, and its estimators, $\regOpOracle\colon \tangentspace{\FmeanXoracle} \ra \tangentspace{\FmeanYoracle}$ and $\regOpEst\colon \tangentspace{\FmeanXest} \ra \tangentspace{\FmeanYest}$, are defined on different tangent spaces, which makes their comparison not so straightforward.} 
For this,  we employ parallel transport, which is a commonly used tool for data on manifolds \citep{yuan:12,lin:19,pete:19:2}.  
For two probability distributions $\distnOne,\distnTwo\in\manifold$, a parallel transport operator $\paraTrans_{\distnOne, \distnTwo}\colon \hilbert_{\distnOne}\ra \hilbert_{\distnTwo}$ can be defined between the entire Hilbert spaces $\hilbert_{\distnOne}$ and $\hilbert_{\distnTwo}$ by
\be\label{eq:paraTrans}
\paraTrans_{\distnOne,\distnTwo}\arbiFctn \coloneqq \arbiFctn \circ \quantileDistnOne\circ \cdfDistnTwo,\quad \text{for }\arbiFctn\in \hilbert_{\distnOne},
\ee
where $\quantileDistnOne$ and $\cdfDistnTwo$ are the quantile function of $\distnOne$ and cdf of $\distnTwo$, respectively. 
Assuming that $\distnOne$ is atomless, restricted to the tangent space $\tangentspace{\distnOne}$, the parallel transport operator $\paraTrans_{\distnOne,\distnTwo}|_{\tangentspace{\distnOne}}$ defines the parallel transport from tangent space $\tangentspace{\distnOne}$ to $\tangentspace{\distnTwo}$. 

Denote by $\hsSpace{\distnOne}{\distnTwo}$ the space of all Hilbert--Schmidt operators from $\tangentspace{\distnOne}$ to $\tangentspace{\distnTwo}$, for $\distnOne,\distnTwo\in\manifold$. 
With $\distnOne,\distnTwo, \distnOneVar, \distnTwoVar\in\manifold$ where $\distnOneVar$ and $\distnTwo$ are atomless, we can define a parallel transport operator $\paraTransOp_{(\distnOne,\distnTwo), (\distnOneVar,\distnTwoVar)}$ \update{from $\hsSpace{\distnOne}{\distnTwo}$ to $\hsSpace{\distnOneVar}{\distnTwoVar}$} by 
\bgt\label{eq:paraTransOp}
(\paraTransOp_{(\distnOne,\distnTwo), (\distnOneVar,\distnTwoVar)} \hsOp)\arbiFctn = \paraTrans_{\distnTwo,\distnTwoVar} (\hsOp( \paraTrans_{\distnOneVar,\distnOne}g ) ),\quad \text{for } g\in\tangentspace{\distnOneVar}\text{ and }\hsOp\in\hsSpace{\distnOne}{\distnTwo}.\egt
Denoting  the Hilbert--Schmidt norm on $\hsSpace{\distnOne}{\distnTwo}$ by $\hsNorm{\hsSpace{\distnOne}{\distnTwo}}{\cdot}$, for $\distnOne,\distnTwo\in\manifold$, 
properties of parallel transport operators $\paraTrans_{\distnOne,\distnTwo}$ and $\paraTransOp_{(\distnOne,\distnTwo), (\distnOneVar,\distnTwoVar)}$ that are relevant for the theory 
are listed in Proposition~\ref{prop:paraTrans} in Section~\ref{sec:propParaTrans} in the Supplementary Material. 
Given atomless distributions $\distnOne,\distnTwo,\distnOneVar,\distnTwoVar\in\manifold$, 
applying Proposition \ref{prop:paraTrans}, 
the discrepancy between operators $\hsOp\in\hsSpace{\distnOne}{\distnTwo}$ and $\hsOpVar\in\hsSpace{\distnOneVar}{\distnTwoVar}$ can be quantified in the space $\hsSpace{\distnOne}{\distnTwo}$ by 
$\hsNorm{\hsSpace{\distnOne}{\distnTwo}}{\paraTransOp_{(\distnOneVar,\distnTwoVar), (\distnOne,\distnTwo)}\hsOpVar - \hsOp}$. 

\subsection{Asymptotic Theory}\label{sec:theory_indep}

Our goal for the theory is to evaluate the performance of the estimated regression operators, $\regOpOracle$ and $\regOpEst$ as per \eqref{eq:regOpOracle} and \eqref{eq:regOpEst}, respectively. 
According to the discussion in Section~\ref{sec:paraTrans}, if the true Fr\'echet means $\FmeanX$ and $\FmeanY$ and their estimators are atomless,  the discrepancy between the estimated and true regression operators can be gauged by $\hsNorm{\hsSpace{\FmeanX}{\FmeanY}}{\paraTransOp_{(\FmeanXoracle,\FmeanYoracle), (\FmeanX,\FmeanY)} \regOpOracle - \regOp}$ and $\hsNorm{\hsSpace{\FmeanX}{\FmeanY}}{\paraTransOp_{(\FmeanXest,\FmeanYest), (\FmeanX,\FmeanY)} \regOpEst - \regOp}$, for $\regOpOracle$ and $\regOpEst$, respectively. 
To guarantee the atomlessness of $\FmeanX$ and $\FmeanY$ and their estimators $\FmeanXoracle$ and $\FmeanYoracle$, we assume 
\ben[label = (A\arabic*), resume = ass]
\item\label{ass:atomless} 
\update{With probability equal to 1, the random distributions $\X$ and $\Y$ are atomless.}
\een

\update{Let $\factor>1$ denote a constant. To derive the convergence rate of the estimators for the regression operator, $\regOpOracle$ and $\regOpEst$,} we require the following conditions regarding the variability of $\X$ and $\Y$, the spacing of the eigenvalues $\egnvalX_j$ and $\egnvalY_k$, and the decay rates of the coefficients $\regCoef_{jk}$. Conditions of this type are standard in traditional functional linear regression \citep[e.g.,][]{hall:07:1}. 
\ben[label = (A\arabic*), resume = ass]
\item\label{ass:variation}
\update{$\expect(\|\Log_{\FmeanX}\X\|_{\FmeanX}^4)<\infty$, and $\expect(\innerprod{\Log_{\FmeanX}\X}{\egnfctnX_j}_{\FmeanX}^4) \le \factor \egnvalX_j^2$, for all $j\ge 1$; $\expect(\|\Log_{\FmeanY}\Y\|_{\FmeanY}^4)<\infty$, and $\expect(\innerprod{\Log_{\FmeanY}\Y}{\egnfctnY_k}_{\FmeanY}^4) \le \factor \egnvalY_k^2$, for all $k\ge 1$.}
\item\label{ass:eigenval_spacingX}
For $j\ge 1$, $\egnvalX_j - \egnvalX_{j+1} \ge \factor\inv j^{-\egnvalSpX - 1}$, where $\egnvalSpX\ge 1$ is a constant.
\item\label{ass:eigenval_spacingY}
For $k\ge 1$, $\egnvalY_k - \egnvalY_{k+1} \ge \factor\inv k^{-\egnvalSpY - 1}$, where $\egnvalSpY>0$ is a constant.
\item\label{ass:coef_decay}
For $j,k\ge 1$, $|\regCoef_{jk}|\le \factor j^{-\coefDecayX} k^{-\coefDecayY}$, where $\coefDecayX > \egnvalSpX+1$ and $\coefDecayY > 1$ are constants.
\een
Note that \ref{ass:coef_decay} implies \eqref{eq:coef_cvg}. 
\update{Furthermore, for $\nFPCX$ and $\nFPCY$ in \eqref{eq:regOpOracle} and \eqref{eq:regOpEst}, we assume 
	\ben[label = (A\arabic*), resume = ass]
	\item\label{ass:nFPC_rate} 
	$n\inv\nFPCX^{2\egnvalSpX+2}\ra 0$, $n\inv\nFPCY^{2\egnvalSpY+2}\ra 0$, as $n\ra\infty$.
	\een}

\update{Let $\distnfam = \distnfam(\factor,\egnvalSpX,\egnvalSpY,\coefDecayX,\coefDecayY)$ denote the set of distributions $\jointdistn$ of $(\X,\Y)$ that satisfy \ref{ass:convex} and \ref{ass:atomless}--\ref{ass:coef_decay}. \revtwo{Defining the sequence 
		\bal\nn
		\enFPCY(n) &= \enFPCY(n;\egnvalSpX,\egnvalSpY,\coefDecayX,\coefDecayY)\\
		&= \left\{\begin{array}{ll}
			\min\left\{n^{\max\{2\coefDecayX/(2\egnvalSpY+3),(4\coefDecayX-1)/(2\egnvalSpY+2\coefDecayY+2)\}/(\egnvalSpX+2\coefDecayX)},\, n^{1/(2\egnvalSpY+3)}\right\}, & \text{if } \coefDecayY-\egnvalSpY\le 1,\\
			\min\left\{n^{\max\{2\coefDecayX/(2\egnvalSpY+3),(4\coefDecayX-1)/(2\egnvalSpY+2\coefDecayY+2)\}/(\egnvalSpX+2\coefDecayX)}, (n/\log n)^{1/(2\egnvalSpY+3)}\right\}, & \text{if } \coefDecayY-\egnvalSpY\in(1,3/2],\\
			\min\left\{n^{\max\{2\coefDecayX/(2\egnvalSpY+3),(4\coefDecayX-1)/(2\egnvalSpY+2\coefDecayY+2)\}/(\egnvalSpX+2\coefDecayX)}, n^{1/(2\coefDecayY)}\right\}, & \text{if } \coefDecayY-\egnvalSpY>3/2,\\
		\end{array}\right.\eal
		then} when distributions $\Xi$ and $\Yi$ are fully observed, we obtain}
\revtwo{
	\bthm\label{thm:rate_oracle}
	Assume \ref{ass:convex} and \ref{ass:atomless}--\ref{ass:nFPC_rate}. If $\nFPCX\sim n^{1/(\egnvalSpX+2\coefDecayX)}$ and $\nFPCY \sim \enFPCY(n)$, as $n\ra\infty$,  then 
	\bgt\label{eq:rate_oracle}
	\lim_{\largeConst\ra\infty} \limsup_{n\ra\infty} \sup_{\jointdistn\in\distnfam} \prob_{\jointdistn}\left(\hsNormSq{\hsSpace{\FmeanX}{\FmeanY}}{\paraTransOp_{(\FmeanXoracle,\FmeanYoracle), (\FmeanX,\FmeanY)} \regOpOracle - \regOp} > \largeConst \indepRate(n)\right) = 0, \egt
	where 
	\bal\label{eq:rate_oracle_special}
	\indepRate(n) = \max\left\{n^{-(2\coefDecayX-1)/(\egnvalSpX+2\coefDecayX)},\enFPCY(n)^{-(2\coefDecayY-1)}\right\}.\eal 
	\ethm
}

\update{We note that $\indepRate(n) = n^{-(2\coefDecayX-1)/(\egnvalSpX+2\coefDecayX)}$ in \eqref{eq:rate_oracle_special} if either of the following holds:  $\coefDecayY-\egnvalSpY\le 1$ and $4\coefDecayX(\egnvalSpY-\coefDecayY+2)\le 2\egnvalSpY+3\le (2\coefDecayY-1)(\egnvalSpX+2\coefDecayX)/(2\coefDecayX-1)$; 
	or $1<\coefDecayY-\egnvalSpY\le 3/2$ and $4\coefDecayX(\egnvalSpY-\coefDecayY+2)\le 2\egnvalSpY+3< (2\coefDecayY-1)(\egnvalSpX+2\coefDecayX)/(2\coefDecayX-1)$;
	or $\coefDecayY-\egnvalSpY>3/2$ and $\coefDecayY\ge\max\{\egnvalSpY+2 - (2\egnvalSpY+3)/(4\coefDecayX),\, (\egnvalSpX+2\coefDecayX)/(2\egnvalSpX+2)\}$.}
\revtwo{In this case, $\regOpOracle$ achieves the same rate as the minimax rate for function-to-scalar linear regression \citep{hall:07:1} and function-to-function linear regression \citep[following similar arguments as in the proof of  Theorem~3 of][]{imai:18}.}



\update{Next, we consider the case where the distributions $\Xi$ and $\Yi$ are not fully observed. In addition, we require an assumption regarding the number of measurements per distribution and a uniform Lipschitz condition on the estimated cdfs to guarantee the atomlessness of the estimated Fr\'echet means $\FmeanXest$ and $\FmeanYest$ and hence to justify the use of $\hsNorm{\hsSpace{\FmeanX}{\FmeanY}}{\paraTransOp_{(\FmeanXest,\FmeanYest), (\FmeanX,\FmeanY)} \regOpEst - \regOp}$ as a measure of the estimation error.} 
\ben[label = (A\arabic*), resume = ass]
\item \label{ass:nObsPerDen_rate}
\update{For $\denRate_{\nDp}$ in \ref{ass:distn_est_rate}, $\denRate_{\nDp} \le \factor \min\{n\inv\nFPCX^{-\egnvalSpX}, n\inv\nFPCY\inv\}$, for all $n$.}
\item \label{ass:atomlessFmeanEst} 
\update{For any atomless distribution $\arbiDistn\in\manifold$, the corresponding estimate $\arbiDistnEst$ based on a sample of measurements drawn according to $\arbiDistn$ is also atomless\note{\ with probability 1}.}
\een
\update{For example, with $\nFPCX\sim n^{1/(\egnvalSpX+2\coefDecayX)}$ and $\nFPCY\sim\enFPCY(n)$ as in Theorem~\ref{thm:rate_oracle}, \ref{ass:nObsPerDen_rate} holds with $\nDp\sim \max\{n^{3(\egnvalSpX+\coefDecayX)/(\egnvalSpX+2\coefDecayX)},\,n^{3/2}\enFPCY(n)^{3/2}\}$ and $\nDp\sim \max\{n^{4(\egnvalSpX+\coefDecayX)/(\egnvalSpX+2\coefDecayX)},\,n^2\enFPCY(n)^2\}$ for the estimators proposed by \citet{pete:16:1} and \citet{pana:16}, respectively. 
	We note that these two estimators also satisfy \ref{ass:atomlessFmeanEst}. 
	Then we find that the data-based estimator $\regOpEst$ achieves the same rate as the estimator $\regOpOracle$ based on fully observed distributions as shown in Theorem~\ref{thm:rate_oracle}. 
}
\bthm\label{thm:rate_realistic}
If \ref{ass:convex}--\ref{ass:atomlessFmeanEst} hold \revtwo{and choosing $\nFPCX$ and $\nFPCY$ as in Theorem~\ref{thm:rate_oracle}}, then 
\bgt\label{eq:rate_realistic}
\lim_{\largeConst\ra\infty} \limsup_{n\ra\infty} \sup_{\jointdistn\in\distnfam} \prob_{\jointdistn}\left(\hsNormSq{\hsSpace{\FmeanX}{\FmeanY}}{\paraTransOp_{(\FmeanXest,\FmeanYest), (\FmeanX,\FmeanY)} \regOpEst - \regOp} > \largeConst \indepRate(n)\right) = 0. \egt
\ethm

\update{We note that while the proposed method is based on function-to-function linear regression, the asymptotic analysis is more involved. The proofs of Theorems~\ref{thm:rate_oracle} and \ref{thm:rate_realistic} are based on the geometry of the Wasserstein space, since we are not dealing with general functions in $\hilbert$ space (with respect to the Lebesgue measure) as in functional data analysis but rather the log maps. 
	In particular, we do not assume additive noise in the proposed model in \eqref{eq:d2dreg}. 
	Furthermore, parallel transport maps are employed to quantify the estimation discrepancy of the estimators of the regression operator, $\regOp$, the covariance and cross-covariance operators, $\covopX$, $\covopY$ and $\crosscovopXY$, and the eigenfunctions, $\egnfctnX_j$ and $\egnfctnY_k$. 
	All of these create additional complexities for the theoretical derivations. 
	For Theorem~\ref{thm:rate_realistic}, the distributions $\Xi$ and $\Yi$ are not be fully observed and instead only data samples  drawn from these distributions are available. Hence, we need to deal with two layers of stochastic mechanisms: The first layer generates random elements $(\Xi,\Yi)$ taking values in $\manifold\times\manifold$; the second layer generates random samples according to $\Xi$ and $\Yi$. Specifically, we need to tackle the discrepancy between  the estimated distributions based on the observed data $\XiEst$ and $\YiEst$ and the actual underlying distributions $\Xi$ and $\Yi$. 
}

Theorems~\ref{thm:rate_oracle} and \ref{thm:rate_realistic} entail the following corollaries on the prediction of $\Y$ based on $\X$, \update{where the target is the conditional Fr\'echet mean of $\Y$ given $\X$, i.e., $\expect_\oplus(\Y|\X)\coloneqq \argmin_{\arbiDistnVar\in\manifold}\expect[\wdist^2(\Y,\arbiDistnVar) \mid \X] = \Exp_{\FmeanY}[\expect(\Log_{\FmeanY}\Y\mid\Log_{\FmeanX}\X)]$}. 
In the following, for any given $\arbiDistn\in\manifold$, the corresponding estimate $\arbiDistnEst$ is assumed to be based on a sample of $\nDp_{\arbiDistn}\ge \nDp$ observations drawn from $\arbiDistn$, where $\nDp$ is the lower bound of the number of observations per distribution as per \ref{ass:nObsPerDens}. 
We denote the prediction of $\Y$ based on fully observed distributions by $\YpredOracle(\arbiDistn) \coloneqq \Exp_{\FmeanYoracle} [\regOpOracle( \Log_{\FmeanXoracle} \arbiDistn)]$, and the prediction based on samples generated from the distributions by $\YpredEst(\arbiDistnEst) \coloneqq \Exp_{\FmeanYest} [\regOpEst( \Log_{\FmeanXest} \arbiDistnEst)]$, where $\regOpOracle$ and $\regOpEst$ are as per \eqref{eq:regOpOracle} and \eqref{eq:regOpEst}, respectively.
\update{
	\begin{cor}\label{cor:pred_oracle} 
		Under the assumptions of Theorem~\ref{thm:rate_oracle}, 
		\bgt\label{eq:pred_oracle}
		\lim_{\largeConst\ra\infty} \limsup_{n\ra\infty} \sup_{\jointdistn\in\distnfam} \prob_{\jointdistn}\left( \wdist^2\left(\YpredOracle(\arbiDistn),  \expect_\oplus(\Y\mid\X = \arbiDistn)\right) > \largeConst\indepRate(n)\right) = 0.\egt
	\end{cor}
	\begin{cor}\label{cor:pred_realistic} 
		Under the assumptions of Theorem~\ref{thm:rate_realistic}, 
		\bgt\nn
		\lim_{\largeConst\ra\infty} \limsup_{n\ra\infty} \sup_{\jointdistn\in\distnfam} \prob_{\jointdistn}\left( \wdist^2\left(\YpredEst(\arbiDistnEst),  \expect_\oplus(\Y\mid\X = \arbiDistn)\right) > \largeConst \indepRate(n)\right) = 0.\egt
	\end{cor}
}
For the proofs, see Section~\ref{sec:proof_indep} in the Supplementary Material.
\update{We further discuss the estimation and theoretical analysis for the distribution-to-scalar regression model as per \eqref{eq:d2sreg} in Section~\ref{sec:d2sEst} in the Supplementary Material, where we show that the estimates of the regression coefficient function $\regKernelDtoS$ achieve the same rate as the minimax rate for the function-to-scalar linear regression based on fully observed predictor functions; see \citet{hall:07:1}. }




\section[Autoregressive Models]{Autoregressive Models for Distribution-Valued Time Series}\label{sec:dts}

\update{Here we consider a distribution-valued time series $\{\Zi\}_{i\in\intgSet}$, each element taking values in $\manifold$. We assume that the random process $\{\Zi\}_{i\in\intgSet}$ is stationary in the sense that
	\ben[(\arabic*)]
	\item $\Zi$ are square integrable, i.e., $\expect\wdist^2(\arbiDistn,\Zi)<\infty$ for some (and thus for all) $\arbiDistn\in\manifold$; 
	\item $\Zi$ have a common Fr\'echet mean $\FmeanZ$ that is atomless, i.e, $\FmeanZ = \argmin_{\arbiDistn\in\manifold}\expect\wdist^2(\arbiDistn,\Zi)$, for all $i\in\intgSet$; 
	\item The autocovariance operators $\expect(\Log_{\FmeanZ}\ZiPluslag\otimes\Log_{\FmeanZ}\Zi)$ do not depend on $i\in\intgSet$, which are hence denoted by $\covopZlagGen$, for all $\lag\in\intgSet$. 
	\een
	For $\{\Zi\}_{i\in\intgSet}$}, we \update{assume a first order autoregressive model which is an extension of the distribution-to-distribution regression model} in \eqref{eq:d2dreg}
\bgt\label{eq:autoreg}
\Log_{\FmeanZ}\ZiPlusOne =\regOp(\Log_{\FmeanZ}\Zi) + \noise_{i+1},\text{ for }i\in\intgSet. \egt
Here, $\regOp\colon \tangentspace{\FmeanZ} \ra \tangentspace{\FmeanZ}$ is a linear operator defined as
\be\label{eq:autoRegOp}
\regOp \arbiFctn(t) = \innerprod{\regKernel(\cdot,t)}{\arbiFctn}_{\FmeanZ},\quad \text{for } t\in\dom, \text{ and } \arbiFctn\in\tangentspace{\FmeanZ},\ee
where $\regKernel:\dom^2\ra\real$ is the auto-regression coefficient kernel lying in $\hilbert_{\FmeanZ\times \FmeanZ}$, and $\{\noise_i\}_{i\in\intgSet}$ are i.i.d. random elements taking values in the tangent space $\tangentspace{\FmeanZ}$ such that $\expect(\noise_i) = 0$ and $\expect\|\noise_i\|_{\FmeanZ}^2 < \infty$. 
Similar models have been previously studied in the seminal work of \citet{bosq:00}. 
\update{To ensure the existence and uniqueness of such a stationary process, we assume 
	\ben[label = (B\arabic*), series = assTS]
	\item\label{ass:autoRegOpConv} There exists an integer $\autoBound\ge 1$ such that $\|\regOp^{\autoBound}\|_{\hilbert_{\FmeanZ}} < 1$. 
	\een
	Here, $\|\cdot\|_{\hilbert_{\FmeanZ}}$ denotes the sup norm for linear operators on $\hilbert_{\FmeanZ}$ and we define $\regOp^{\autoBound}$ by induction, $\regOp^{k}(\cdot) = \regOp[\regOp^{k-1}(\cdot)]$, for any integer $k>1$. 
	We note that under \ref{ass:autoRegOpConv}, \eqref{eq:autoreg} has a unique stationary solution given by
	\bal\label{eq:autoRegSol}
	\Log_{\FmeanZ}\Zi = \sum_{\lag=0}^\infty\regOp^{\lag}(\noise_{i-\lag}),\eal
	where $\regOp^0(\noise_i) \coloneqq \noise_i$ and the right hand side converges in mean square,  $\lim_{n\ra\infty}\expect\|\sum_{\lag=n}^\infty\regOp^{\lag}(\noise_{i-\lag})\|_{\FmeanZ}^2$ $=0$, and also almost surely, i.e., $\lim_{n\ra\infty}\|\sum_{\lag=n}^\infty\regOp^{\lag}(\noise_{i-\lag})\|_{\FmeanZ}=0$ with probability 1 \citep[Theorem~3.1,][]{bosq:00}. 
}
Furthermore, we assume
\ben[label = (B\arabic*), resume = assTS]
\item\label{ass:convex_ts} With probability 1, $\sum_{\lag=0}^\infty\regOp^{\lag}(\noise_{-\lag}) + \id$ is non-decreasing. 
\een
\update{Assumption~\ref{ass:convex_ts} guarantees that the right hand side of \eqref{eq:autoRegSol} lies in $\Log_{\FmeanZ}\manifold$ a.s. We further provide a fully detailed example of a stationary process $\{\Zi\}_{i\in\intgSet}$ that satisfies the autoregressive model as per \eqref{eq:autoreg} in Section~\ref{sec:eg-WAR} in the Supplementary Material.} 

As in Section~\ref{sec:d2dEst}, we have $\expect (\Log_{\FmeanZ}\Zone) = 0$, $\FmeanZ$-almost surely. 
The operator $\covopZlagZero$ admits the eigendecomposition
\bgt\nn
\covopZlagZero = \sum_{j=1}^\infty \egnvalZ_j \egnfctnZ_j \otimes \egnfctnZ_j, \egt
with eigenvalues $\egnvalZ_1\ge \egnvalZ_2\ge\cdots\ge 0$ and orthonormal eigenfunctions $\{\egnfctnZ_j\}_{j=1}^\infty$ in $\tangentspace{\FmeanZ}$. 
With probability 1, the logarithmic transforms $\Log_{\FmeanZ}\Zi$ admit the expansion
\bgt\nn
\Log_{\FmeanZ}\Zi = \sum_{j=1}^{\infty} \innerprod{\Log_{\FmeanZ}\Zi}{\egnfctnZ_j}_{\FmeanZ} \egnfctnZ_j, \quad i\in\intgSet,\egt
and hence 
$\covopZlagOne = \sum_{l=1}^\infty \sum_{j=1}^\infty \crovalZlagOne_{jl} \egnfctnZ_l \otimes \egnfctnZ_j$, 
where $\crovalZlagOne_{jl} = \expect(\innerprod{\Log_{\FmeanZ}\Zone}{\egnfctnZ_j}_{\FmeanZ} \innerprod{\Log_{\FmeanZ}\Ztwo}{\egnfctnZ_l}_{\FmeanZ})$. 
With $\regCoef_{jl} = \egnvalZ_j\inv \crovalZlagOne_{jl}$, the auto-regression coefficient function can then be expressed as 
\be
\regKernel = \sum_{l=1}^\infty \sum_{j=1}^\infty \regCoef_{jl} \egnfctnZ_{l}\otimes  \egnfctnZ_j.\nn
\ee

For the estimation of the operator $\regOp$ in \eqref{eq:autoRegOp}, first considering a fully observed sequence of length $n$, $\Zone,\Ztwo,\ldots,\Zn$, with the oracle estimator of the Fr\'echet mean $\FmeanZoracle$ defined analogously to \eqref{eq:oracleFrechetMean}, 
the autocovariance operators $\covopZlagZero$ and $\covopZlagOne$ can be estimated by their empirical counterparts
$\covopZlagZeroOracle = n\inv\sum_{i=1}^n \Log_{\FmeanZoracle} \Zi \otimes \Log_{\FmeanZoracle} \Zi$ and $\covopZlagOneOracle= (n-1)\inv\sum_{i=1}^{n-1} \Log_{\FmeanZoracle} \ZiPlusOne \otimes \Log_{\FmeanZoracle} \Zi$. We denote the eigenvalues and eigenfunctions of $\covopZlagZeroOracle$ by $\egnvalZoracle_j$ and $\egnfctnZoracle_j$, respectively, where the eigenvalues $\egnvalZoracle_j$ are in non-ascending order. 
Then \update{oracle estimators for the auto-regression coefficient function $\regKernel$ and operator $\regOp$ in \eqref{eq:autoRegOp} are }
\bgt\label{eq:autoRegOpOracle}
\regKernelOracle = \sum_{l=1}^{\nFPCZ}\sum_{j=1}^{\nFPCZ} \regCoefOracle_{jl} \egnfctnZoracle_l \otimes \egnfctnZoracle_j, \quad\text{and}\quad 
\regOpOracle \arbiFctn(t) = \innerprod{\arbiFctn}{\regKernelOracle(\cdot,t)}_{\FmeanZoracle},\ \text{for }\arbiFctn\in \Log_{\FmeanZoracle}\manifold,\ t\in\dom,\egt
where $\regCoefOracle_{jl} = \egnvalZoracle_j\inv \crovalZlagOneOracle_{jl}$, $\crovalZlagOneOracle_{jl} = (n-1)\inv\sum_{i=1}^{n-1} \innerprod{\Log_{\FmeanZoracle}\Zi}{\egnfctnZoracle_j}_{\FmeanZoracle} \innerprod{\Log_{\FmeanZoracle}\ZiPlusOne}{\egnfctnZoracle_l}_{\FmeanZoracle}$, and $\nFPCZ$ is the truncation bound. 

As discussed for the independent case in Section~\ref{sec:coef_est_indep}, a realistic estimator $\regKernelEst$ for $\regKernel$ based on the distribution estimation discussed in Section~\ref{sec:denEst} can be obtained by replacing $\Zi$ and $\FmeanZ$ with the corresponding estimates $\ZiEst$ and $\FmeanZest$, the latter analogous to \eqref{eq:quantile_practicalFrechetMean}. 
Specifically, estimates for the autocovariance operators with corresponding decompositions are given by 
$\covopZlagZeroEst = n\inv\sum_{i=1}^n \Log_{\FmeanZest} \ZiEst \otimes \Log_{\FmeanZest} \ZiEst$ and 
$\covopZlagOneEst = (n-1)\inv\sum_{i=1}^{n-1} \Log_{\FmeanZest} \ZiPlusOneEst \otimes \Log_{\FmeanZest} \ZiEst$. 
We denote the eigenvalues and eigenfunctions of $\covopZlagZeroEst$ by $\egnvalZest_j$ and $\egnfctnZest_j$, respectively, where the eigenvalues $\egnvalZest_j$ are in non-ascending order. 
With $\regCoefEst_{jl} = \egnvalZest_j\inv \crovalZlagOneEst_{jl}$ and $\crovalZlagOneEst_{jl} = (n-1)\inv\sum_{i=1}^{n-1} \innerprod{\Log_{\FmeanZest}\ZiEst}{\egnfctnZest_j}_{\FmeanZest} \innerprod{\Log_{\FmeanZest}\ZiPlusOneEst}{\egnfctnZest_l}_{\FmeanZest}$, data-based estimators for the auto-regression coefficient function $\regKernel$ and operator $\regOp$ in \eqref{eq:autoRegOp} are then given by
\bgt\label{eq:autoRegOpEst} 
\regKernelEst = \sum_{l=1}^\nFPCZ \sum_{j=1}^\nFPCZ \regCoefEst_{jl}  \egnfctnZest_{l} \otimes \egnfctnZest_j,\quad\text{and}\quad
\regOpEst \arbiFctn(t) = \innerprod{\arbiFctn}{\regKernelEst(\cdot,t)}_{\FmeanZest},\ \text{for }\arbiFctn\in\Log_{\FmeanZest}\manifold,\ t\in\dom.\egt

We first focus on the case where the distributions are fully observed. To derive the convergence rate of the estimator $\regOpOracle$ in \eqref{eq:autoRegOpOracle}, we require the following assumptions analogous to the independent case in Section~\ref{sec:d2dEst}. Let $\factor>1$ be a constant. 
\ben[label = (B\arabic*), resume = assTS]
\item\label{ass:atomless_ts}
\update{With probability 1, the distributions $\Zi$ are all atomless.}
\item\label{ass:variation_ts}
\update{$\expect(\|\Log_{\FmeanZ}\Zi\|_{\FmeanZ}^4)<\infty$, and $\expect(\innerprod{\Log_{\FmeanZ}\Zi}{\egnfctnZ_j}_{\FmeanZ}^4) \le \factor \egnvalZ_j^2$, for all $j\ge 1$.}
\item\label{ass:eigenval_spacingZ}
For $j\ge 1$, $\egnvalZ_j - \egnvalZ_{j+1} \ge \factor\inv j^{-\egnvalSpZ - 1}$, where $\egnvalSpZ\ge 1/2$ is a constant.
\item\label{ass:coef_decay_ts}
For $j,l\ge 1$, $|\regCoef_{jl}|\le \factor j^{-\coefDecayZx} l^{-\coefDecayZy}$, where $\coefDecayZx > \egnvalSpZ+1$ and $\coefDecayZy >1$ are constants.
\item\label{ass:nFPC_rate_ts}
\update{$n\inv\nFPCZ^{2\egnvalSpZ+2}\ra 0$, as $n\ra\infty$.}
\een

\update{Let $\vdistnfam = \vdistnfam(\factor,\egnvalSpZ,\coefDecayZx,\coefDecayZy)$ denote the set of distributions $\vjointdistn$ of the process $\{\Zi\}$ that satisfy \ref{ass:autoRegOpConv}--\ref{ass:coef_decay_ts}. Then we obtain 
	\revtwo{\bthm\label{thm:rate_oracle_ts}
		Assume \ref{ass:autoRegOpConv}--\ref{ass:nFPC_rate_ts}. If $\nFPCZ\sim\min\{n^{1/(2\egnvalSpZ+2\coefDecayZx+2\max\{2-\coefDecayZy,\,0\})},\,n^{1/(2\egnvalSpZ+2\max\{\coefDecayZy,\,2\})}\}$, then 
		\bal\nn
		\lim_{\largeConst\ra\infty} \limsup_{n\ra\infty} \sup_{\vjointdistn\in\vdistnfam} \prob_{\vjointdistn}\left( \hsNormSq{\hsSpace{\FmeanZ}{\FmeanZ}}{\paraTransOp_{(\FmeanZoracle,\FmeanZoracle), (\FmeanZ,\FmeanZ)} \regOpOracle - \regOp} > \largeConst \tsRate(n) \right) = 0,\eal
		where 
		\bal\label{eq:rate_oracle_ts_special}
		\tsRate(n) = \max\left\{n^{-(2\coefDecayZx-1)/(2\egnvalSpZ+2\coefDecayZx+2\max\{2-\coefDecayZy,\,0\})},\,n^{-(2\coefDecayZy-1)/(2\egnvalSpZ+2\max\{\coefDecayZy,\,2\})}\right\}.\eal
		\ethm}
	The convergence rate obtained for the estimator $\regOpOracle$ in Theorem~\ref{thm:rate_oracle_ts} is slower than the rate obtained for the independent case as per Theorem~\ref{thm:rate_oracle}. This is due to the serial dependence among $\Zi$ and with the special choice of $\nFPCZ$ as above is manifested by the fact that $\indepRate(n)$ as per \eqref{eq:rate_oracle_special} with $\egnvalSpX=\egnvalSpY$ is always smaller than $\tsRate(n)$ as per \eqref{eq:rate_oracle_ts_special}.
}
\update{Furthermore, regarding the estimator $\regOpEst$ in \eqref{eq:autoRegOpEst} where only samples drawn from the distributions $\Zi$ are available, we in addition make the following assumption of the numbers of measurements observed per distribution.}
\ben[label = (B\arabic*), resume = assTS]
\item \label{ass:nObsPerDens_ts}
There exists a sequence $\nDp = \nDp(n)$ such that for the number of measurements per distribution $\nDpZi$, $\min\{\nDpZi: i=1,2,\ldots,n\} \ge \nDp$ and $\nDp\ra\infty$ as $n\ra \infty$.
\item \label{ass:nObsPerDen_rate_ts}
\update{$\denRate_{\nDp}\le\factor n\inv$, for all $n$, where $\denRate_{\nDp}$ is as per \ref{ass:distn_est_rate}.}
\een
\update{For example, if distributions $\Zi$ are estimated via the methods used by \citet{pana:16} and \citet{pete:16:1}, in order to ensure \ref{ass:nObsPerDen_rate_ts}, it suffices to take $\nDp\sim n^2$ and $\nDp\sim n^{3/2}$, respectively. 
	Then we show that the estimator $\regOpEst$ in \eqref{eq:autoRegOpEst} converges with the same rate as $\regOpOracle$,  as shown in Theorem~\ref{thm:rate_oracle_ts}. 
	\bthm\label{thm:rate_realistic_ts}
	If \ref{ass:distn_est_rate}, 
	\ref{ass:atomlessFmeanEst} and \ref{ass:autoRegOpConv}--\ref{ass:nObsPerDen_rate_ts} hold \revtwo{and choosing $\nFPCZ$ as in Theorem~\ref{thm:rate_oracle_ts}}, then 
	\bal\nn 
	\lim_{\largeConst\ra\infty} \limsup_{n\ra\infty} \sup_{\vjointdistn\in\vdistnfam} \prob_{\vjointdistn}\left( \hsNormSq{\hsSpace{\FmeanZ}{\FmeanZ}}{\paraTransOp_{(\FmeanZest,\FmeanZest), (\FmeanZ,\FmeanZ)} \regOpEst - \regOp} > \largeConst \tsRate(n) \right) = 0.\eal
	\ethm
}

\update{As for the independent case, Theorems~\ref{thm:rate_oracle_ts} and \ref{thm:rate_realistic_ts} entail the following asymptotic results for the one-on-one prediction of $\ZnPlusOne$ given $\Zn$, where the target is the conditional Fr\'echet mean of $\ZnPlusOne$ given $\Zn$ by
	$\expect_\oplus(\ZnPlusOne\mid\Zn)\coloneqq\argmin_{\arbiDistnVar}\expect[\wdist^2(\ZnPlusOne,\arbiDistnVar)\mid\Zn] = \Exp_{\FmeanZ}[\expect(\Log_{\FmeanZ}\ZnPlusOne\mid \Log_{\FmeanZ}\Zn)]$. 
	For any given $\arbiDistn\in\manifold$, the corresponding estimate $\arbiDistnEst$ is assumed to be based on a sample of $\nDp_{\arbiDistn}\ge \nDp$ observations drawn from $\arbiDistn$, where $\nDp$ is the lower bound of the number of observations per distribution as per \ref{ass:nObsPerDens_ts}. 
	The prediction of $\ZnPlusOne$ based on fully observed distributions is given by $\Exp_{\FmeanZoracle}[\regOpOracle(\Log_{\FmeanZoracle}\arbiDistn)]$ and the prediction based on samples generated from the distributions by $\Exp_{\FmeanZest}[\regOpEst(\Log_{\FmeanZest}\arbiDistnEst)]$, where $\regOpOracle$ and $\regOpEst$ are as per \eqref{eq:autoRegOpOracle} and \eqref{eq:autoRegOpEst}, respectively. 
	Then these predictions achieve the same rate as the estimates of the regression operators in  Theorems~\ref{thm:rate_oracle_ts} and \ref{thm:rate_realistic_ts}. 
	\begin{cor}\label{cor:pred_oracle_ts} 
		Under the assumptions of Theorem~\ref{thm:rate_oracle_ts}, 
		\bgt\nn
		\lim_{\largeConst\ra\infty} \limsup_{n\ra\infty} \sup_{\vjointdistn\in\vdistnfam} \prob_{\vjointdistn}\left( \wdist^2(\Exp_{\FmeanZoracle}[\regOpOracle(\Log_{\FmeanZoracle}\arbiDistn)], \expect_\oplus(\Y\mid\X = \arbiDistn)) > \largeConst \tsRate(n) \right) = 0.\egt
	\end{cor}
	\begin{cor}\label{cor:pred_realistic_ts} 
		Under the assumptions of Theorem~\ref{thm:rate_realistic_ts}, 
		\bgt\nn
		\lim_{\largeConst\ra\infty} \limsup_{n\ra\infty} \sup_{\vjointdistn\in\vdistnfam} \prob_{\vjointdistn}\left(
		\wdist^2(\Exp_{\FmeanZest}[\regOpEst(\Log_{\FmeanZest}\arbiDistnEst)], \expect_\oplus(\ZnPlusOne\mid\Zn=\arbiDistn))> \largeConst \tsRate(n) \right) = 0.\egt
	\end{cor}
}
Proofs and auxiliary lemmas for this section are in Section~\ref{sec:proof_ts} in the Supplementary Material.

\section{Simulations}\label{sec:simu}
In practice, the fit of the logarithmic response may not fall in the logarithmic space with base point $\FmeanYest$, i.e.,
\be\label{eq:out-of-set-issue} \regOpEst(\Log_{\FmeanXest} \XiEst)\notin \Log_{\FmeanYest}\manifold,\ee
with $\regOpEst$ given in \eqref{eq:regOpEst}. This problem was already recognized by \citet{bigo:17:2}. 
If \eqref{eq:out-of-set-issue} happens, we employ a boundary projection method described in Section~\ref{sec:projection} in the Supplementary Material. 
We compared the performance of the proposed method implemented with boundary projection (referred to as projection method) with two other approaches. 
The first of these  is to employ an alternative to the proposed boundary projection for those situations where  the event \eqref{eq:out-of-set-issue} takes place,  which was proposed by \citet{caze:18} in the context of principal component analysis (PCA). 
This alternative to handle the problem extends the domains of the distributions. 
We use this method by fitting the proposed distribution-to-distribution regression model with distributions on an extended domain when the event \eqref{eq:out-of-set-issue} happens, and then normalize the fitted distributions by restricting them back to the original domain. We refer to this as the domain-extension method in the following. 
\update{The second alternative approach is the log quantile density (LQD) method \citep{pete:16:1}, where we apply function-to-function linear regression to the LQD transformations of distributions and map the fitted responses back to the Wasserstein space $\manifold$ through the inverse LQD transformation \citep{chen:19}. Specifically, we use the R package \texttt{fdadensity} \citep{fdadensity} for implementations of the LQD transformations. 
	To generate data for simulations, we provide the following framework to construct explicit examples, which also demonstrates the feasibility of the proposed model in \eqref{eq:d2dreg}.} 

\paragraph{Framework for Explicit Construction.}
\update{For $\dom = [0,1]$, we consider Fr\'echet mean distributions $\FmeanX,\FmeanY\in\manifold$ with bounded density functions, i.e., $\sup_{s\in\dom}\denFmeanX(s) < \infty$ and $\sup_{t\in\dom}\denFmeanY(t) < \infty$. 
	We consider a set of orthonormal functions $\{\basisL_j\}_{j=1}^\infty$ in the Lebesgue-square-integrable function space on $[0,1]$, $\hilbert([0,1])$,} such that the $\basisL_j$ are continuously differentiable with bounded derivatives, and $\basisL_j(0) = \basisL_j(1)$, for all $j\in\pint$. 
In particular, $\basisL_j$ can be taken as \bgt\label{eq:egbases}
\basisL_j(\argu) = \update{\sqrt{2}\sin(2\pi j\argu)}, \quad \text{for } \argu\in[0,1],\text{ and }j\in\pint.\egt

\update{Suppose 
	$\Log_{\FmeanX}\X$ admits the expansion $\Log_{\FmeanX}\X = \sum_{j=1}^\infty \frcoef_j\basisL_j\circ\cdfFmeanX$, where $\frcoef_j$ are uncorrelated random variables with zero mean such that 
	$\sum_{j=1}^\infty \frcoef_j^2 <\infty$ almost surely. 
	We define the regression operator $\regOp$ as 
	$\regOp\arbiFctn = \sum_{k=1}^{\infty} \sum_{j=1}^{\infty} \linCoef_{jk} \innerprod{\arbiFctn}{\basisL_j\circ\cdfFmeanX}_{\FmeanX}\basisL_k\circ\cdfFmeanY$, for $\arbiFctn\in\tangentspace{\FmeanX}$, with $\linCoef_{jk}\in\real$ such that $\sum_{j=1}^\infty\sum_{k=1}^\infty {\linCoef_{jk}}^2<\infty$. 
	Hence, $\regOp(\Log_{\FmeanX}\X) = \sum_{k=1}^\infty\sum_{j=1}^\infty \linCoef_{jk} \frcoef_j\basisL_k\circ\cdfFmeanY$. 
	To guarantee $\sum_{j=1}^\infty \frcoef_j\basisL_j\circ\cdfFmeanX\in\Log_{\FmeanX}\manifold$ and $\sum_{k=1}^\infty\sum_{j=1}^\infty \linCoef_{jk} \frcoef_j\basisL_k\circ\cdfFmeanY \in \Log_{\FmeanY}\manifold$,} it suffices to require 
\bgt\label{eq:mod_req} 
\left\{\begin{aligned}
	&\sum_{j=1}^\infty\frcoef_j\basisL'_j(\cdfFmeanX(s))\denFmeanX(s) + 1 \ge 0, \text{ for all }s\in\dom,\\
	&\sum_{k=1}^\infty\sum_{j=1}^\infty \linCoef_{jk} \frcoef_j\basisL'_k(\cdfFmeanY(t))\denFmeanY(t) + 1 \ge 0, \text{ for all }t\in\dom,\\
	&\sum_{j=1}^\infty\frcoef_j(\basisL_j\circ\cdfFmeanX)'\text{ and } \sum_{k=1}^\infty\sum_{j=1}^\infty \linCoef_{jk} \frcoef_j(\basisL_k\circ\cdfFmeanY)' \text{ uniformly converge},\end{aligned}\right. \quad\update{\text{a.s.}}\egt
Requirement \eqref{eq:mod_req} is satisfied, e.g., when  
\update{$|\frcoef_j|\le \convSeries_{1j} /(\sup_{\argu\in[0,1]}|\basisL'_j(\argu)| \sup_{s\in\dom}\denFmeanX(s) \sum_{j'=1}^\infty \convSeries_{1j'}) $ and $|\linCoef_{jk}\frcoef_j|\le \convSeries_{1j} \convSeries_{2k} / (\sup_{\argu\in[0,1]}|\basisL'_k(\argu)| \sup_{t\in\dom}\denFmeanY(t) \sum_{j'=1}^\infty \convSeries_{1j'} \sum_{k'=1}^\infty \convSeries_{2k'})$, a.s.,}
where $\{\convSeries_{1j}\}_{j=1}^\infty$ and $\{\convSeries_{2k}\}_{k=1}^\infty$ are two non-negative sequences such that $\sum_{j=1}^\infty \convSeries_{1j} < \infty$ and $\sum_{k=1}^\infty \convSeries_{2k} < \infty$, 
examples including $\{a^{-j}\}_{j=1}^\infty$ and $\{j^{-a}\}_{j=1}^\infty$, for any given $a>1$. 

With $\regOp(\Log_{\FmeanX}\X)$ and $\FmeanY$, the distributional response $\Y$ can be generated by adding distortions to $\Exp_{\FmeanY}(\regOp(\Log_{\FmeanX}\X))$ through push-forward maps, 
\update{i.e., $\Y = \distortmap\# \Exp_{\FmeanY}(\regOp(\Log_{\FmeanX}\X))$, where $\distortmap\colon \dom\ra\dom$ is a random distortion function independent of $\X$, such that $\distortmap$ is non-decreasing almost surely, and that $\expect[\distortmap(t)] = t$ almost everywhere on $\dom$. 
	This is a valid method to provide random distortions for distributions \citep{pana:16} in the sense that the conditional Fr\'echet mean of $\Y$ is on target, i.e., $\expect_\oplus(\Y|\X)\coloneqq\argmin_{\arbiDistn\in\manifold}\expect[\wdist^2(\Y,\arbiDistn)\mid \X] = \Exp_{\FmeanY}(\regOp(\Log_{\FmeanX}\X))$. 
	Furthermore, the pair $(\X,\Y)$ generated in this way satisfies our model in \eqref{eq:d2dreg}. 
	An example \citep{pete:19} of the random distortion function is $\distortmap = \distortmap_{A}$}, where $A$ is a random variable such that $\prob(A \le \argu) = \prob(A \ge -\argu)$ for any $\argu\in\real$ and $\prob(A = 0) = 0$, and $\distortmap_a$ 
is defined as
\bgt\label{eq:distort}
\distortmap_a(\argu) = \left\{\begin{array}{ll}
	\argu - |a|\inv\sin(a\argu), & \text{if } a\ne 0,\\
	\argu, & \text{if } a=0,
\end{array}\right.
\quad \text{for } \argu\in\dom. \egt 

\update{Specifically, for our simulation studies, with $\dom=[0,1]$, we consider two cases with different choices of the Fr\'echet means $\FmeanX$ and $\FmeanY$:
	\renewcommand{\thecase}{1.\arabic{case}}
	\begin{case}\label{case:d2d_tgaus}
		$\FmeanX = TN_{\dom}(0.5,0.2^2)$, and $\FmeanY = TN_{\dom}(0.75,0.3^2)$, where $TN_{\dom}(\mu,\sigma^2)$ denotes the Gaussian distribution $N(\mu,\sigma^2)$ truncated on $\dom$. 
	\end{case}
	\begin{case}\label{case:d2d_beta}
		$\FmeanX = \mathrm{Beta}(6,2)$, and $\FmeanY = \mathrm{Beta}(2,4)$.
	\end{case}
	Taking $\usedFPCX = \usedFPCY = 20$, for $j,k\in\pint$, we set $\linCoef_{jk} = 2^{-k}\maxDbasisL_k\inv\maxDenFmeanY\inv \maxDbasisL_j\maxDenFmeanX$ if $j\le\usedFPCX$ and $k\le\usedFPCY$, and set $\linCoef_{jk}=0$ otherwise, where $\maxDbasisL_l = \sup_{\argu\in[0,1]}|\basisL'_l(\argu)| = 2\sqrt{2}\pi l$, for $l\in\pint$, $\maxDenFmeanX = \sup_{s\in\dom}\denFmeanX(s)$ and $\maxDenFmeanY = \sup_{t\in\dom}\denFmeanY(t)$.} 
Taking $\convSeries_{1j} = 2^{-j}$, data were generated as follows: 
\ben[label = Step \arabic*:, itemindent=1.1em]
\item \update{Generate $\frcoef_{ij} \sim \mathrm{Unif}(-\convSeries_{1j}(\maxDbasisL_j\maxDenFmeanX\sum_{l=1}^\infty \convSeries_{1l})\inv, \convSeries_{1j}(\maxDbasisL_j\maxDenFmeanX\sum_{l=1}^\infty \convSeries_{1l})\inv)$ independently for $i=1,\dots,n$ and $j=1,\dots,\usedFPCX$, 
	whence $\Log_{\FmeanX}\Xi = \sum_{j=1}^{\usedFPCX} \frcoef_{ij} \basisL_j\circ\cdfFmeanX$, with the basis functions $\basisL_j$ as per \eqref{eq:egbases},  $\regOp(\Log_{\FmeanX}\Xi) = \sum_{k=1}^{\usedFPCY} \sum_{j=1}^{\usedFPCX} \linCoef_{jk}\frcoef_{ij}\basisL_k\circ\cdfFmeanY$, and $\Xi 
	= \Exp_{\FmeanX}(\sum_{j=1}^{\usedFPCX} \frcoef_{ij} \basisL_j\circ\cdfFmeanX)$.} 
\item Generate $\Yi$ by adding distortion to $\regOp(\Log_{\FmeanX}\Xi)$: Sample $A_i\overset{\text{iid}}\sim \text{Unif}\{\pm \pi,\pm 2\pi, \pm 3\pi\}$; let $\Yi =\distortmap_{A_i}\# \Exp_{\FmeanY}[\regOp(\Log_{\FmeanX}\Xi)]$, with function $\distortmap_a$ defined as per \eqref{eq:distort}. 
\item Draw an i.i.d. sample of size $\nDp$ from each of the distributions $\{\Xi\}_{i=1}^n$ and $\{\Yi\}_{i=1}^n$.
\een

Four scenarios were considered with $n\in\{20, 200\}$ and $\nDp\in \{50, 500\}$ for each case. We simulated 500 runs for each $(n,\nDp)$ pair. 
For the domain-extension method, the distribution domain is expanded from $[0,1]$ to $[-0.5,1.5]$ and $[-1,2]$. 
To compare the three methods, we computed the out-of-sample average Wasserstein discrepancy (AWD) based on observations for 200 new predictors $\{\Xi\}_{i=n+1}^{n+200}$, for each Monte Carlo run. Denoting the fitted response distributions by $\Yi^\natural$, the out-of-sample AWD is given by  
\bgt\label{eq:AWD}
\mathrm{AWD}(n,m) = \frac{1}{200} \sum_{i=n+1}^{n+200} \wdist(\update{\expect_\oplus(\Yi|\Xi)}, \Yi^\natural), \egt
\update{with $\expect_\oplus(\Yi|\Xi)$ being the conditional Fr\'echet mean of $\Yi$ given $\Xi$ as defined above \eqref{eq:distort}}. 

\begin{figure}[hbt!]
	\centering
	\begin{subfigure}[c]{.4\textwidth}
		\includegraphics[width=.95\linewidth]{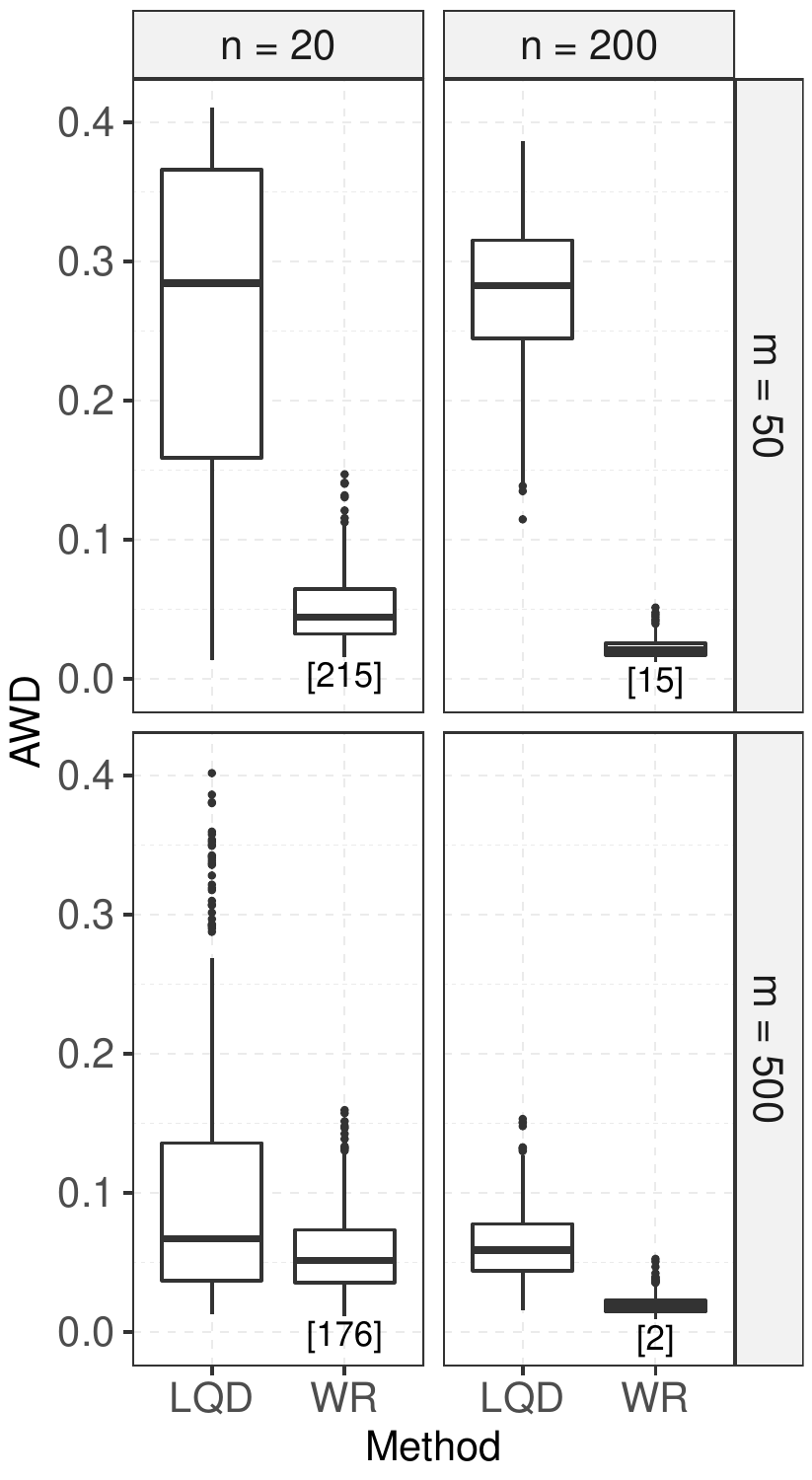}
		\caption{Case~\ref{case:d2d_tgaus}.}\label{fig:simu_methodComp_tgaus}
	\end{subfigure}
	\begin{subfigure}[c]{.4\textwidth}
		\includegraphics[width=.95\linewidth]{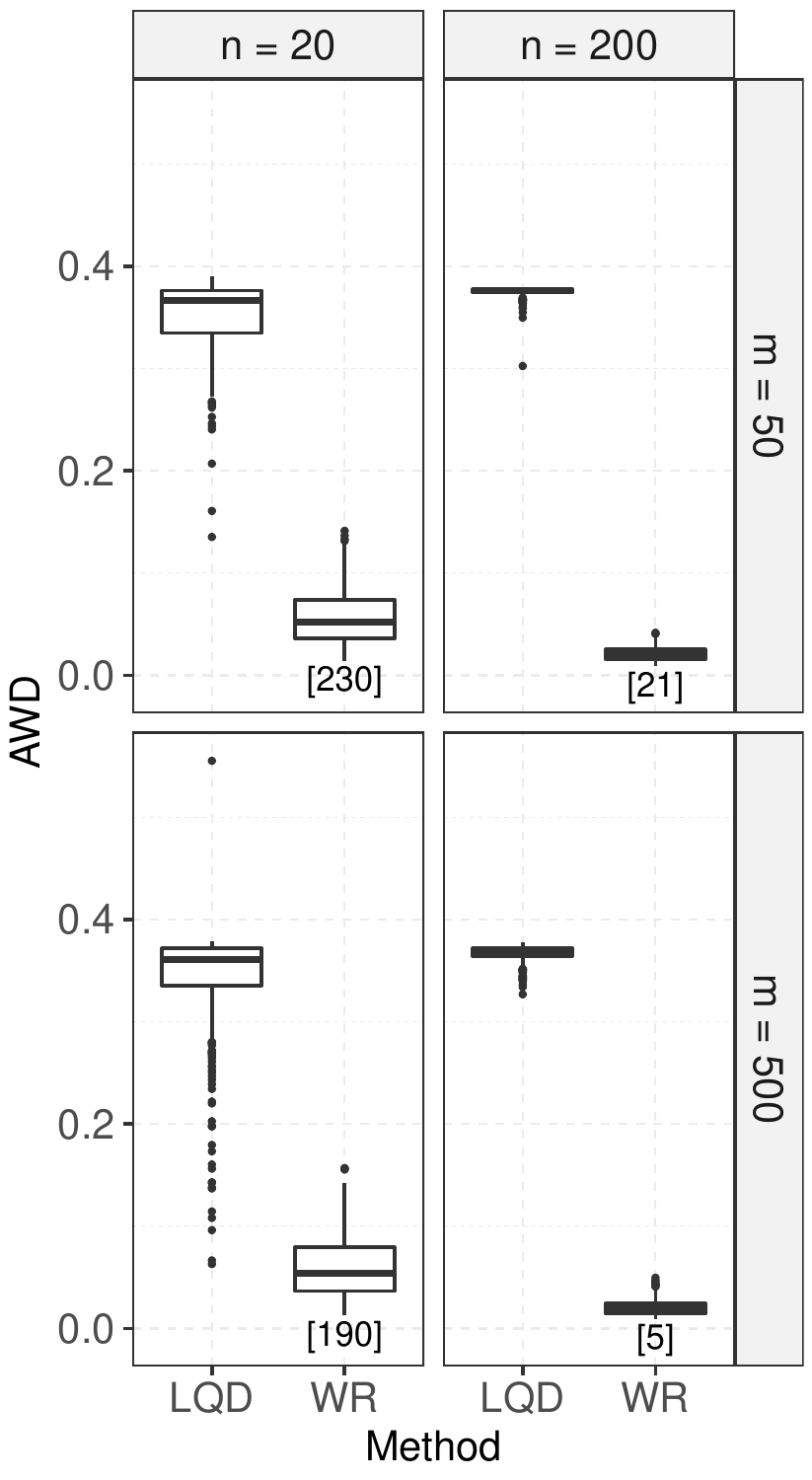}
		\caption{Case~\ref{case:d2d_beta}.}\label{fig:simu_methodComp_beta}
	\end{subfigure}
	\caption{Boxplots of the out-of-sample AWDs as per \eqref{eq:AWD} for the four simulation setups with $(n,\nDp)\in\{20,200\}\times \{50,500\}$, where  ``LQD'' denotes the LQD method and ``WR'' denotes the proposed Wasserstein regression method. The numbers in brackets ``[ ]'' below the boxplots for WR indicate for how many runs event \eqref{eq:out-of-set-issue} happened and boundary projection became necessary.}  \label{fig:simu_methodComp_box} 
\end{figure}

We found that the domain-extension method often failed to force the fit $\regOpEst(\Log_{\FmeanXest} \XiEst)$ to fall in the log space $\Log_{\FmeanYest}\manifold$.  
\update{In particular, this failure occurred in around 15--25\% of the Monte Carlo runs where \eqref{eq:out-of-set-issue} happened when $n=20$}; therefore we do not report the results for this method. 
The results of the LQD method and the proposed Wasserstein regression method with boundary projection (WR) are summarized in the boxplots of Figure~\ref{fig:simu_methodComp_box}. 

The proposed method outperforms the LQD method in all the scenarios considered. 
\update{In fact, the log maps are isometries between the Wasserstein space and the log image spaces. 
	This provides support for the proposed approach. In contrast, the LQD transformation is not an isometry and the ensuing distortions likely contribute to its inferior behavior.  
	In particular, in Case~\ref{case:d2d_beta} where the Fr\'echet mean distributions are beta distributions and  the density functions are not bounded away from zero on $\dom$, the LQD method suffers from  bias issues.} 
When the number of distributions $n$ increases, \eqref{eq:out-of-set-issue} is seen to happen less frequently 
and  boundary projection is seldom needed when the sample size is large ($n=200$).

\update{Additional simulations illustrating the asymptotic result in Theorem~\ref{thm:rate_oracle}, regarding the robustness of the proposed distribution-to-distribution regression method and comparing the proposed distribution-to-scalar regression method with a Gaussian process regression approach \citep{bach:17} can be found in Section~\ref{sec:simu_supp} in the Supplementary Material.}

\section{Applications}\label{sec:app}
\subsection{Mortality Data}
There has been continuing interest in the nature of  human longevity and the analysis of mortality data across countries and calendar years  has provided some of the key data to study it  \citep[e.g.,][]{chio:09,ouel:11,hynd:13,shan:17}. 
Of particular interest is how patterns of mortality of specific populations evolve over calendar time. 
Going  beyond summary statistics such as life expectancy,  viewing  the entire age-at-death distributions as data objects is expected to lead to deeper insights into the secular evolution of human longevity and its dynamics. 
The Human Mortality Database (\url{http://www.mortality.org}) provides yearly life tables for 38 countries,  which yield histograms for the distributions of age-at-death. Smooth densities can then be obtained by applying local linear regression \citep{fan:96}. We obtained these densities on the domain $[0, 100]$ (years of age). 

\begin{figure}[hbt!]
	\centering
	\includegraphics[width=0.95\textwidth]{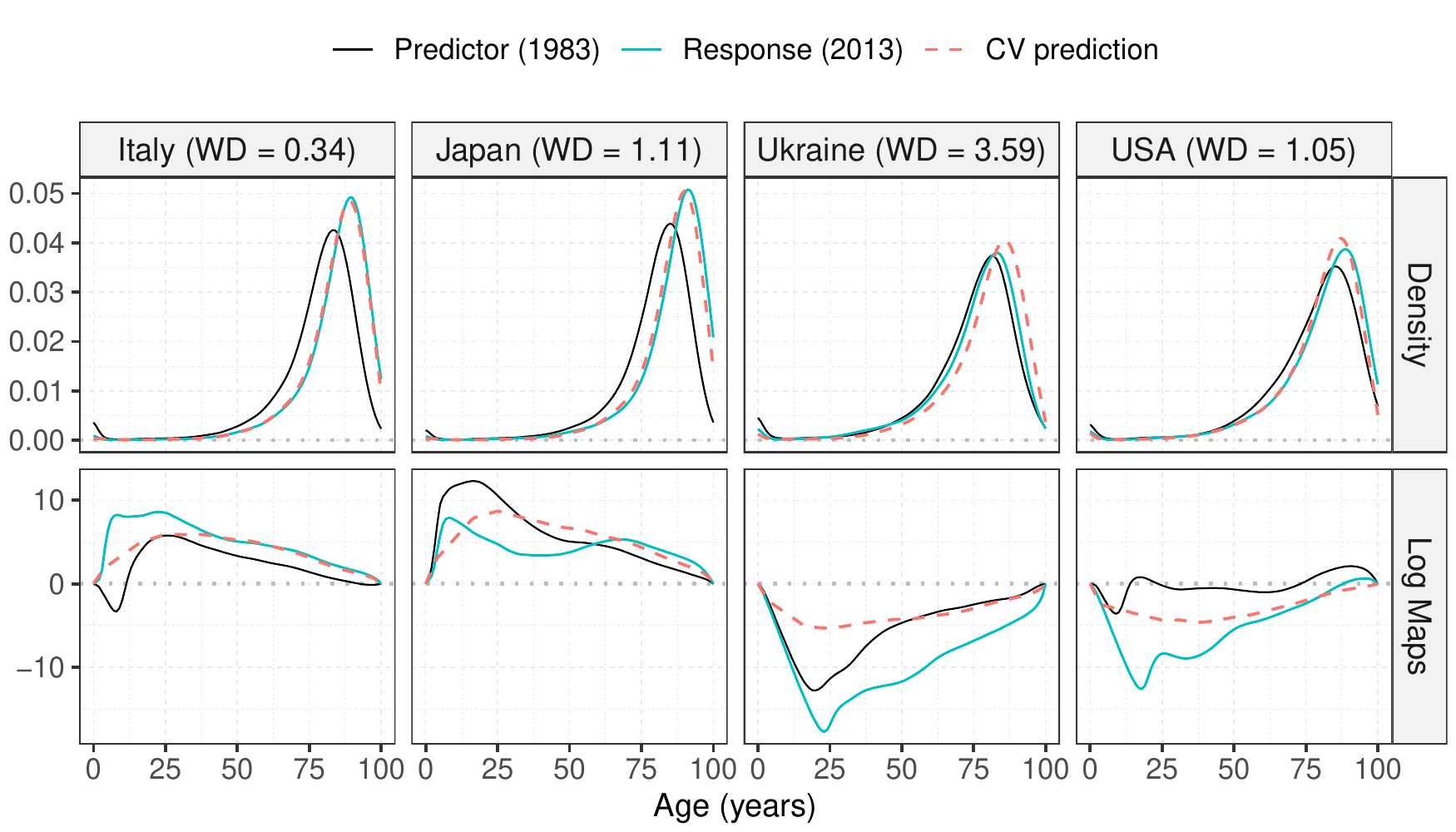}
	\caption{\update{Age-at-death distributions of females in Italy, Japan, Ukraine, and the USA for 1983 and 2013, and the leave-one-out cross validation prediction based on the proposed distribution-to-distribution regression model, where the predictors are the distributions for 1983 and the responses are the distributions 30 years later. 
			Top row: Observed densities for 1983 and 2013 and the leave-one-out predicted densities $\Exp_{\FmeanYest}(\regOpEst(\Log_{\FmeanXest} \XiEst))$ for 2013; 
			Bottom row: Log-mapped predictors and responses, $\Log_{\FmeanXest} \XiEst$ and $\Log_{\FmeanYest} \YiEst$, and leave-one-out prediction for log responses $\regOpEst(\Log_{\FmeanXest} \XiEst)$, where the estimated regression operator $\regOpEst$ is defined in \eqref{eq:regOpEst} and no boundary projection is needed for these four countries. The Wasserstein discrepancies (WDs) between the observed distributions and the corresponding leave-one-out prediction are indicated for each country.}}\label{fig:d2d_fit}
\end{figure}

In a first analysis, we focused on the $n=32$ countries for which data are available for the years 1983 and 2013. We applied the proposed distribution-to-distribution regression model with mortality distributions for an earlier year (1983) as the predictor and a later year (2013) as the response to compare the temporal evolution of age-at-death distributions among different countries. 
\update{We show the leave-one-out prediction results together with the observed distributional predictors and responses for females in Figure~\ref{fig:d2d_fit} for Japan, Ukraine, Italy and the USA, which showcase different patterns of mortality change between 1983 and 2013.} 
In addition to the graphical comparisons, Wasserstein discrepancies (WD) between the observed and leave-one-out predicted distributions are also listed. 
For all four countries, the observed and predicted distributions for 2013 are seen to be shifted to the right from the corresponding distributions in 1983, indicating increased longevity.

\update{The top row of Figure~\ref{fig:d2d_fit} shows a comparison between the model anticipation and the actual observed distributions in 2013 in terms of density functions. 
	Specifically, for Japan and the USA, the rightward mortality shift is seen to be more expressed than suggested by the leave-one-out prediction, indicating that longevity extension is more than anticipated, 
	while the mortality distribution for Ukraine seems to shift to the right at a slower pace than the model prediction would suggest, leading to a relatively large WD with a value of 3.59 between the observed and predicted response. 
	In contrast, the regression fit for Italy almost perfectly matches the observed distribution in 2013.} 

\update{The log maps shown in the bottom row of Figure~\ref{fig:d2d_fit} indicate the shifts of the distributions relative to the Fr\'echet mean across countries for the corresponding year. 
	For Japan, the log maps for the observed predictors and responses and also the model prediction are all positive across the age domain, indicating that the distributions for Japan shift to the right from the Fr\'echet mean across countries, and Japanese females live longer compared to the average across countries at all the ages, while the magnitude of these log maps vary between 1983 and 2013 and also between observed and predicted distributions for 2013. 
	The observed mortality distribution for 2013 has a bigger rightward shift relative to the Fr\'echet mean distribution for older females and minors and a smaller one for younger adults than the model prediction. 
	In contrast, Ukraine has a leftward shift from the Fr\'echet mean for females of all ages, and for 2013 the shift exceeds the model anticipation. 
	For Italy, the log transformed predictor is negative before 15 and positive after, whence the predicted log response becomes positive throughout and also expands in size, meaning the relative standing of Italy in terms of longevity is anticipated to be improved in 2013 by the model prediction. The predicted distribution of Italy in 2013 is shifted to the right from the Fr\'echet mean for all ages, and such rightward shift is more expressed in the actual distribution in 2013. 
	For the USA, the predicted log-mapped response for 2013 is entirely negative and consequently the mortality distribution moved to the left of the Fr\'echet mean, i.e., its relative standing in terms of longevity is anticipated to become worse, while the actual observation is a mixture of a rightward shift for more than 88 years of age and a leftward shift for the other ages.}

We also illustrated the proposed autoregressive model for distribution-valued time series with the mortality data for Sweden, and the results are summarized in Section~\ref{sec:dts_sweden} in the Supplementary Material. 

\subsection{House Price Data}
A question of continuing  interest to economists is how house prices change over time \citep[e.g.,][]{oika:18,bogi:19}. 
We fitted the temporal evolution of house price distributions via the autoregressive distribution time series model described in Section~\ref{sec:dts}, where we downloaded 
house  price data  from \url{http://www.zillow.com}.  These data included 
bimonthly median house prices after inflation adjustment for  $\nDp=306$ cities in the US from June 1996 to August 2015, for which the distribution of median house prices across the cities was constructed for every second month. The autoregressive model was trained on data up to April 2007 and  predictions were computed for the remaining period, where we successively predicted the distribution of each month  based on the prediction two months prior, i.e., by running the distribution time series model as estimated from the training period. 

\begin{figure}[hbt!]
	\centering
	\includegraphics[width=\textwidth]{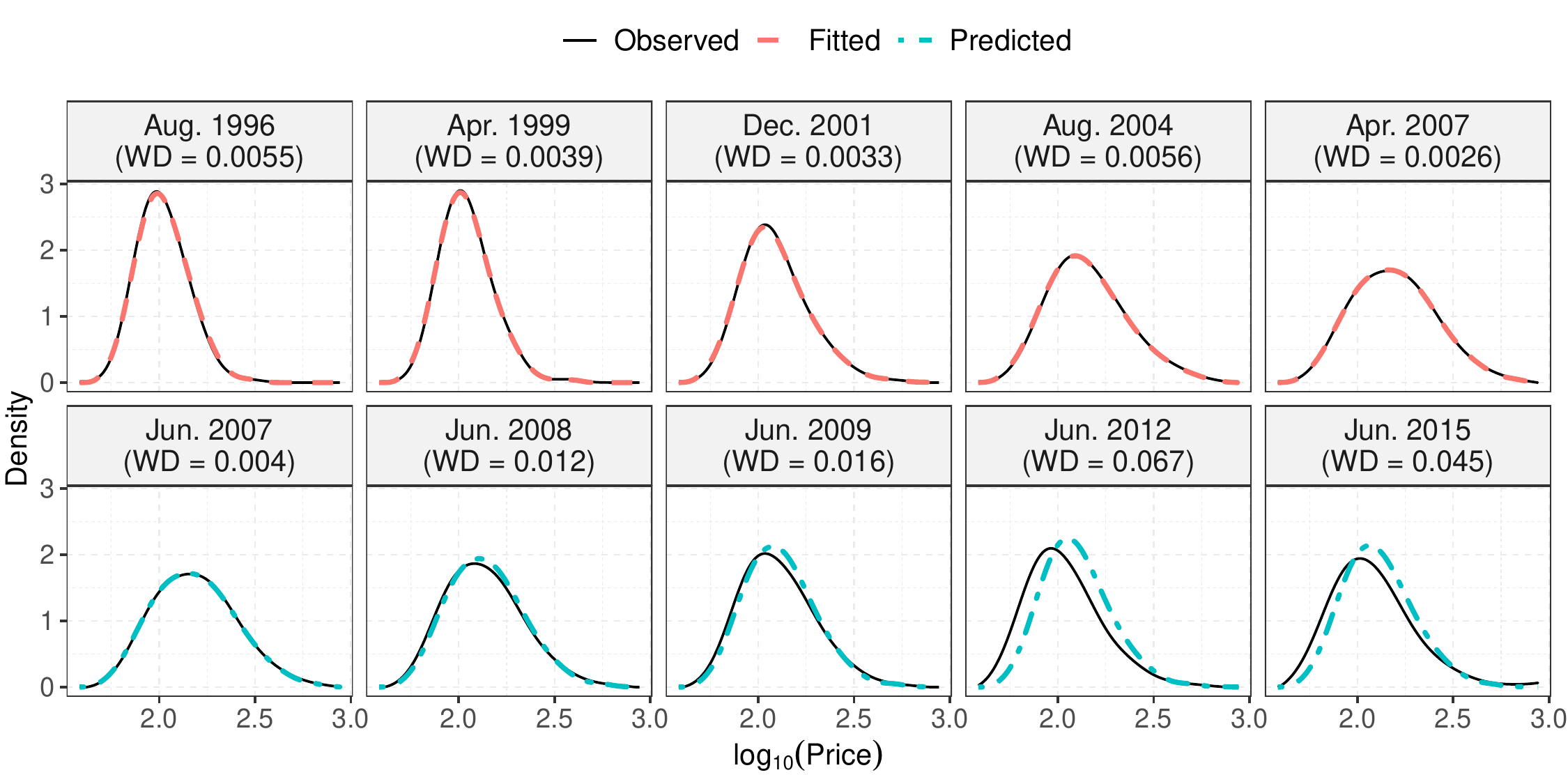}
	\caption{Observed and fitted (top row) / predicted (bottom row) densities of the house price distributions. Training period: August 1996 to April 2007. Prediction period: June 2007 to August 2015. Five representative months are depicted for each of the training and prediction periods in time order, where the Wasserstein discrepancies (WDs) are also listed.}\label{fig:dts_house_time}
\end{figure}

\begin{table}[hbt!]
	\centering
	\caption{Five-number summary of the Wasserstein discrepancies in training and prediction periods.}\label{tab:dts_house_WD}
	\begin{tabular}{rrrrrr}
		\toprule
		& Min & $Q_{0.25}$ & Median & $Q_{0.75}$ & Max \\ 
		\midrule
		Training & 0.0020 & 0.0035 & 0.0047 & 0.0066 & 0.017 \\
		Prediction & 0.0040 & 0.016 & 0.042 & 0.054 & 0.068 \\
		\bottomrule
	\end{tabular}
\end{table}

Figure~\ref{fig:dts_house_time} shows the fitting and prediction results for training and prediction periods, where selected months are ordered in time, while a five-number summary of the fitting and prediction WDs is given in Table~\ref{tab:dts_house_WD}.  
The house price densities are found to be mostly uni-modal, and the peak shifts gradually to the right over time. 
Within the training period, the fitted densities are initially very close to the observed densities
and then gradually are situated to the left of the observed densities, which means that the house price evolution overall accelerates during this period.  
\update{For the prediction period, the predicted densities almost coincide with the observed distributions in 2007, fall behind the actual distribution in 2008, and then continue shifting to the right of the observed distributions}. We find  that the discrepancy between the predicted and observed house price distributions increases from 2007 to 2012 and then decreases afterwards. 
These findings are in line with  the overheating of the housing market before 2006, the crash in 2007--2008, and the lingering effects of the financial crisis, followed by a recovery after 2012. 


\clearpage

\centerline{\LARGE Supplement to ``Wasserstein Regression''}
\setcounter{section}{0}
\setcounter{equation}{0}
\setcounter{lem}{0}
\setcounter{thm}{0}
\setcounter{cor}{0}
\setcounter{prop}{0}
\setcounter{figure}{0}
\renewcommand{\theequation}{S.\arabic{equation}}
\renewcommand{\thethm}{S\arabic{thm}}
\renewcommand{\thelem}{S\arabic{lem}}
\renewcommand{\thecor}{S\arabic{cor}}
\renewcommand{\theprop}{S\arabic{prop}}
\renewcommand{\thesection}{S.\arabic{section}}
\renewcommand{\thefigure}{S.\arabic{figure}}

\section{Proofs and Ancillary Results}

Throughout the proof, given any $\arbiDistn\in\manifold$, we denote the space of all Hilbert--Schmidt operators from $\tangentspace{\arbiDistn}$ to $\tangentspace{\arbiDistn}$ by $\hsSpaceAuto{\arbiDistn}\coloneqq\hsSpace{\arbiDistn}{\arbiDistn}$.

\subsection{Properties of Parallel Transport Operators}\label{sec:propParaTrans}
\begin{prop}\label{prop:paraTrans}
	With probability measures $\distnOne,\distnTwo, \distnOneVar, \distnTwoVar\in\manifold$, 
	the parallel transport operators  $\paraTrans_{\distnOne,\distnTwo}$ and  $\paraTransOp_{(\distnOne,\distnTwo), (\distnOneVar,\distnTwoVar)}$,  as defined in 
	\eqref{eq:paraTrans} and \eqref{eq:paraTransOp},  have the following properties: 	
	
	\ben
	\item If $\distnOne$ is atomless, $\paraTrans_{\distnOne,\distnTwo}$ is unitary, i.e.,
	\be \innerprod{\paraTrans_{\distnOne,\distnTwo}\arbiFctn} {\paraTrans_{\distnOne,\distnTwo}\arbiFctnTwo}_{\distnTwo} = \innerprod{\arbiFctn}{\arbiFctnTwo}_{\distnOne},\quad\text{ for }\arbiFctn,\arbiFctnTwo\in \hilbert_{\distnOne}. \nn\ee
	\item If $\distnOne$ and $\distnTwo$ are both atomless, then the parallel transport $\paraTrans_{\distnTwo,\distnOne}$ from $\hilbert_{\distnTwo}$ to $\hilbert_{\distnOne}$ is the adjoint operator of $\paraTrans_{\distnOne,\distnTwo}$, i.e., 
	\be\innerprod{\paraTrans_{\distnOne,\distnTwo}\arbiFctn_1} {\arbiFctn_2}_{\distnTwo} = \innerprod{\arbiFctn_1} {\paraTrans_{\distnTwo,\distnOne}\arbiFctn_2}_{\distnOne}, \quad\text{for }\arbiFctn_1\in \hilbert_{\distnOne}\text{ and }\arbiFctn_2\in \hilbert_{\distnTwo}.\nn\ee
	\item If $\distnOne, \distnTwo$, and $\distnOneVar$ are atomless, given any positive integers $N_1$ and $N_2$, for $\arbiFctn_{1j}\in\tangentspace{\distnOne}$, $j=1,\cdots,N_1$ and $\arbiFctn_{2k}\in\tangentspace{\distnTwo}$, $k=1,\cdots,N_2$,
	\beq\aligned 
	\paraTransOp_{(\distnOne,\distnTwo), (\distnOneVar,\distnTwoVar)} \left(\sum_{j=1}^{N_1}\sum_{k=1}^{N_2} c_{jk} \arbiFctn_{2k}\otimes \arbiFctn_{1j} \right)
	&= \sum_{j=1}^{N_1}\sum_{k=1}^{N_2} c_{jk} (\paraTrans_{\distnTwo, \distnTwoVar}\arbiFctn_{2k}) \otimes (\paraTrans_{\distnOne, \distnOneVar}\arbiFctn_{1j}).\nn 
	\endaligned\nn\eeq
	\item If $\distnOne, \distnTwo$, $\distnOneVar$, and $\distnTwoVar$ are all atomless, for $\hsOp\in \hsSpace{\distnOne}{\distnTwo}$ and $\hsOpVar\in\hsSpace{\distnOneVar}{\distnTwoVar}$, 
	\be \hsNorm{\hsSpace{\distnOneVar}{\distnTwoVar}} {\paraTransOp_{(\distnOne,\distnTwo), (\distnOneVar,\distnTwoVar)}\hsOp - \hsOpVar} = \hsNorm{\hsSpace{\distnOne}{\distnTwo}}{\paraTransOp_{(\distnOneVar,\distnTwoVar), (\distnOne,\distnTwo)}\hsOpVar - \hsOp}. \nn\ee
	\een
\end{prop}

Related results for finite-dimensional Riemannian manifolds are in Proposition~2 of \citet{lin:18}. 
These four statements can be easily checked by the definition of parallel transport operators in \eqref{eq:paraTrans} and \eqref{eq:paraTransOp}; we omit the proof. 

\subsection{\update{Proofs for Section~\ref{sec:theory_indep}}}\label{sec:proof_indep}

We start with the definitions of notations to be used in the proofs. 
For random variables $X_n$ and a sequence of positive constants $c_n$, we will write $X_n=\Op(c_n)$ if
\bal\nn
\lim_{\largeConst\ra\infty} \limsup_{n\ra\infty} \sup_{\jointdistn\in\distnfam} \prob_{\jointdistn}(|X_n|>\largeConst c_n) =0,\eal
and $X_n =\op(c_n)$ if there exists $\largeConst_0>0$ such that
\bal\nn
\lim_{n\ra\infty} \sup_{\jointdistn\in\distnfam} \prob_{\jointdistn}(|X_n|>\largeConst_0 c_n) =0.\eal
For a sequence of deterministic quantities $a_n = a_n(\jointdistn)$, we will write $a_n = O(c_n)$ if
\bal\nn
\sup_{n\ge 1} c_n\inv \sup_{\jointdistn\in\distnfam} |a_n(\jointdistn)| < \infty.\eal
In addition, for simplicity, we denote $\paraTrans_{\FmeanXoracle,\FmeanX}\arbiFctn$, $\paraTrans_{\FmeanYoracle,\FmeanY}\arbiFctn$, $\paraTrans_{\FmeanXest,\FmeanX}\arbiFctn$, and $\paraTrans_{\FmeanYest,\FmeanY}\arbiFctn$ by $\paraTrans\arbiFctn$  for $\arbiFctn$ in $\tangentspace{\FmeanXoracle}$, $\tangentspace{\FmeanYoracle}$, $\tangentspace{\FmeanXest}$, and $\tangentspace{\FmeanYest}$, respectively; 
we define $\pCovopXoracle \coloneqq \paraTransOp_{(\FmeanXoracle,\FmeanXoracle),(\FmeanX,\FmeanX)}\covopXoracle$, 
$\pCovopYoracle \coloneqq \paraTransOp_{(\FmeanYoracle,\FmeanYoracle),(\FmeanY,\FmeanY)}\covopYoracle$, 
$\pCrosscovopXYoracle\coloneqq \paraTransOp_{(\FmeanXoracle,\FmeanYoracle),(\FmeanX,\FmeanY)}\crosscovopXYoracle$, 
$\pCovopXest \coloneqq \paraTransOp_{(\FmeanXest,\FmeanXest),(\FmeanX,\FmeanX)}\covopXest$, 
$\pCovopYest \coloneqq \paraTransOp_{(\FmeanYest,\FmeanYest),(\FmeanY,\FmeanY)}\covopYest$, 
$\pCrosscovopXYest\coloneqq \paraTransOp_{(\FmeanXest,\FmeanYest),(\FmeanX,\FmeanY)}\crosscovopXYest$, 
\bal\nn
\pRegKernelOracle \coloneqq \sum_{k=1}^{\nFPCY} \sum_{j=1}^{\nFPCX} \regCoefOracle_{jk} \pEgnfctnYoracle_k \otimes \pEgnfctnXoracle_j \qaq
\pRegKernelEst \coloneqq \sum_{k=1}^{\nFPCY} \sum_{j=1}^{\nFPCX} \regCoefEst_{jk} \pEgnfctnYest_k \otimes \pEgnfctnXest_j.\eal

By the third statement in Proposition~\ref{prop:paraTrans}, under \ref{ass:atomless}, $\pRegKernelOracle$ is the kernel of $\paraTransOp_{(\FmeanXoracle,\FmeanYoracle), (\FmeanX,\FmeanY)}\regOpOracle$; under \ref{ass:atomless} and \ref{ass:atomlessFmeanEst}, $\pRegKernelEst$ is the kernel of $\paraTransOp_{(\FmeanXest,\FmeanYest), (\FmeanX,\FmeanY)}\regOpEst$. 
Hence, 
\bal\label{eq:measureOfError}
\hsNormSq{\hsSpace{\FmeanX}{\FmeanY}}{\paraTransOp_{(\FmeanXoracle,\FmeanYoracle), (\FmeanX,\FmeanY)} \regOpOracle - \regOp} &=\int_{\dom}\int_{\dom}\left(\pRegKernelOracle(s,t) - \regKernel(s,t)\right)^2 \diffop\FmeanX(s) \diffop\FmeanY(t),\\
\hsNormSq{\hsSpace{\FmeanX}{\FmeanY}}{\paraTransOp_{(\FmeanXest,\FmeanYest), (\FmeanX,\FmeanY)} \regOpEst - \regOp} &=\int_{\dom}\int_{\dom}\left(\pRegKernelEst(s,t) - \regKernel(s,t)\right)^2 \diffop\FmeanX(s) \diffop\FmeanY(t).\eal
Thus, for the proofs of Theorems~\ref{thm:rate_oracle} and \ref{thm:rate_realistic}, we will derive the asymptotic order of the right hand sides in \eqref{eq:measureOfError}.  To this end, we need to study the asymptotic properties of the estimators of the covariance and cross-covariance operators, i.e., $\covopXoracle$, $\covopYoracle$ and $\crosscovopXYoracle$ when the distributions $\Xi$ and $\Yi$ are fully observed, and $\covopXest$, $\covopYest$ and $\crosscovopXYest$ when only samples of observations drawn from the $\Xi$ and $\Yi$ are available. 
We use the convention that $\innerprod{\pEgnfctnXoracle_j}{\egnfctnX_j}_{\FmeanX}\ge 0$,  $\innerprod{\pEgnfctnYoracle_k}{\egnfctnY_k}_{\FmeanY}\ge 0$, $\innerprod{\pEgnfctnXest_j}{\egnfctnX_j}_{\FmeanX}\ge 0$ and $\innerprod{\pEgnfctnYest_k}{\egnfctnY_k}_{\FmeanY}\ge 0$ to determine the signs of the estimated eigenfunctions, $\egnfctnXoracle_j$, $\egnfctnYoracle_k$, $\egnfctnXest_j$ and $\egnfctnYest_k$ where choice of the signs may impact the validity of the results. 

We first focus on the case where $\Xi$ and $\Yi$ are fully observed. 

\blem\label{lem:cov_oracle}
Assume \ref{ass:atomless} and \ref{ass:variation}. Furthermore, assume that the eigenvalues $\{\egnvalX_j\}_{j=1}^\infty$ and $\{\egnvalY_k\}_{k=1}^\infty$ are distinct, respectively. Then
\bgt\nn
\hsNormSq{\hsSpaceAuto{\FmeanX}}{\pCovopXoracle - \covopX} = \Op(n\inv),\quad
\hsNormSq{\hsSpaceAuto{\FmeanY}}{\pCovopYoracle - \covopY} = \Op(n\inv),\\
\text{and}\quad \hsNormSq{\hsSpace{\FmeanX}{\FmeanY}}{\pCrosscovopXYoracle-\crosscovopXY} = \Op(n\inv). \egt
Furthermore, 
\begin{gather*}
\sup_{j\ge 1} |\egnvalXoracle_j - \egnvalX_j| \le \hsNorm{\hsSpaceAuto{\FmeanX}}{\pCovopXoracle - \covopX},\quad 
\sup_{k\ge 1} |\egnvalYoracle_k - \egnvalY_k| \le 
\hsNorm{\hsSpaceAuto{\FmeanY}}{\pCovopYoracle - \covopY},\\
\|\pEgnfctnXoracle_j - \egnfctnX_j\|_{\FmeanX}
\le 2\sqrt 2 \hsNorm{\hsSpaceAuto{\FmeanX}}{\pCovopXoracle - \covopX} / 
\min_{1\le j'\le j}\{\egnvalX_{j'} - \egnvalX_{j'+1}\},\text{ for all }j\ge 1,\\
\|\pEgnfctnYoracle_k - \egnfctnY_k\|_{\FmeanY} \le 2\sqrt 2 \hsNorm{\hsSpaceAuto{\FmeanY}}{\pCovopYoracle - \covopY} / 
\min_{1\le k'\le k}\{\egnvalY_{k'} - \egnvalY_{k'+1}\}, \text{ for all }k\ge 1.
\end{gather*}
\elem
\bpf We only prove the results for $\covopXoracle$; those for $\covopYoracle$ and $\crosscovopXYoracle$ can be shown analogously. 
We first note that under \ref{ass:atomless}, $\FmeanX$ and $\FmeanXoracle$ are atomless, the latter with probability 1. 
By the third statement in Proposition~\ref{prop:paraTrans}, 
\begin{align*}
\pCovopXoracle - \covopX 
&= n\inv \sum_{i=1}^n (\pLogXoracle\Xi) \otimes (\pLogXoracle\Xi)  - \covopX\\
&= n\inv \sum_{i=1}^n (\Log_{\FmeanX}\Xi) \otimes (\Log_{\FmeanX}\Xi) - \covopX\\
&\quad + n\inv \sum_{i=1}^n (\pLogXoracle\Xi - \Log_{\FmeanX}\Xi) \otimes (\Log_{\FmeanX}\Xi)\\
&\quad + n\inv \sum_{i=1}^n (\Log_{\FmeanX}\Xi)\otimes (\pLogXoracle\Xi - \Log_{\FmeanX}\Xi)\\
&\quad + n\inv \sum_{i=1}^n (\pLogXoracle\Xi - \Log_{\FmeanX}\Xi) \otimes (\pLogXoracle\Xi - \Log_{\FmeanX}\Xi)\\
&\eqqcolon \hsOp_1 + \hsOp_2 + \hsOp_3 + \hsOp_4.
\end{align*}
For $\hsOp_2$, it can be observed that
\bal\nn
\hsNormSq{\hsSpaceAuto{\FmeanX}}{\hsOp_2}
&\le \left(n\inv\sum_{i=1}^n \|\Log_{\FmeanX}\Xi\|_{\FmeanX}^2\right) 
\left(n\inv\sum_{i=1}^n \|\pLogXoracle\Xi - \Log_{\FmeanX}\Xi\|_{\FmeanX}^2\right).
\eal
For the first term, since $\X$ is square integrable,
\bal\label{eq:log_oracle_rate}
\expect\left(n\inv\sum_{i=1}^n \|\Log_{\FmeanX}\Xi\|_{\FmeanX}^2\right) 
&= \expect\left(n\inv\sum_{i=1}^n \wdist^2(\FmeanX,\Xi)\right)
= \expect\left(\wdist^2(\FmeanX,\X)\right) < \infty, \eal
whence $n\inv\sum_{i=1}^n \|\Log_{\FmeanX}\Xi\|_{\FmeanX}^2 = \Op(1)$.
For the second term, by \eqref{eq:log}, \eqref{eq:paraTrans} and the atomlessness of $\FmeanXoracle$, $\pLogXoracle\Xi = \Log_{\FmeanX}\Xi - \Log_{\FmeanX}\FmeanXoracle$. 
In conjunction with \eqref{eq:quantile_practicalFrechetMean}, we have 
\bal\label{eq:oracleFmean_rate}
\expect\left(\frac{1}{n}\sum_{i=1}^n \|\pLogXoracle\Xi - \Log_{\FmeanX}\Xi\|_{\FmeanX}^2\right) &=\expect\left(\|\Log_{\FmeanX}\FmeanXoracle\|_{\FmeanX}^2\right)
= \frac{1}{n}\expect\wdist^2(\X,\FmeanX), \eal
where the last equality is due to the fact that $\{\Xi\}_{i=1}^n$ are independent realizations of $\X$. 
Thus, $\hsNormSq{\hsSpaceAuto{\FmeanX}}{\hsOp_2} = \Op(n\inv)$. 
Similarly, it can be shown that $\hsNormSq{\hsSpaceAuto{\FmeanX}}{\hsOp_3} =  \Op(n\inv)$, and $\hsNormSq{\hsSpaceAuto{\FmeanX}}{\hsOp_4}= \Op(n^{-2})$. 
For $\hsOp_1$, 
\bal\nn
&\expect\left(\hsNormSq{\hsSpaceAuto{\FmeanX}}{\hsOp_1}\right)\\
&= \frac{1}{n} \expect\left\{\int_{\dom}\int_{\dom} \left[\Log_{\FmeanX}\X(s)\Log_{\FmeanX}\X(t) - \expect\left(\Log_{\FmeanX}\X(s)\Log_{\FmeanX}\X(t)\right)\right]^2\diffop\FmeanX(s)\diffop\FmeanX(t)\right\}\\
&\le \frac{2}{n}\left\{\expect\left(\|\Log_{\FmeanX}\X\|_{\FmeanX}^4\right) + \hsNormSq{\hsSpaceAuto{\FmeanX}}{\covopX}\right\}.
\eal
In conjunction with \ref{ass:variation} and the fact that $\sum_{j=1}^\infty\egnvalX_j = \expect\wdist^2(\X,\FmeanX)<\infty$, this implies  
\bgt\nn
\hsNormSq{\hsSpaceAuto{\FmeanX}}{\hsOp_1} = \Op(n\inv).\egt
Therefore, $\hsNormSq{\hsSpaceAuto{\FmeanX}}{\pCovopXoracle - \covopX} = \Op(n\inv)$. 
By the atomlessness of $\FmeanX$ and $\FmeanXoracle$ and Proposition~\ref{prop:paraTrans}, $\{\egnvalXoracle_j\}_{j=1}^\infty$ and $\{\pEgnfctnXoracle_j\}_{j=1}^\infty$ are the eigenvalues and eigenfunctions of $\pCovopXoracle$, for which the results follow from Lemmas~4.2 and 4.3 of \citet{bosq:00}.
\epf

\bpf[Proof of Theorem~\ref{thm:rate_oracle}]
Defining 
\bal\label{eq:regOpOracle_decom} 
\wt{A}_1 &= \sum_{k=1}^{\nFPCY}\sum_{j=1}^{\nFPCX} (\regCoefOracle_{jk} - \regCoef_{jk}) \pEgnfctnYoracle_k \otimes \pEgnfctnXoracle_j,\\  
\wt{A}_2 &= \sum_{k=1}^{\nFPCY}\sum_{j=1}^{\nFPCX} \regCoef_{jk} \left( \pEgnfctnYoracle_k \otimes \pEgnfctnXoracle_j - \egnfctnY_k \otimes \egnfctnX_j\right),\\
A_3 &= \sum_{k=1}^\infty \sum_{j=1}^\infty \regCoef_{jk} \egnfctnY_k \otimes \egnfctnX_j - \sum_{k=1}^{\nFPCY} \sum_{j=1}^{\nFPCX} \regCoef_{jk} \egnfctnY_k \otimes \egnfctnX_j,\eal 
we observe that by \eqref{eq:measureOfError}, 
\begin{align*}
\left\hsNormSq{\hsSpace{\FmeanX}{\FmeanY}}{\paraTransOp_{(\FmeanXoracle,\FmeanYoracle), (\FmeanX,\FmeanY)} \regOpOracle - \regOp\right} 
&\le 3 \|\wt{A}_1\|_{\jointmean}^2 + 3\|\wt{A}_2\|_{\jointmean}^2 + 3\|A_3\|_{\jointmean}^2,
\end{align*}
where and hereafter, $\|A\|_{\jointmean}^2 \coloneqq \int_{\dom}\int_{\dom} A(s,t)^2 \diffop\FmeanX(s)\diffop\FmeanY(t)$ for $A\in\hilbert_{\FmeanX\times\FmeanY}$. 

For $\wt{A}_1$, 
\bal\label{eq:regCoefOracle1}
\|\wt{A}_1\|_{\jointmean}^2 
&= \sum_{k=1}^{\nFPCY}\sum_{j=1}^{\nFPCX} (\regCoefOracle_{jk} - \regCoef_{jk})^2 
= \sum_{k=1}^{\nFPCY}\sum_{j=1}^{\nFPCX} (\egnvalXoracle_j\inv \crovalOracle_{jk} - \egnvalX_j\inv\croval_{jk})^2\\
&\le 2\sum_{k=1}^{\nFPCY}\sum_{j=1}^{\nFPCX} \egnvalXoracle_j^{-2}(\crovalOracle_{jk} - \croval_{jk})^2 + 2\sum_{k=1}^{\nFPCY}\sum_{j=1}^{\nFPCX} (\egnvalXoracle_j\inv - \egnvalX_j\inv)^2 \croval_{jk}^2.\eal
Using the same technique as in the proof of \citet{hall:07:1}, we define events
\bgt\nn \wt\event_{1\nFPCX} = \wt\event_{1\nFPCX}(n) = \left\{ \egnvalX_{\nFPCX}\ge 2 \hsNorm{\hsSpaceAuto{\FmeanX}}{\pCovopXoracle - \covopX}\right\}.\egt
On $\wt\event_{1\nFPCX}$, Lemma~\ref{lem:cov_oracle} entails $\egnvalXoracle_j \ge \egnvalX_j/2$, for all $j=1,\ldots,\nFPCX$. 
Note that $n^{-1/2} \egnvalX_{\nFPCX}\inv = O(n^{-1/2}\nFPCX^{\egnvalSpX}) = o(1)$, as $n\ra\infty$, under the assumptions of Theorem~\ref{thm:rate_oracle}.
Thus, Lemma~\ref{lem:cov_oracle} entails that $\prob(\wt\event_{1\nFPCX})\ra 1$ as $n\ra \infty$. 
Hence, on $\wt\event_{1\nFPCX}$, it holds for the last two terms in \eqref{eq:regCoefOracle1} that 
\bal
\sum_{k=1}^{\nFPCY}\sum_{j=1}^{\nFPCX} \egnvalXoracle_j^{-2}(\crovalOracle_{jk} - \croval_{jk})^2
&\le 4\sum_{k=1}^{\nFPCY}\sum_{j=1}^{\nFPCX} \egnvalX_j^{-2}(\crovalOracle_{jk} - \croval_{jk})^2,\\
\sum_{k=1}^{\nFPCY}\sum_{j=1}^{\nFPCX} (\egnvalXoracle_j\inv - \egnvalX_j\inv)^2 \croval_{jk}^2
&=\sum_{k=1}^{\nFPCY}\sum_{j=1}^{\nFPCX} (\egnvalXoracle_j\inv\egnvalX_j - 1)^2 \regCoef_{jk}^2
\le 4\sum_{k=1}^{\nFPCY}\sum_{j=1}^{\nFPCX} \egnvalX_j^{-2}(\egnvalXoracle_j - \egnvalX_j)^2 \regCoef_{jk}^2\\
&\le \const \hsNormSq{\hsSpaceAuto{\FmeanX}}{\pCovopXoracle-\covopX} \sum_{j=1}^{\nFPCX} j^{2\egnvalSpX-2\coefDecayX} 
= \Op(n\inv).\note{\text{ since }2\egnvalSpX-2\coefDecayX<-2} \label{eq:regCoefOracle1-2}
\eal
We will show later that \note{if $\coefDecayX>\egnvalSpX+1$, $\coefDecayY>1$, w.o. $p$ series assumption of $\egnvalX_j$,}
\bal\label{eq:regCoefOracle1-1}
&\sum_{k=1}^{\nFPCY}\sum_{j=1}^{\nFPCX}\egnvalX_j^{-2}(\crovalOracle_{jk}-\croval_{jk})^2\\
&= \Op\left(n\inv\nFPCX^{\egnvalSpX+1}\right)  +\Op\left(n^{-2}\nFPCX^{2\egnvalSpX+1}\nFPCY^{2\egnvalSpY+3}\right) +\Op\left(n^{-2}\nFPCX^{4\egnvalSpX+3} (1+n\inv\nFPCY^{2\egnvalSpY+3})\right)\\
&\quad
+\Op\left((1+n\inv\nFPCY^{2\egnvalSpY+3})n\inv\sum_{k=1}^{\nFPCY}k^{2\egnvalSpY-2\coefDecayY+2}\right)\\
&\quad + 
\left\{\begin{array}{ll}
	\Op\left((1+n\inv\nFPCY^{2\egnvalSpY+3})n^{-2}\nFPCY^{4\egnvalSpY-2\coefDecayY+5}(\log \nFPCY)^2\right), &\text{if } \coefDecayY\le\egnvalSpY+1,\\
	\Op\left((1+n\inv\nFPCY^{2\egnvalSpY+3})n^{-2}\nFPCY^{2\egnvalSpY+3}\right), &\text{if }\coefDecayY>\egnvalSpY+1.
\end{array}\right.\eal

For $\wt{A}_2$, we observe that
$\|\wt{A}_2\|_{\jointmean}^2
\le 3\|\wt{A}_{21}\|_{\jointmean}^2 +3\|\wt{A}_{22}\|_{\jointmean}^2 +3\|\wt{A}_{23}\|_{\jointmean}^2$, where
\bal\nn
\wt{A}_{21}
&= \sum_{k=1}^{\nFPCY}\sum_{j=1}^{\nFPCX} \regCoef_{jk} \egnfctnY_k \otimes (\pEgnfctnXoracle_j-\egnfctnX_j),\\
\wt{A}_{22}
&= \sum_{k=1}^{\nFPCY}\sum_{j=1}^{\nFPCX} \regCoef_{jk} (\pEgnfctnYoracle_k-\egnfctnY_k) \otimes \egnfctnX_j,\\
\wt{A}_{23}
&= \sum_{k=1}^{\nFPCY}\sum_{j=1}^{\nFPCX} \regCoef_{jk} (\pEgnfctnYoracle_k-\egnfctnY_k) \otimes (\pEgnfctnXoracle_j-\egnfctnX_j).\eal
We note that \ref{ass:eigenval_spacingX} and Lemma~\ref{lem:cov_oracle} entail \note{(w.o. $p$ series assumption of $\egnvalX_j$)}
\bal\label{eq:egnfctn_oracle_rate}
\|\pEgnfctnXoracle_j-\egnfctnX_j\|_{\FmeanX}^2 
&\le \const j^{2\egnvalSpX+2} \hsNormSq{\hsSpaceAuto{\FmeanX}}{\pCovopXoracle-\covopX},\\
\|\pEgnfctnYoracle_k-\egnfctnY_k\|_{\FmeanY}^2 
&\le \const k^{2\egnvalSpY+2} \hsNormSq{\hsSpaceAuto{\FmeanY}}{\pCovopYoracle-\covopY},\eal
uniformly in $j,k$, which implies \note{(w.o. $p$ series assumption of  $\egnvalX_j$)}
\bal\nn
\|\wt{A}_{21}\|_{\jointmean}^2
&
\le\nFPCX\sum_{k=1}^{\nFPCY}\sum_{j=1}^{\nFPCX}\regCoef_{jk}^2\|\pEgnfctnXoracle_j-\egnfctnX_j\|_{\FmeanX}^2 
=\Op\left(n\inv\nFPCX\sum_{j=1}^{\nFPCX}j^{2\egnvalSpX-2\coefDecayX+2}\right)\\
\|\wt{A}_{22}\|_{\jointmean}^2
&
\le\nFPCY\sum_{k=1}^{\nFPCY}\sum_{j=1}^{\nFPCX}\regCoef_{jk}^2\|\pEgnfctnYoracle_k-\egnfctnY_k\|_{\FmeanY}^2 
=\Op\left(n\inv\nFPCY\sum_{k=1}^{\nFPCY}k^{2\egnvalSpY-2\coefDecayY+2}\right),\\
\|\wt{A}_{23}\|_{\jointmean}^2
&\le \nFPCX\nFPCY \sum_{k=1}^{\nFPCY}\sum_{j=1}^{\nFPCX} \regCoef_{jk}^2\|\pEgnfctnXoracle_j-\egnfctnX_j\|_{\FmeanX}^2 \|\pEgnfctnYoracle_k-\egnfctnY_k\|_{\FmeanY}^2 \\
&=\Op\left(n^{-2}\nFPCX\nFPCY\sum_{j=1}^{\nFPCX}j^{2\egnvalSpX-2\coefDecayX+2}\sum_{k=1}^{\nFPCY}k^{2\egnvalSpY-2\coefDecayY+2}\right). \eal
Hence, 
\bal\label{eq:regCoefOracle2}
\|\wt{A}_2\|_{\jointmean}^2 
&=\Op\left(n\inv\nFPCX\sum_{j=1}^{\nFPCX}j^{2\egnvalSpX-2\coefDecayX+2} +n\inv\nFPCY \sum_{k=1}^{\nFPCY}k^{2\egnvalSpY-2\coefDecayY+2}\right). \eal

For $A_3$, we observe that by \ref{ass:coef_decay}, 
\bgt\label{eq:regCoefOracle3}
\|A_3\|_{\jointmean}^2 
= \sum_{k=1}^\nFPCY \sum_{j=\nFPCX+1}^\infty \regCoef_{jk}^2 + \sum_{k=\nFPCY+1}^\infty \sum_{j=1}^\nFPCX \regCoef_{jk}^2 +  \sum_{k=\nFPCY+1}^\infty \sum_{j=\nFPCX+1}^\infty \regCoef_{jk}^2= O\left(\nFPCX^{-2\coefDecayX+1} + \nFPCY^{-2\coefDecayY + 1}\right).\egt
Under \ref{ass:nFPC_rate}, combining \eqref{eq:regCoefOracle1-2}, \eqref{eq:regCoefOracle1-1}, \eqref{eq:regCoefOracle2} and \eqref{eq:regCoefOracle3} yields \note{(w.o. $p$ series assumption of $\egnvalX_j$)}
\bal\nn
&\left\hsNormSq{\hsSpace{\FmeanX}{\FmeanY}}{\paraTransOp_{(\FmeanXoracle,\FmeanYoracle), (\FmeanX,\FmeanY)} \regOpOracle - \regOp\right} \note{\text{ if }\coefDecayX>\egnvalSpX+1,\, \egnvalSpX\ge 1,\, \coefDecayY>1,\, n\inv\nFPCX^{2\egnvalSpX+2}=o(1),\, n\inv\nFPCY^{2\egnvalSpY+2}=o(1),}\\
&= 
\Op\left(n\inv\nFPCX^{\egnvalSpX+1}\right) 
+\Op\left(n\inv\nFPCY \sum_{k=1}^{\nFPCY}k^{2\egnvalSpY-2\coefDecayY+2}\right) +O\left(\nFPCX^{-2\coefDecayX+1} + \nFPCY^{-2\coefDecayY+1}\right)\\
&\quad +\Op\left(n^{-2}\nFPCX^{4\egnvalSpX+3}(1+n\inv\nFPCY^{2\egnvalSpY+3})\right) +\Op\left(n^{-2}\nFPCX^{2\egnvalSpX+1}\nFPCY^{2\egnvalSpY+3}\right)\\
&\quad +\left\{\begin{array}{ll}
	\Op\left((1+n\inv\nFPCY^{2\egnvalSpY+3})n^{-2}\nFPCY^{4\egnvalSpY-2\coefDecayY+5}(\log \nFPCY)^2\right), &\text{if } \coefDecayY\le\egnvalSpY+1,\\
	\Op\left((1+n\inv\nFPCY^{2\egnvalSpY+3})n^{-2}\nFPCY^{2\egnvalSpY+3}\right), &\text{if }\coefDecayY>\egnvalSpY+1.
\end{array}\right.
\eal
Observe that $n\inv\nFPCX\sum_{j=1}^{\nFPCX}j^{2\egnvalSpX-2\coefDecayX+2} = O(n\inv\nFPCX^2(\nFPCX/\log\nFPCX)^{2(\egnvalSpX-\coefDecayX+1)}) =o(n\inv\nFPCX^{\egnvalSpX+1})$, since $\egnvalSpX+1<\coefDecayX$ and $\egnvalSpX\ge 1$, and that $n^{-2}\nFPCX^{4\egnvalSpX+3}n\inv\nFPCY^{2\egnvalSpY+3} / (n^{-2}\nFPCX^{2\egnvalSpX+1}\nFPCY^{2\egnvalSpY+3}) = n\inv\nFPCX^{2\egnvalSpX+2}\ra 0$ as $n\ra\infty$. 
Also observe that $\sum_{k=1}^{\nFPCY}k^{2\egnvalSpY-2\coefDecayY+2} \sim \nFPCY^{2\egnvalSpY-2\coefDecayY+3}\mbf{1}_{\{\coefDecayY-\egnvalSpY\le 1\}} +\nFPCY(\nFPCY/\log\nFPCY)^{2\egnvalSpY-2\coefDecayY+2}\mbf{1}_{\{1<\coefDecayY-\egnvalSpY<3/2\}} +(\log\nFPCY)\mbf{1}_{\{\coefDecayY-\egnvalSpY=3/2\}} +\mbf{1}_{\{\coefDecayY-\egnvalSpY>3/2\}}$, whence $n^{-2}\nFPCY^{4\egnvalSpY-2\coefDecayY+5}(\log \nFPCY)^2/(n\inv\nFPCY\sum_{k=1}^{\nFPCY}k^{2\egnvalSpY-2\coefDecayY+2}) \sim n\inv\nFPCY^{2\egnvalSpY+1}$ 
$\cdot(\log\nFPCY)^2 \ra 0$ as $n\ra\infty$, if $\coefDecayY\le\egnvalSpY+1$, and  $n^{-2}\nFPCY^{2\egnvalSpY+3}/(n\inv\nFPCY\sum_{k=1}^{\nFPCY}k^{2\egnvalSpY-2\coefDecayY+2}) \le\const n\inv\nFPCY^{2\egnvalSpY+2}\ra 0$ as $n\ra\infty$, if $\coefDecayY>\egnvalSpY+1$. 
Therefore, $\hsNormSq{\hsSpace{\FmeanX}{\FmeanY}}{\paraTransOp_{(\FmeanXoracle,\FmeanYoracle), (\FmeanX,\FmeanY)} \regOpOracle - \regOp} =\indepRate(n)$, where 
\bal\nn
\indepRate(n)
&= n\inv\nFPCX^{\egnvalSpX+1} +n^{-2}\nFPCX^{4\egnvalSpX+3} +n^{-2}\nFPCX^{2\egnvalSpX+1}\nFPCY^{2\egnvalSpY+3}
+\nFPCX^{-2\coefDecayX+1} + \nFPCY^{-2\coefDecayY+1}
+n\inv\nFPCY \left(\nFPCY^{2\egnvalSpY-2\coefDecayY+3}\mbf{1}_{\{\coefDecayY-\egnvalSpY\le 1\}}\right.\\ &\quad+\left.\nFPCY(\nFPCY/\log\nFPCY)^{2\egnvalSpY-2\coefDecayY+2}\mbf{1}_{\{1<\coefDecayY-\egnvalSpY<3/2\}} +(\log\nFPCY)\mbf{1}_{\{\coefDecayY-\egnvalSpY=3/2\}} +\mbf{1}_{\{\coefDecayY-\egnvalSpY>3/2\}}\right) \\ 
&\quad +n\inv\nFPCY^{2\egnvalSpY+3}\left(n^{-2}\nFPCY^{4\egnvalSpY-2\coefDecayY+5}(\log\nFPCY)^2\mbf{1}_{\{\coefDecayY-\egnvalSpY\le 1\}} +n^{-2}\nFPCY^{2\egnvalSpY+3}\mbf{1}_{\{\coefDecayY-\egnvalSpY>1\}}\right),\eal
whence \eqref{eq:rate_oracle} follows. 
\note{
	\bi
	\item $n\inv\nFPCX^{\egnvalSpX+1}\sim \nFPCX^{1-2\coefDecayX}$ implies $\nFPCX\sim n^{1/(\egnvalSpX+2\coefDecayX)}$, whence $n^{-2}\nFPCX^{4\egnvalSpX+3}/(n\inv\nFPCX^{\egnvalSpX+1}) =n\inv\nFPCX^{3\egnvalSpX+2} \sim n^{2(\egnvalSpX-\coefDecayX+1)/(\egnvalSpX+2\coefDecayX)} \ra 0$ as $n\ra\infty$ since $\egnvalSpX+1<\coefDecayX$.
	\item $n\inv\nFPCY^{2\egnvalSpY+2}\ra 0 \Leftarrow \nFPCY = O((n/\log n)^{1/(2\egnvalSpY+2)})$.
	\item $n^{-2}\nFPCX^{2\egnvalSpX+1}\nFPCY^{2\egnvalSpY+3} \le \max\{\nFPCX^{1-2\coefDecayX},\nFPCY^{1-2\coefDecayY}\}$ requires $\nFPCY\le n^{\max\{2\coefDecayX/(2\egnvalSpY+3),\, (4\coefDecayX-1)/(2\egnvalSpY+2\coefDecayY+2)\}/(\egnvalSpX+2\coefDecayX)}$. 
	If $\nFPCY\sim n^{\max\{2\coefDecayX/(2\egnvalSpY+3),\, (4\coefDecayX-1)/(2\egnvalSpY+2\coefDecayY+2)\}/(\egnvalSpX+2\coefDecayX)}$, $\nFPCY^{1-2\coefDecayY}\sim n^{\max\{2\coefDecayX(1-2\coefDecayY)/(2\egnvalSpY+3),\, (4\coefDecayX-1)(1-2\coefDecayY)/(2\egnvalSpY+2\coefDecayY+2)\}/(\egnvalSpX+2\coefDecayX)}$. In this case, 
	\bal\nn
	&\frac{2\coefDecayX(1-2\coefDecayY)}{2\egnvalSpY+3} - (1-2\coefDecayX)
	= \frac{2\coefDecayX(2\egnvalSpY-2\coefDecayY+4)}{2\egnvalSpY+3}-1\le 0 \Leftrightarrow 2\coefDecayX(\egnvalSpY-\coefDecayY+2)\le\egnvalSpY+3/2,\\
	&\frac{(4\coefDecayX-1)(1-2\coefDecayY)}{2\egnvalSpY+2\coefDecayY+2} - (1-2\coefDecayX)
	=\frac{4\coefDecayX(\egnvalSpY-\coefDecayY+2) - (2\egnvalSpY+3)}{2\egnvalSpY+2\coefDecayY+2} \le 0 \Leftrightarrow 2\coefDecayX(\egnvalSpY-\coefDecayY+2)\le\egnvalSpY+3/2,\eal
	whence $\nFPCY^{1-2\coefDecayY}\le \nFPCX^{1-2\coefDecayX} \Leftrightarrow 2\coefDecayX(\egnvalSpY-\coefDecayY+2)\le\egnvalSpY+3/2$.
	\item If $\coefDecayY-\egnvalSpY\le 1$, we want
	\bal\nn
	\left\{\begin{array}{l}
		n\inv\nFPCY^{2\egnvalSpY-2\coefDecayY+4}\le \nFPCY^{1-2\coefDecayY} \Leftrightarrow \nFPCY\le n^{1/(2\egnvalSpY+3)}\\
		n^{-3}\nFPCY^{6\egnvalSpY-2\coefDecayY+8}(\log\nFPCY)^2\le \nFPCY^{1-2\coefDecayY}
	\end{array}\right. 
	\Leftrightarrow \nFPCY\le n^{1/(2\egnvalSpY+3)}.\eal
	Note that with $\nFPCY\le n^{1/(2\egnvalSpY+3)}$, $n^{-3}\nFPCY^{6\egnvalSpY-2\coefDecayY+8}(\log\nFPCY)^2/\nFPCY^{1-2\coefDecayY} = n^{-3}\nFPCY^{6\egnvalSpY+7}(\log\nFPCY)^2 \le (n\inv\nFPCY^{2\egnvalSpY+3})^3 \le \const$ 
	Hence, we take $\nFPCY\sim \enFPCY(n) = \min\{n^{\max\{2\coefDecayX/(2\egnvalSpY+3),\, (4\coefDecayX-1)/(2\egnvalSpY+2\coefDecayY+2)\}/(\egnvalSpX+2\coefDecayX)},\, n^{1/(2\egnvalSpY+3)}\}$, whence 
	\bal\nn
	\nFPCY^{1-2\coefDecayY}\le \nFPCX^{1-2\coefDecayX} &\Leftrightarrow 
	\left\{\begin{array}{l}
		2\coefDecayX(\egnvalSpY-\coefDecayY+2)\le\egnvalSpY+3/2\\
		(1-2\coefDecayY)/(2\egnvalSpY+3)\le (1-2\coefDecayX)/(\egnvalSpX+2\coefDecayX)
	\end{array}\right.\\
	&\Leftrightarrow 
	4\coefDecayX(\egnvalSpY-\coefDecayY+2)\le 2\egnvalSpY+3\le (2\coefDecayY-1)(\egnvalSpX+2\coefDecayX)/(2\coefDecayX-1). \eal
	Here, 
	\bal\nn
	\frac{(2\coefDecayY-1)(\egnvalSpX+2\coefDecayX)}{2\coefDecayX-1} -4\coefDecayX(\egnvalSpY-\coefDecayY+2) 
	&=\frac{(2\coefDecayY-1)(\egnvalSpX+2\coefDecayX) - 4\coefDecayX(\egnvalSpY-\coefDecayY+2)(2\coefDecayX-1) }{2\coefDecayX-1}\\
	&=\frac{2\coefDecayY(\egnvalSpX+4\coefDecayX^2) -4\egnvalSpY\coefDecayX(2\coefDecayX-1) - (\egnvalSpX-6\coefDecayX+16\coefDecayX^2)}{2\coefDecayX-1}\\
	&= \frac{(2\coefDecayY-1)\egnvalSpX+(4\egnvalSpY+6)\coefDecayX +(8\coefDecayY-8\egnvalSpY-16)\coefDecayX^2}{2\coefDecayX-1},\eal
	with maximum equal to $[(2\coefDecayY-1)\egnvalSpX + (2\egnvalSpY+3)^2/(2\egnvalSpY+4-2\coefDecayY)]/(2\coefDecayX-1)>0$. 
	\item If $1<\coefDecayY-\egnvalSpY<3/2$, we want
	\bal\nn
	\left\{\begin{array}{l}
		n\inv\nFPCY^2(\nFPCY/\log\nFPCY)^{2\egnvalSpY-2\coefDecayY+2}\le \nFPCY^{1-2\coefDecayY} \Leftrightarrow \nFPCY^{2\egnvalSpY+3}(\log\nFPCY)^{2(\coefDecayY-\egnvalSpY-1)}\le n\\
		n^{-3}\nFPCY^{4\egnvalSpY+6}\le \nFPCY^{1-2\coefDecayY} \Leftrightarrow n^{-3}\nFPCY^{4\egnvalSpY+2\coefDecayY+5} \le\const
	\end{array}\right. 
	\Leftarrow \nFPCY\sim (n/\log n)^{1/(2\egnvalSpY+3)}.\eal
	Note that with $\nFPCY\sim (n/\log n)^{1/(2\egnvalSpY+3)}$, $n^{-3}\nFPCY^{4\egnvalSpY+2\coefDecayY+5}\sim n^{-2(\egnvalSpY+2-\coefDecayY)/[(2\egnvalSpY+3)(4\egnvalSpY+2\coefDecayY+5)]}(\log n)^{-1/(2\egnvalSpY+3)} \ra 0$, as $n\ra\infty$. 
	Hence, we take $\nFPCY\sim \min\{n^{\max\{2\coefDecayX/(2\egnvalSpY+3),\, (4\coefDecayX-1)/(2\egnvalSpY+2\coefDecayY+2)\}/(\egnvalSpX+2\coefDecayX)},\, (n/\log n)^{1/(2\egnvalSpY+3)}\}$, whence 
	\bal\nn
	\nFPCY^{1-2\coefDecayY}\le \nFPCX^{1-2\coefDecayX} &\Leftrightarrow 
	\left\{\begin{array}{l}
		2\coefDecayX(\egnvalSpY-\coefDecayY+2)\le\egnvalSpY+3/2\\
		(1-2\coefDecayY)/(2\egnvalSpY+3)< (1-2\coefDecayX)/(\egnvalSpX+2\coefDecayX)
	\end{array}\right.\\
	&\Leftrightarrow 
	4\coefDecayX(\egnvalSpY-\coefDecayY+2)\le 2\egnvalSpY+3 < (2\coefDecayY-1)(\egnvalSpX+2\coefDecayX)/(2\coefDecayX-1).\eal 
	\item If $\coefDecayY-\egnvalSpY=3/2$, we want
	\bal\nn
	\left\{\begin{array}{l}
		n\inv\nFPCY\log\nFPCY\le \nFPCY^{1-2\coefDecayY} \Leftrightarrow \nFPCY^{2\egnvalSpY+3}\log\nFPCY\le n\\
		n^{-3}\nFPCY^{4\egnvalSpY+6}\le \nFPCY^{1-2\coefDecayY} \Leftrightarrow n^{-3}\nFPCY^{4\egnvalSpY+2\coefDecayY+5} \le\const
	\end{array}\right. 
	\Leftarrow \nFPCY\sim (n/\log n)^{1/(2\egnvalSpY+3)}.\eal
	\item If $\coefDecayY-\egnvalSpY>3/2$, we want
	\bal\nn
	\left\{\begin{array}{l}
		n\inv\nFPCY\le \nFPCY^{1-2\coefDecayY} \Leftrightarrow \nFPCY^{2\coefDecayY}\le n\\
		n^{-3}\nFPCY^{4\egnvalSpY+6}\le \nFPCY^{1-2\coefDecayY} \Leftrightarrow n^{-3}\nFPCY^{4\egnvalSpY+2\coefDecayY+5} \le\const
	\end{array}\right. 
	\Leftarrow \nFPCY\sim n^{1/(2\coefDecayY)},\eal
	where $3/(4\egnvalSpY+2\coefDecayY+5) - 1/(2\coefDecayY) = 4(\coefDecayY-\egnvalSpY-5/4)/[(4\egnvalSpY+2\coefDecayY+5)(2\coefDecayY)] > 0$. 
	Hence, we take $\nFPCY\sim \min\{n^{\max\{2\coefDecayX/(2\egnvalSpY+3),\, (4\coefDecayX-1)/(2\egnvalSpY+2\coefDecayY+2)\}/(\egnvalSpX+2\coefDecayX)},\, n^{1/(2\coefDecayY)}\}$, whence 
	\bal\nn
	\nFPCY^{1-2\coefDecayY}\le \nFPCX^{1-2\coefDecayX} &\Leftrightarrow 
	\left\{\begin{array}{l}
		2\coefDecayX(\egnvalSpY-\coefDecayY+2)\le\egnvalSpY+3/2\\
		(1-2\coefDecayY)/(2\coefDecayY)\le (1-2\coefDecayX)/(\egnvalSpX+2\coefDecayX)
	\end{array}\right.\\
	&\Leftrightarrow 
	\left\{\begin{array}{l}
		\coefDecayY\ge\egnvalSpY+2 - (2\egnvalSpY+3)/(4\coefDecayX)\\
		\coefDecayY\ge(\egnvalSpX+2\coefDecayX)/(2\egnvalSpX+2).
	\end{array}\right. \eal 
	\ei
} 
Furthermore, choosing $\nFPCX\sim n^{1/(\egnvalSpX+2\coefDecayX)}$, $\nFPCY\sim\enFPCY(n)$ with $\enFPCY(n)$ as defined in Theorem~\ref{thm:rate_oracle}, we have 
\bal\nn \hsNormSq{\hsSpace{\FmeanX}{\FmeanY}}{\paraTransOp_{(\FmeanXoracle,\FmeanYoracle), (\FmeanX,\FmeanY)} \regOpOracle - \regOp} = \Op\left(\max\left\{n^{-(2\coefDecayX-1)/(\egnvalSpX+2\coefDecayX)},\enFPCY(n)^{-(2\coefDecayY-1)}\right\}\right).\eal

Regarding \eqref{eq:regCoefOracle1-1}, by the atomlessness of $\FmeanX$, $\FmeanY$, $\FmeanXoracle$ and $\FmeanYoracle$ and Proposition~\ref{prop:paraTrans}, 
\bal\nn
\crovalOracle_{jk} - \croval_{jk}
&= \innerprod{\pCrosscovopXYoracle\pEgnfctnXoracle_j}{\pEgnfctnYoracle_k}_{\FmeanY} - \innerprod{\crosscovopXY\egnfctnX_j}{\egnfctnY_k}_{\FmeanY}\\
&= \innerprod{(\pCrosscovopXYoracle - \crosscovopXY)\egnfctnX_j}{\egnfctnY_k}_{\FmeanY} 
+\innerprod{\crosscovopXY(\pEgnfctnXoracle_j-\egnfctnX_j)}{\egnfctnY_k}_{\FmeanY} \\
&\quad+ \innerprod{\crosscovopXY\egnfctnX_j}{\pEgnfctnYoracle_k-\egnfctnY_k}_{\FmeanY} 
+\innerprod{(\pCrosscovopXYoracle - \crosscovopXY)\egnfctnX_j}{\pEgnfctnYoracle_k-\egnfctnY_k}_{\FmeanY} \\
&\quad+ \innerprod{(\pCrosscovopXYoracle - \crosscovopXY)(\pEgnfctnXoracle_j-\egnfctnX_j)}{\egnfctnY_k}_{\FmeanY} 
+ \innerprod{\crosscovopXY(\pEgnfctnXoracle_j-\egnfctnX_j)}{\pEgnfctnYoracle_k-\egnfctnY_k}_{\FmeanY} \\
&\quad+ \innerprod{(\pCrosscovopXYoracle - \crosscovopXY)(\pEgnfctnXoracle_j-\egnfctnX_j)}{\pEgnfctnYoracle_k-\egnfctnY_k}_{\FmeanY}\\
&\eqqcolon \wt{I}_{jk1} + \wt{I}_{jk2} + \wt{I}_{jk3} + \wt{I}_{jk4} + \wt{I}_{jk5} + \wt{I}_{jk6} + \wt{I}_{jk7},\eal
whence 
\bal\nn
\sum_{k=1}^{\nFPCY}\sum_{j=1}^{\nFPCX} \egnvalX_j^{-2}(\crovalOracle_{jk} - \croval_{jk})^2
&\le 7\sum_{l=1}^7 \sum_{k=1}^{\nFPCY}\sum_{j=1}^{\nFPCX} \egnvalX_j^{-2} \wt{I}_{jkl}^2.\eal
For $\wt{I}_{jk4},\dots,\wt{I}_{jk7}$, applying Lemma~\ref{lem:cov_oracle} and \eqref{eq:egnfctn_oracle_rate} yields \note{(w.o. $p$ series assumption of $\egnvalX_j$)}
\bal\label{eq:regCoefOracle1-1-4to7}
\sum_{k=1}^{\nFPCY}\sum_{j=1}^{\nFPCX} \egnvalX_j^{-2} \wt{I}_{jk4}^2
&\le \hsNormSq{\hsSpace{\FmeanX}{\FmeanY}}{\pCrosscovopXYoracle-\crosscovopXY} \sum_{j=1}^{\nFPCX}\egnvalX_j^{-2}  \sum_{k=1}^{\nFPCY}\|\pEgnfctnYoracle_k-\egnfctnY_k\|_{\FmeanY}^2\\
&= \Op\left(n^{-2} \nFPCX^{2\egnvalSpX+1}\nFPCY^{2\egnvalSpY+3}\right), \\
\sum_{k=1}^{\nFPCY}\sum_{j=1}^{\nFPCX} \egnvalX_j^{-2} \wt{I}_{jk5}^2
&\le \hsNormSq{\hsSpace{\FmeanX}{\FmeanY}}{\pCrosscovopXYoracle-\crosscovopXY} \sum_{j=1}^{\nFPCX}\egnvalX_j^{-2}  \|\pEgnfctnXoracle_j-\egnfctnX_j\|_{\FmeanX}^2\\
&=\Op\left(n^{-2} \nFPCX^{4\egnvalSpX+3}\right),\\
\sum_{k=1}^{\nFPCY}\sum_{j=1}^{\nFPCX} \egnvalX_j^{-2} \wt{I}_{jk6}^2
&\le \sum_{j=1}^{\nFPCX} \egnvalX_j^{-2}\|\crosscovopXY(\pEgnfctnXoracle_j-\egnfctnX_j)\|_{\FmeanY}^2 \sum_{k=1}^{\nFPCY}\|\pEgnfctnYoracle_k-\egnfctnY_k\|_{\FmeanY}^2 \\
&= \left(\sum_{k=1}^{\nFPCY}\sum_{j=1}^{\nFPCX} \egnvalX_j^{-2} \wt{I}_{jk3}^2\middle) \middle(\sum_{k=1}^{\nFPCY}\|\pEgnfctnYoracle_k-\egnfctnY_k\|_{\FmeanY}^2\right), \\
\sum_{k=1}^{\nFPCY}\sum_{j=1}^{\nFPCX} \egnvalX_j^{-2} \wt{I}_{jk7}^2
&\le\hsNormSq{\hsSpace{\FmeanX}{\FmeanY}}{\pCrosscovopXYoracle-\crosscovopXY} \sum_{k=1}^{\nFPCY} \|\pEgnfctnYoracle_k-\egnfctnY_k\|_{\FmeanY}^2\sum_{j=1}^{\nFPCX} \egnvalX_j^{-2}\|\pEgnfctnXoracle_j-\egnfctnX_j\|_{\FmeanX}^2\\
&=\Op\left(n^{-3}\nFPCX^{4\egnvalSpX+3}\nFPCY^{2\egnvalSpY+3}\right).\eal

For $\wt{I}_{jk1}$, we observe that 
\bal\label{eq:croval_oracle_decom}
\wt{I}_{jk1}
&= \innerprod{(\pCrosscovopXYoracle - \crosscovopXY)\egnfctnX_j}{\egnfctnY_k}_{\FmeanY}\\
&= \frac{1}{n}\sum_{i=1}^n \innerprod{\Log_{\FmeanX}\Xi}{\egnfctnX_j}_{\FmeanX} \innerprod{\Log_{\FmeanY}\Yi}{\egnfctnY_k}_{\FmeanY} -\innerprod{\Log_{\FmeanX}\FmeanXoracle}{\egnfctnX_j}_{\FmeanX} \innerprod{\Log_{\FmeanY}\FmeanYoracle}{\egnfctnY_k}_{\FmeanY}\\
&\quad -\expect\left(\innerprod{\Log_{\FmeanX}\X}{\egnfctnX_j}_{\FmeanX} \innerprod{\Log_{\FmeanY}\Y}{\egnfctnY_k}_{\FmeanY}\right). \eal
In conjunction with \ref{ass:variation} and the fact that $\{(\Xi,\Yi)\}_{i=1}^n$ are independent realizations of $(\X,\Y)$, it follows that 
\bal\label{eq:croval_oracle_rate1}
\expect(\wt{I}_{jk1}^2)
&\le \frac{2}{n}\var\left(\innerprod{\Log_{\FmeanX}\X}{\egnfctnX_j}_{\FmeanX} \innerprod{\Log_{\FmeanY}\Y}{\egnfctnY_k}_{\FmeanY}\right)\\
&\quad + 2\expect\left[\left(\frac{1}{n}\sum_{i=1}^n\innerprod{\Log_{\FmeanX}\Xi}{\egnfctnX_j}_{\FmeanX}\right)^2 \left(\frac{1}{n}\sum_{i=1}^n\innerprod{\Log_{\FmeanY}\Yi}{\egnfctnY_k}_{\FmeanY}\right)^2\right]\\
&\le \const n\inv \egnvalX_j\egnvalY_k,\eal
uniformly in $j,k$, whence we have
\bal\label{eq:regCoefOracle1-1-1}
\sum_{k=1}^{\nFPCY}\sum_{j=1}^{\nFPCX} \egnvalX_j^{-2} \wt{I}_{jk1}^2 &= \Op\left(n\inv\nFPCX^{\egnvalSpX+1}\right). \eal 

For $\wt{I}_{jk2}$, note that $\covopXoracle$ has at most $(n-1)$ non-zero eigenvalues, i.e., $\egnvalXoracle_j = 0$ for all $j\ge n$. In conjunction with the atomlessness of $\FmeanX$ and $\FmeanXoracle$ and Proposition~\ref{prop:paraTrans}, we have 
\bal\nn
\pCovopXoracle = \sum_{j=1}^{n-1} \egnvalXoracle_j \pEgnfctnXoracle_j\otimes \pEgnfctnXoracle_j 
= \sum_{j=1}^\infty \egnvalXoracle_j \pEgnfctnXoracle_j\otimes \pEgnfctnXoracle_j, \eal
and $\{\egnvalXoracle_j\}_{j=1}^{\infty}$ and $\{\pEgnfctnXoracle_j\}_{j=1}^{\infty}$ are the eigenvalues and eigenfunctions of $\pCovopXoracle$. 
Hence, applying Lemma~5.1 of \citet{hall:07:1} yields 
\bal\nn
\wt{I}_{jk2}
&=\innerprod{\crosscovopXY(\pEgnfctnXoracle_j-\egnfctnX_j)}{\egnfctnY_k}_{\FmeanY}\\
&=\croval_{jk}\innerprod{\pEgnfctnXoracle_j-\egnfctnX_j}{\egnfctnX_j}_{\FmeanX} + \sum_{j':\,j'\ne j} \croval_{j'k} (\egnvalXoracle_{j'}-\egnvalX_j)\inv \innerprod{(\pCovopXoracle-\covopX)\pEgnfctnXoracle_j}{\egnfctnX_{j'}}_{\FmeanX}\\
&= \wt{I}_{jk2}^{(1)} + \wt{I}_{jk2}^{(2)} + \wt{I}_{jk2}^{(3)} + \wt{I}_{jk2}^{(4)},\eal
where 
\bal\nn
\wt{I}_{jk2}^{(1)} &\coloneqq \croval_{jk}\innerprod{\pEgnfctnXoracle_j-\egnfctnX_j}{\egnfctnX_j}_{\FmeanX},\\
\wt{I}_{jk2}^{(2)} &\coloneqq \sum_{j':\,j'\ne j} \croval_{j'k} (\egnvalX_{j'}-\egnvalX_j)\inv \innerprod{(\pCovopXoracle-\covopX)\egnfctnX_j}{\egnfctnX_{j'}}_{\FmeanX},\\
\wt{I}_{jk2}^{(3)} &\coloneqq \sum_{j':\,j'\ne j} \croval_{j'k} \left((\egnvalXoracle_{j'}-\egnvalX_j)\inv - (\egnvalX_{j'}-\egnvalX_j)\inv\right)\ \innerprod{(\pCovopXoracle-\covopX)\egnfctnX_j}{\egnfctnX_{j'}}_{\FmeanX},\\
\wt{I}_{jk2}^{(4)} &\coloneqq \sum_{j':\,j'\ne j} \croval_{j'k} (\egnvalXoracle_{j'}-\egnvalX_j)\inv \innerprod{(\pCovopXoracle-\covopX)(\pEgnfctnXoracle_j-\egnfctnX_j)}{\egnfctnX_{j'}}_{\FmeanX}.\eal

We define events 
\bal\nn
\wt\event_{2\nFPCX} = \wt\event_{2\nFPCX}(n)
= \left\{(\egnvalXoracle_j-\egnvalX_{j'})^{-2} \le 2(\egnvalX_j-\egnvalX_{j'})^{-2},\text{ for all } j,j'=1,\dots,\nFPCX\text{ s.t. }j\ne j' \right\}.\eal
Note that by \ref{ass:eigenval_spacingX},
\bal\label{eq:egnval_diff_rate}
|\egnvalX_j - \egnvalX_{j'}| \ge 
\left\{\begin{array}{ll}
	|\egnvalX_j - \egnvalX_{2j}| \ge C_1\sum_{l=j}^{2j-1} l^{-\egnvalSpX-1} \ge C_2 j^{-\egnvalSpX}, &\text{if }j'\ge 2j,\\
	|\egnvalX_{j'} - \egnvalX_{2j'}| \ge C_2 {j'}^{-\egnvalSpX}, &\text{if }j'\le j/2,\\
	C_1 \sum_{j\wedge j' \le l < j\vee j'} l^{-\egnvalSpX-1} \ge C_2 |j-j'| j^{-\egnvalSpX-1}, &\text{if }j/2<j'<j \text{ or } j<j'<2j,
\end{array}\right.\eal
where $C_1,C_2>0$ are constants that do not depend on $j,j'$. 
Hence, $|\egnvalX_j-\egnvalX_{j'}|\inv = O(\nFPCX^{\egnvalSpX+1})$, uniformly in distinct $j,j'=1,\dots,\nFPCX$, and in conjunction with Lemma~\ref{lem:cov_oracle}, this implies that with some $t_{j,j'}\in(0,1)$, 
\bal\nn
\frac{\left|(\egnvalXoracle_j-\egnvalX_{j'})^{-2} - (\egnvalX_j-\egnvalX_{j'})^{-2}\right|}{(\egnvalX_j-\egnvalX_{j'})^{-2}}
&= \frac{2|\egnvalX_j-\egnvalX_{j'} + t_{j,j'}(\egnvalXoracle_j-\egnvalX_j)|^{-3} |\egnvalXoracle_j-\egnvalX_j|}{(\egnvalX_j-\egnvalX_{j'})^{-2}}\\
&= 2\left|1 + t_{j,j'}\frac{\egnvalXoracle_j-\egnvalX_j}{\egnvalX_j-\egnvalX_{j'}}\right|^{-3} |\egnvalXoracle_j-\egnvalX_j||\egnvalX_j-\egnvalX_{j'}|\inv\\
&=\Op\left(n^{-1/2}\nFPCX^{\egnvalSpX+1}\right),\eal
uniformly in $j,j'$. 
Hence, under the assumptions of Theorem~\ref{thm:rate_oracle}, $\prob(\wt\event_{2\nFPCX})\ra 1$ as $n\ra\infty$\note{, since $n^{-1/2}\nFPCX^{\egnvalSpX+1} = o(1)$}. 

By \ref{ass:coef_decay} and \eqref{eq:egnfctn_oracle_rate}, \note{(w.o. $p$ series assumption of $\egnvalX_j$)}
\bal\label{eq:regCoefOracle1-1-2-1}
\sum_{k=1}^{\nFPCY}\sum_{j=1}^{\nFPCX} \egnvalX_j^{-2} (\wt{I}_{jk2}^{(1)})^2
&\le \sum_{k=1}^{\nFPCY}\sum_{j=1}^{\nFPCX} \regCoef_{jk}^2 \|\pEgnfctnXoracle_j-\egnfctnX_j\|_{\FmeanX}^2
\le \const\sum_{j=1}^{\nFPCX} j^{-2\coefDecayX} \|\pEgnfctnXoracle_j-\egnfctnX_j\|_{\FmeanX}^2\\
&= \Op\left(n\inv\sum_{j=1}^{\nFPCX} j^{2\egnvalSpX-2\coefDecayX+2}\right) 
= \op\left(n\inv \nFPCX^{\egnvalSpX+1}\right). \note{\text{ if }\egnvalSpX+1<\coefDecayX}\eal

For $\wt{I}_{jk2}^{(2)}$, using similar arguments to \eqref{eq:croval_oracle_decom}, it can be shown that 
\bal\nn
n\expect(\wt{I}_{jk2}^{(2)})^2
&\le\const
\expect\left(\sum_{j':\,j'\ne j} \croval_{j'k} (\egnvalX_{j'}-\egnvalX_j)\inv \innerprod{\Log_{\FmeanX}\X}{\egnfctnX_j}_{\FmeanX}
\innerprod{\Log_{\FmeanX}\X}{\egnfctnX_{j'}}_{\FmeanX}\right)^2\\
&= \const
\expect\left(\innerprod{\Log_{\FmeanX}\X}{\egnfctnX_j}_{\FmeanX}^2\right)
\sum_{j':\,j'\ne j} \regCoef_{j'k}^2\egnvalX_{j'}^2 (\egnvalX_{j'}-\egnvalX_j)^{-2} \expect\left(\innerprod{\Log_{\FmeanX}\X}{\egnfctnX_{j'}}_{\FmeanX}^2\right)\\
&= \const \egnvalX_j \sum_{j':\,j'\ne j} \egnvalX_{j'}^3 \regCoef_{j'k}^2 (\egnvalX_{j'}-\egnvalX_j)^{-2}\\
&\le \const \egnvalX_j k^{-2\coefDecayY} \sum_{j':\,j'\ne j} \egnvalX_{j'}^3 {j'}^{-2\coefDecayX} (\egnvalX_{j'}-\egnvalX_j)^{-2}.\eal
By \eqref{eq:egnval_diff_rate}, under \ref{ass:eigenval_spacingX} and \ref{ass:coef_decay}, $\sum_{j':\,j'\ne j} \egnvalX_{j'}^3 {j'}^{-2\coefDecayX} (\egnvalX_{j'}-\egnvalX_j)^{-2}$ is bounded by a multiple of \note{(w.o. $p$ series assumption of $\egnvalX_j$)}
\bal\nn
&\left(j^{2\egnvalSpX}\sum_{j':\,j'\ge 2j} \egnvalX_{j'}^3 {j'}^{-2\coefDecayX}\right)
+ \left(j^{2\egnvalSpX+2} \sum_{\substack{j':\,j'\ne j,\, j/2<j'<2j}} \egnvalX_{j'}^3 {j'}^{-2\coefDecayX} |j-j'|^{-2}\right)
+ \left(\sum_{j':\,j'\le j/2} \egnvalX_{j'}^3 {j'}^{2\egnvalSpX-2\coefDecayX}\right)\\
&\le \const \left(j^{2\egnvalSpX-2\coefDecayX+1} + j^{2\egnvalSpX-2\coefDecayX+2} + \max_{j':\,j'\le j/2}{j'}^{2\egnvalSpX-2\coefDecayX}\right)
\le \const, \note{\text{ if } \egnvalSpX+1<\coefDecayX}  \eal
uniformly in $j$. Therefore, 
\bal\nn
\expect(\wt{I}_{jk2}^{(2)})^2\le\const n\inv\egnvalX_j k^{-2\coefDecayY},\eal
and hence 
\bal\label{eq:regCoefOracle1-1-2-2}
\sum_{k=1}^{\nFPCY}\sum_{j=1}^{\nFPCX} \egnvalX_j^{-2} (\wt{I}_{jk2}^{(2)})^2
&= \Op\left(n\inv\sum_{k=1}^{\nFPCY}\sum_{j=1}^{\nFPCX} \egnvalX_j\inv k^{-2\coefDecayY}\right) 
= \Op\left(n\inv \nFPCX^{\egnvalSpX+1}\right).\eal

For $\wt{I}_{jk2}^{(3)}$, on $\wt\event_{2\nFPCX}$, we have 
\bal\nn
(\wt{I}_{jk2}^{(3)})^2 
&\le \left[\sum_{j':\,j'\ne j} |\croval_{j'k}| \left|(\egnvalXoracle_{j'}-\egnvalX_j)\inv - (\egnvalX_{j'}-\egnvalX_j)\inv\right| \left|\innerprod{(\pCovopXoracle-\covopX)\egnfctnX_j}{\egnfctnX_{j'}}_{\FmeanX}\right|\right]^2\\
&\le \const \left[\sum_{j':\,j'\ne j} \egnvalX_{j'} |\regCoef_{j'k}| \frac{|\egnvalXoracle_{j'}-\egnvalX_{j'}|}{(\egnvalX_{j'}-\egnvalX_j)^2} \left|\innerprod{(\pCovopXoracle-\covopX)\egnfctnX_j}{\egnfctnX_{j'}}_{\FmeanX}\right|\right]^2\\
&\le \const k^{-2\coefDecayY}  \|(\pCovopXoracle-\covopX)\egnfctnX_j\|_{\FmeanX}^2 \sum_{j':\,j'\ne j} \egnvalX_{j'}^2 {j'}^{-2\coefDecayX}  \frac{|\egnvalXoracle_{j'}-\egnvalX_{j'}|^2}{(\egnvalX_{j'}-\egnvalX_j)^4} \\
&\le \const k^{-2\coefDecayY}  \hsNorm{\hsSpaceAuto{\FmeanX}}{\pCovopXoracle-\covopX}^4 \sum_{j':\,j'\ne j} \egnvalX_{j'}^2 {j'}^{-2\coefDecayX}(\egnvalX_{j'}-\egnvalX_j)^{-4}. \eal
Under \ref{ass:eigenval_spacingX},  \eqref{eq:egnval_diff_rate} implies \note{(w.o. $p$ series assumption of $\egnvalX_j$)}
\bal\nn
\sum_{j':\,j'\ge 2j}\egnvalX_{j'}^2 {j'}^{-2\coefDecayX}(\egnvalX_{j'}-\egnvalX_j)^{-4}
&\le\const j^{4\egnvalSpX} \sum_{j':\,j'\ge 2j}\egnvalX_{j'}^2 {j'}^{-2\coefDecayX}
\le \const j^{4\egnvalSpX-2\coefDecayX},\\
\sum_{j':\,j'\le j/2}\egnvalX_{j'}^2{j'}^{-2\coefDecayX}(\egnvalX_{j'}-\egnvalX_j)^{-4}
&\le\const \sum_{j':\,j'\le j/2}\egnvalX_{j'}^2 {j'}^{4\egnvalSpX-2\coefDecayX}
\le \const \max_{j':\,j'\le j/2}{j'}^{4\egnvalSpX-2\coefDecayX},\\
\sum_{\substack{j':\,j'\ne j,\\ j/2<j'<2j}} \egnvalX_{j'}^2{j'}^{-2\coefDecayX}(\egnvalX_{j'}-\egnvalX_j)^{-4}
&\le\const j^{4\egnvalSpX+4}\sum_{\substack{j':\,j'\ne j,\\ j/2<j'<2j}}\egnvalX_{j'}^2 {j'}^{-2\coefDecayX}|j-j'|^{-4} 
\le \const j^{4\egnvalSpX-2\coefDecayX+4}.\eal
Therefore, \note{(w.o. $p$ series assumption of $\egnvalX_j$)}
\bal\nn
\sum_{j':\,j'\ne j} \egnvalX_{j'}^2 {j'}^{-2\coefDecayX}(\egnvalX_{j'}-\egnvalX_j)^{-4}
\le\const \left(1+ j^{4\egnvalSpX-2\coefDecayX+4}\right),\eal
whence \note{(w.o. $p$ series assumption of $\egnvalX_j$)} 
\bal\label{eq:regCoefOracle1-1-2-3}
\sum_{k=1}^{\nFPCY}\sum_{j=1}^{\nFPCX} \egnvalX_j^{-2}(\wt{I}_{jk2}^{(3)})^2 
&= \Op\left(n^{-2} \sum_{j=1}^{\nFPCX}j^{2\egnvalSpX} \left(1+ j^{4\egnvalSpX-2\coefDecayX+4}\right)\right)
= \Op\left(n^{-2}\left(\nFPCX^{2\egnvalSpX+1} + \nFPCX^{6\egnvalSpX-2\coefDecayX+5} \right)\right).\eal

For $\wt{I}_{jk2}^{(4)}$, on $\wt\event_{2\nFPCX}$, we have 
\bal\nn
|\wt{I}_{jk2}^{(4)}|
&\le \const k^{-\coefDecayY}\sum_{j':\,j'\ne j} {j'}^{-\coefDecayX} \egnvalX_{j'} |\egnvalX_{j'}-\egnvalX_j|\inv |\innerprod{(\pCovopXoracle-\covopX)(\pEgnfctnXoracle_j-\egnfctnX_j)}{\egnfctnX_{j'}}_{\FmeanX}|\\
&\le \const k^{-\coefDecayY} \hsNorm{\hsSpaceAuto{\FmeanX}}{\pCovopXoracle-\covopX} \|\pEgnfctnXoracle_j-\egnfctnX_j\|_{\FmeanX} \sum_{j':\,j'\ne j} {j'}^{-\coefDecayX} \egnvalX_{j'} |\egnvalX_{j'}-\egnvalX_j|\inv, \eal
where by \eqref{eq:egnval_diff_rate}, $\sum_{j':\,j'\ne j} {j'}^{-\coefDecayX} \egnvalX_{j'} |\egnvalX_{j'}-\egnvalX_j|\inv$ is bounded by a multiple of \note{(w.o. $p$ series assumption of $\egnvalX_j$)} 
\bal\nn
&\left(j^{\egnvalSpX}\sum_{j':\,j'\ge 2j} \egnvalX_{j'} {j'}^{-\coefDecayX}\right)
+ \left(j^{\egnvalSpX+1} \sum_{\substack{j':\,j'\ne j,\, j/2<j'<2j}} \egnvalX_{j'} {j'}^{-\coefDecayX} |j-j'|^{-1}\right)
+ \left(\sum_{j':\,j'\le j/2} \egnvalX_{j'} {j'}^{\egnvalSpX-\coefDecayX}\right)\\
&\le \const\left(j^{\egnvalSpX-\coefDecayX} + j^{\egnvalSpX-\coefDecayX+1} + \max_{j':\,j'\le j/2}{j'}^{\egnvalSpX-\coefDecayX}\right) 
\le \const \note{\text{ if } \egnvalSpX+1<\coefDecayX} \eal
Therefore, \note{(w.o. $p$ series assumption of $\egnvalX_j$)} 
\bal\label{eq:regCoefOracle1-1-2-4}
\sum_{k=1}^{\nFPCY}\sum_{j=1}^{\nFPCX} \egnvalX_j^{-2}(\wt{I}_{jk2}^{(4)})^2 
&=\Op\left(n^{-2}\sum_{j=1}^{\nFPCX}j^{4\egnvalSpX+2}\right)
= \Op\left(n^{-2}\nFPCX^{4\egnvalSpX+3}\right).\eal
Combining \eqref{eq:regCoefOracle1-1-2-1}--\eqref{eq:regCoefOracle1-1-2-4} yields \note{(w.o. $p$ series assumption of $\egnvalX_j$)}
\bal\label{eq:regCoefOracle1-1-2}
\sum_{k=1}^{\nFPCY}\sum_{j=1}^{\nFPCX} \egnvalX_j^{-2} \wt{I}_{jk2}^2 
&= \Op\left(n\inv \nFPCX^{\egnvalSpX+1}\right) 
+ \Op\left(n^{-2}\nFPCX^{4\egnvalSpX+3}\right). \note{\text{ if } \egnvalSpX+1<\coefDecayX}\eal

For $\wt{I}_{jk3} =\innerprod{\crosscovopXY\egnfctnX_j}{\pEgnfctnYoracle_k-\egnfctnY_k}_{\FmeanY}
= \wt{I}_{jk3}^{(1)} + \wt{I}_{jk3}^{(2)} + \wt{I}_{jk3}^{(3)} + \wt{I}_{jk3}^{(4)}$, 
where 
\bal\nn
\wt{I}_{jk3}^{(1)} &\coloneqq \croval_{jk}\innerprod{\pEgnfctnYoracle_k-\egnfctnY_k}{\egnfctnY_k}_{\FmeanY},\\
\wt{I}_{jk3}^{(2)} &\coloneqq \sum_{k':\,k'\ne k} \croval_{jk'} (\egnvalY_{k'}-\egnvalY_k)\inv \innerprod{(\pCovopYoracle-\covopY)\egnfctnY_k}{\egnfctnY_{k'}}_{\FmeanY},\\
\wt{I}_{jk3}^{(3)} &\coloneqq \sum_{k':\,k'\ne k} \croval_{jk'} \left((\egnvalYoracle_{k'}-\egnvalY_k)\inv - (\egnvalY_{k'}-\egnvalY_k)\inv\right) \innerprod{(\pCovopYoracle-\covopY)\egnfctnY_k}{\egnfctnY_{k'}}_{\FmeanY},\\
\wt{I}_{jk3}^{(4)} &\coloneqq \sum_{k':\,k'\ne k} \croval_{jk'} (\egnvalYoracle_{k'}-\egnvalY_k)\inv \innerprod{(\pCovopYoracle-\covopY)(\pEgnfctnYoracle_k-\egnfctnY_k)}{\egnfctnY_{k'}}_{\FmeanY}.\eal
Define events 
\bal\nn
\wt\event_{3\nFPCY} = \wt\event_{3\nFPCY}(n)
= \left\{(\egnvalYoracle_k-\egnvalY_{k'})^{-2} \le 2(\egnvalY_k-\egnvalY_{k'})^{-2},\text{ for all } k,k'=1,\dots,\nFPCY\text{ s.t. }k\ne k' \right\}.\eal
Using similar arguments to the proof of \eqref{eq:regCoefOracle1-1-2}, it can be shown that $\prob(\wt\event_{3\nFPCY})\ra 1$, as $n\ra\infty$\note{, since $n^{-1/2}\nFPCY^{\egnvalSpY+1}\ra 0$} and that on $\wt\event_{3\nFPCY}$, \note{(w.o. $p$ series assumption of $\egnvalX_j$)}
\bal\label{eq:regCoefOracle1-1-3-1to4}
\sum_{k=1}^{\nFPCY}\sum_{j=1}^{\nFPCX} \egnvalX_j^{-2} (\wt{I}_{jk3}^{(1)})^2
&\le \sum_{k=1}^{\nFPCY} \sum_{j=1}^{\nFPCX} \regCoef_{jk}^2 \|\pEgnfctnYoracle_k-\egnfctnY_k\|_{\FmeanY}^2 
=\Op\left(n\inv\sum_{k=1}^{\nFPCY}k^{2\egnvalSpY-2\coefDecayY+2}\right),\\
\sum_{k=1}^{\nFPCY}\sum_{j=1}^{\nFPCX} \egnvalX_j^{-2} (\wt{I}_{jk3}^{(2)})^2
&= \Op\left(n\inv\sum_{k=1}^{\nFPCY} \egnvalY_k \sum_{k':\,k'\ne k} {k'}^{-2\coefDecayY} (\egnvalY_{k'}-\egnvalY_k)^{-2} \egnvalY_{k'}\right)\\
&=\Op\left(n\inv \max_{1\le k\le\nFPCY}k^{2\egnvalSpY-2\coefDecayY+2} \right),\\
\sum_{k=1}^{\nFPCY}\sum_{j=1}^{\nFPCX} \egnvalX_j^{-2} (\wt{I}_{jk3}^{(3)})^2
&=\Op\left(n^{-2}\sum_{k=1}^{\nFPCY} \sum_{k':\,k'\ne k} {k'}^{-2\coefDecayY}(\egnvalY_{k'}-\egnvalY_k)^{-4} \right)\\
&=\Op\left(n^{-2}\nFPCY+ n^{-2}\sum_{k=1}^{\nFPCY} k^{4\egnvalSpY-2\coefDecayY+4}\right)
=\Op\left(n^{-2}\nFPCY + n^{-2}\nFPCY^{4\egnvalSpY-2\coefDecayY+5}\right),\\
\sum_{k=1}^{\nFPCY}\sum_{j=1}^{\nFPCX} \egnvalX_j^{-2} (\wt{I}_{jk3}^{(4)})^2
&=\Op\left(n^{-2}\sum_{k=1}^{\nFPCY}  k^{2\egnvalSpY+2} \left(\sum_{k':\,k'\ne k} {k'}^{-\coefDecayY}|\egnvalY_{k'}-\egnvalY_k|\inv\right)^2\right)\\
&=\left\{\begin{array}{ll}
	\Op\left(n^{-2}\nFPCY^{4\egnvalSpY-2\coefDecayY+5}(\log \nFPCY)^2\right), &\text{if }\egnvalSpY-\coefDecayY\ge -1,\\
	\Op\left(n^{-2}\nFPCY^{2\egnvalSpY+3}\right), &\text{if }\egnvalSpY-\coefDecayY<-1.
\end{array}\right. \eal
Here, under \ref{ass:eigenval_spacingY}, similar arguments to \eqref{eq:egnval_diff_rate} imply that  $\sum_{k':\,k'\ne k} {k'}^{-2\coefDecayY} (\egnvalY_{k'}-\egnvalY_k)^{-2} \egnvalY_{k'}$ is bounded by a multiple of \note{(w.o. $p$ series assumption of $\egnvalX_j$)} 
\bal\label{eq:sum-2-2-1}
&k^{2\egnvalSpY}\left(\sum_{k':\,k'\ge 2k} {k'}^{-2\coefDecayY} \egnvalY_{k'}\right) 
+ k^{2\egnvalSpY+2}\left(\sum_{\substack{k':\,k'\ne k,\, k/2<k'<2k}} |k'-k|^{-2} {k'}^{-2\coefDecayY} \egnvalY_{k'}\right) + \left(\sum_{k':\,k'\le k/2} {k'}^{2\egnvalSpY-2\coefDecayY} \egnvalY_{k'}\right)\\
&\le \const \left(k^{2\egnvalSpY-2\coefDecayY+2} + \max_{k':\,k'\le k/2} {k'}^{2\egnvalSpY-2\coefDecayY}\right)\\
&\le\const \left\{\begin{array}{ll}
	k^{2\egnvalSpY-2\coefDecayY+2}, &\text{if }\egnvalSpY-\coefDecayY\ge 0,\\
	k^{2\egnvalSpY-2\coefDecayY+2}+1, &\text{if }\egnvalSpY-\coefDecayY<0,
\end{array}\right.\eal
that $\sum_{k':\,k'\ne k} {k'}^{-2\coefDecayY}(\egnvalY_{k'}-\egnvalY_k)^{-4}$  is bounded by a multiple of 
\bal\nn
&k^{4\egnvalSpY}\left(\sum_{k':\,k'\ge 2k} {k'}^{-2\coefDecayY} \right) 
+ k^{4\egnvalSpY+4}\left(\sum_{\substack{k':\,k'\ne k,\, k/2<k'<2k}} |k'-k|^{-4} {k'}^{-2\coefDecayY} \right) + \left(\sum_{k':\,k'\le k/2} {k'}^{4\egnvalSpY-2\coefDecayY} \right)\\
&\le \const \left\{\begin{array}{ll}
	k^{4\egnvalSpY-2\coefDecayY+4}, &\text{if }2\egnvalSpY-\coefDecayY\ge -1/2,\\
	k^{4\egnvalSpY-2\coefDecayY+4}+ 1, &\text{if }2\egnvalSpY-\coefDecayY<-1/2,\\
\end{array}\right. \eal
and that $\sum_{k':\,k'\ne k} {k'}^{-\coefDecayY}|\egnvalY_{k'}-\egnvalY_k|\inv$ is bounded by a multiple of 
\bal\nn
&k^{\egnvalSpY}\left(\sum_{k':\,k'\ge 2k} {k'}^{-\coefDecayY} \right) 
+ k^{\egnvalSpY+1}\left(\sum_{\substack{k':\,k'\ne k,\, k/2<k'<2k}} |k'-k|\inv {k'}^{-\coefDecayY} \right) + \left(\sum_{k':\,k'\le k/2} {k'}^{\egnvalSpY-\coefDecayY} \right)\\
&\le \const \left\{\begin{array}{ll}
	k^{\egnvalSpY-\coefDecayY+1} + k^{\egnvalSpY-\coefDecayY+1}\log k, &\text{if }\egnvalSpY-\coefDecayY\in[0,\infty)\cup\{-1\},\\
	k^{\egnvalSpY-\coefDecayY+1}\log k, &\text{if }\egnvalSpY-\coefDecayY\in(-1,0),\\
	1 + k^{\egnvalSpY-\coefDecayY+1}\log k, &\text{if }\egnvalSpY-\coefDecayY<-1.
\end{array}\right. \note{\text{ if }\coefDecayY>1} \eal
Therefore, \note{(w.o. $p$ series assumption of $\egnvalX_j$)} 
\bal\label{eq:regCoefOracle1-1-3}
\sum_{k=1}^{\nFPCY}\sum_{j=1}^{\nFPCX}\egnvalX_j^{-2}\wt{I}_{jk3}^2 
&=\Op\left(n\inv\sum_{k=1}^{\nFPCY}k^{2\egnvalSpY-2\coefDecayY+2}\right)\\
&\quad + 
\left\{\begin{array}{ll}
	\Op\left(n^{-2}\nFPCY^{4\egnvalSpY-2\coefDecayY+5}(\log \nFPCY)^2\right), &\text{if }\egnvalSpY-\coefDecayY\ge -1,\\
	\Op\left(n^{-2}\nFPCY^{2\egnvalSpY+3}\right), &\text{if }\egnvalSpY-\coefDecayY<-1.
\end{array}\right.\eal
Combining \eqref{eq:regCoefOracle1-1-4to7}, \eqref{eq:regCoefOracle1-1-1}, \eqref{eq:regCoefOracle1-1-2} and \eqref{eq:regCoefOracle1-1-3} yields \eqref{eq:regCoefOracle1-1}, which completes the proof. 
\epf

\bpf[Proof of Corollary~\ref{cor:pred_oracle}]
First note that for two distributions $\distnOne, \distnTwo \in\manifold$, if $\distnTwo$ is atomless, then 
$\wdist^2(\Exp_{\distnOne}\arbiFctn, \Exp_{\distnTwo}\paraTrans_{\distnOne,\distnTwo}\arbiFctn)= \wdist^2(\distnOne, \distnTwo)$. 
Thus, 
\bal\nn
&\wdist^2\left(\YpredOracle(\arbiDistn), \expect_\oplus(\Y|\X = \arbiDistn)\right)\\
&=\wdist^2\left(\Exp_{\FmeanYoracle} \int_{\dom} \regKernelOracle(s,\cdot) \Log_{\FmeanXoracle} \arbiDistn(s)\diffop \FmeanXoracle(s),
\Exp_{\FmeanY}[\expect(\Log_{\FmeanY}\Y|\Log_{\FmeanX}\X=\Log_{\FmeanX}\arbiDistn)]\right)\\
&\le 2\wdist^2\left(\Exp_{\FmeanY} \paraTrans_{\FmeanYoracle,\FmeanY} \int_{\dom} \regKernelOracle(s,\cdot) \Log_{\FmeanXoracle} \arbiDistn(s)\diffop \FmeanXoracle(s),\Exp_{\FmeanY}\int_{\dom}\regKernel(s,\cdot)\Log_{\FmeanX}\arbiDistn(s)\diffop\FmeanX(s)\right)\\ 
&\qquad + 2\wdist^2(\FmeanYoracle,\FmeanY)\\
&= 2\left\|\paraTrans_{\FmeanYoracle,\FmeanY} \int_{\dom} \regKernelOracle(s,\cdot) \Log_{\FmeanXoracle} \arbiDistn(s)\diffop \FmeanXoracle(s) - \int_{\dom}\regKernel(s,\cdot)\Log_{\FmeanX}\arbiDistn(s)\diffop\FmeanX(s)\right\|_{\FmeanY}^2 
+ 2\wdist^2(\FmeanYoracle,\FmeanY). \eal
Note that by \ref{ass:atomless} and Proposition~\ref{prop:paraTrans},
\be\nn \paraTrans_{\FmeanYoracle,\FmeanY} \int_{\dom} \regKernelOracle(s,\cdot) \Log_{\FmeanXoracle} \arbiDistn (s)\diffop \FmeanXoracle(s) = \int_{\dom} \pRegKernelOracle(s,\cdot) \pLogXoracle \arbiDistn (s)\diffop \FmeanX(s).\ee
Hence,
\bal\nn
&\wdist^2\left(\YpredOracle(\arbiDistn), \expect_\oplus(\Y|\X = \arbiDistn)\right)\\
&\le 2\left\|\int_{\dom} \pRegKernelOracle(s,\cdot) \pLogXoracle \arbiDistn (s)\diffop \FmeanX(s) - \int_{\dom}\regKernel(s,\cdot)\Log_{\FmeanX}\arbiDistn(s)\diffop\FmeanX(s)\right\|_{\FmeanY}^2 
+ 2\wdist^2(\FmeanYoracle,\FmeanY)\\
&\le 4\left\|\int_{\dom} \left[\pRegKernelOracle(s,\cdot) - \regKernel(s,\cdot)\right] \pLogXoracle \arbiDistn (s)\diffop \FmeanX(s)\right\|_{\FmeanY}^2\\
&\quad + 4\left\|\int_{\dom} \regKernel(s,\cdot) \left[ \pLogXoracle \arbiDistn(s)-\Log_{\FmeanX}\arbiDistn(s)\right] \diffop\FmeanX(s)\right\|_{\FmeanY}^2 
+ 2\wdist^2(\FmeanYoracle,\FmeanY)\\
&\le 4\|\pLogXoracle\arbiDistn\|_{\FmeanX}^2 \|\pRegKernelOracle - \regKernel\|_{\jointmean}^2
+ 4\|\pLogXoracle\arbiDistn - \Log_{\FmeanX}\arbiDistn\|_{\FmeanX}^2 \|\regKernel\|_{\jointmean}^2 + 2\wdist^2(\FmeanYoracle,\FmeanY). \eal 
Furthermore, 
\bal\nn
&\|\pLogXoracle\arbiDistn\|_{\FmeanX}^2 
\le 2\|\pLogXoracle\arbiDistn - \Log_{\FmeanX}\arbiDistn\|_{\FmeanX}^2 + 2\wdist^2(\FmeanX,\arbiDistn),\\
&\|\pLogXoracle\arbiDistn - \Log_{\FmeanX}\arbiDistn\|_{\FmeanX}^2
=\wdist^2(\FmeanX,\FmeanXoracle) 
= \Op(n\inv), \eal
where the latter follows from \eqref{eq:oracleFmean_rate}. Analogously, $\wdist^2(\FmeanY,\FmeanYoracle) = \Op(n\inv)$. 
By \ref{ass:coef_decay}, $\|\regKernel\|_{\jointmean}^2 < \infty$. 
In conjunction with the fact that $\|\pRegKernelOracle - \regKernel\|_{\jointmean}^2 = \hsNormSq{\hsSpace{\FmeanX}{\FmeanY}}{\paraTransOp_{(\FmeanXoracle,\FmeanYoracle), (\FmeanX,\FmeanY)} \regOpOracle - \regOp}$, \eqref{eq:pred_oracle} follows from Theorem~\ref{thm:rate_oracle}. 
\epf

Next, we move on to the case where the distributions $\Xi$ and $\Yi$ are not fully observed and hence need to be estimated from the corresponding samples $\{\dpX_{il}\}_{l=1}^{\nDpXi}$ and $\{\dpY_{il}\}_{l=1}^{\nDpYi}$ generated from $\Xi$ and $\Yi$, respectively. For the proof of Theorem~\ref{thm:rate_realistic}, we need to study the asymptotic properties of the covariance operators, $\covopXest$ and $\covopYest$. 

\blem\label{lem:cov_realistic}
Assume \ref{ass:distn_est_rate}--\ref{ass:variation} and \ref{ass:atomlessFmeanEst}. 
Furthermore, assume that the eigenvalues $\{\egnvalX_j\}_{j=1}^\infty$ and $\{\egnvalY_k\}_{k=1}^\infty$ are distinct, respectively. Then
\bgt\nn
\hsNormSq{\hsSpaceAuto{\FmeanX}}{\pCovopXest - \covopX} = \Op(\denRate_\nDp+n\inv),\quad
\hsNormSq{\hsSpaceAuto{\FmeanY}}{\pCovopYest - \covopY} = \Op(\denRate_\nDp+n\inv),\\
\text{and}\quad \hsNormSq{\hsSpace{\FmeanX}{\FmeanY}}{\pCrosscovopXYest-\crosscovopXY} = \Op(\denRate_\nDp+n\inv).\egt
Furthermore, 
\begin{gather*}
\sup_{j\ge 1} |\egnvalXest_j - \egnvalX_j| \le \hsNorm{\hsSpaceAuto{\FmeanX}}{\pCovopXest - \covopX},\quad 
\sup_{k\ge 1} |\egnvalYest_k - \egnvalY_k| \le 
\hsNorm{\hsSpaceAuto{\FmeanY}}{\pCovopYest - \covopY},\\
\|\pEgnfctnXest_j - \egnfctnX_j\|_{\FmeanX}
\le 2\sqrt 2 \hsNorm{\hsSpaceAuto{\FmeanX}}{\pCovopXest - \covopX} / 
\min_{1\le j'\le j}\{\egnvalX_{j'} - \egnvalX_{j'+1}\},\text{ for all }j\ge 1,\\
\|\pEgnfctnYest_k - \egnfctnY_k\|_{\FmeanY} \le 2\sqrt 2 \hsNorm{\hsSpaceAuto{\FmeanY}}{\pCovopYest - \covopY} / 
\min_{1\le k'\le k}\{\egnvalY_{k'} - \egnvalY_{k'+1}\}, \text{ for all }k\ge 1.
\end{gather*}
\elem
\bpf
\redundant{
	We only prove the results for $\covopXest$; those for $\covopYest$ can be shown analogously. 
	By the third statement in Proposition~\ref{prop:paraTrans}, under \ref{ass:atomless} and \ref{ass:atomlessFmeanEst}, 
	\begin{align*}
	\pCovopXest - \covopX 
	&= n\inv \sum_{i=1}^n (\pLogXest\XiEst) \otimes (\pLogXest\XiEst)  - \covopX\\
	&= n\inv \sum_{i=1}^n (\Log_{\FmeanX}\Xi) \otimes (\Log_{\FmeanX}\Xi) - \covopX\\
	&\quad + n\inv \sum_{i=1}^n (\pLogXest\XiEst - \Log_{\FmeanX}\Xi) \otimes (\Log_{\FmeanX}\Xi)\\
	&\quad + n\inv \sum_{i=1}^n (\Log_{\FmeanX}\Xi)\otimes (\pLogXest\XiEst - \Log_{\FmeanX}\Xi)\\
	&\quad + n\inv \sum_{i=1}^n (\pLogXest\XiEst - \Log_{\FmeanX}\Xi) \otimes (\pLogXest\XiEst - \Log_{\FmeanX}\Xi)\\
	&\eqqcolon \hsOp_1 + \hsOp_2 + \hsOp_3 + \hsOp_4.
	\end{align*}
	For $\hsOp_2$, it can be observed that
	\bal\nn
	\hsNormSq{\hsSpaceAuto{\FmeanX}}{\hsOp_2}
	&\le \left(n\inv\sum_{i=1}^n \|\Log_{\FmeanX}\Xi\|_{\FmeanX}^2\right) 
	\left(n\inv\sum_{i=1}^n \|\pLogXest\XiEst - \Log_{\FmeanX}\Xi\|_{\FmeanX}^2\right).
	\eal
	For the second term, 
} 
We note that by \eqref{eq:log}, \eqref{eq:paraTrans} and \ref{ass:atomlessFmeanEst}, 
\redundant{
	$\pLogXest\XiEst = \Log_{\FmeanX}\XiEst - \Log_{\FmeanX}\FmeanXest$. Hence,
}
\bal\nn 
&\expect\left(\frac{1}{n}\sum_{i=1}^n \|\pLogXest\XiEst - \Log_{\FmeanX}\Xi\|_{\FmeanX}^2\right)\\
&\quad =\expect\left(\frac{1}{n}\sum_{i=1}^n \|\Log_{\FmeanX}\XiEst - \Log_{\FmeanX}\FmeanXest - \Log_{\FmeanX}\Xi\|_{\FmeanX}^2\right)\\
&\quad\le 3\expect\left(\frac{1}{n}\sum_{i=1}^n \|\Log_{\FmeanX}\XiEst - \Log_{\FmeanX}\Xi\|_{\FmeanX}^2\right) + 3\expect\left(\|\Log_{\FmeanX}\FmeanXest - \Log_{\FmeanX}\FmeanXoracle\|_{\FmeanX}^2\right)\\
&\qquad + 3\expect\left(\|\Log_{\FmeanX}\FmeanXoracle\|_{\FmeanX}^2\right).\eal
Moreover, by \eqref{eq:log}, \eqref{eq:quantile_oracleFrechetMean} and \eqref{eq:quantile_practicalFrechetMean}, 
\bal\nn
&\expect\left[\frac{1}{n}\sum_{i=1}^n \|\Log_{\FmeanX}\XiEst - \Log_{\FmeanX}\Xi\|_{\FmeanX}^2\right]
=\frac{1}{n}\sum_{i=1}^n\expect\wdist^2(\XiEst,\Xi),\\
&\expect\left[\|\Log_{\FmeanX}\FmeanXest - \Log_{\FmeanX}\FmeanXoracle\|_{\FmeanX}^2\right] 
=\expect\int_0^1\left[\frac{1}{n}\sum_{i=1}^n\left(\quantileXiEst(p) - \quantileXi(p)\right)\right]^2\diffop p
\le \frac{1}{n}\sum_{i=1}^n\expect\wdist^2(\XiEst,\Xi).
\eal
Furthermore, in conjunction with \ref{ass:distn_est_rate}--\ref{ass:nObsPerDens} and \eqref{eq:oracleFmean_rate}, this entails 
\bal\label{eq:log_est_rate}
\expect\left(n\inv\sum_{i=1}^n \|\pLogXest\XiEst - \Log_{\FmeanX}\Xi\|_{\FmeanX}^2 \right)
= O(\denRate_\nDp+n\inv),\eal
whence $n\inv\sum_{i=1}^n \|\pLogXest\XiEst - \Log_{\FmeanX}\Xi\|_{\FmeanX}^2 = \Op(\denRate_\nDp+n\inv)$. 
\redundant{
	Thus, $\hsNormSq{\hsSpaceAuto{\FmeanX}}{\hsOp_2} = \Op(\denRate_\nDp+n\inv)$. 
	Similarly, it can be shown that $\hsNormSq{\hsSpaceAuto{\FmeanX}}{\hsOp_3} =  \Op(\denRate_\nDp+n\inv)$, and $\hsNormSq{\hsSpaceAuto{\FmeanX}}{\hsOp_4}= \Op(\denRate_\nDp^2+n^{-2})$. 
	In conjunction with \eqref{eq:OracleCovWithTrueMean_rate}, we have
} 
Using similar arguments to the proof of Lemma~\ref{lem:cov_oracle}, it can be shown that $\hsNormSq{\hsSpaceAuto{\FmeanX}}{\pCovopXest - \covopX} = \Op(\denRate_\nDp+n\inv)$. 
Results for $\covopYest$ and $\crosscovopXYest$ can be shown analogously. 
By \ref{ass:atomless} and the atomlessness of $\FmeanXest$ following from \ref{ass:atomlessFmeanEst}, Proposition~\ref{prop:paraTrans} implies that $\{\egnvalXest_j\}_{j=1}^\infty$ and $\{\pEgnfctnXest_j\}_{j=1}^\infty$ are the eigenvalues and eigenfunctions of $\pCovopXest$, for which the results follow from Lemmas~4.2 and 4.3 of \citet{bosq:00}.
\epf

\bpf[Proof of Theorem~\ref{thm:rate_realistic}]
Defining
\bal\nn 
\wh{A}_1 &= \sum_{k=1}^{\nFPCY}\sum_{j=1}^{\nFPCX} (\regCoefEst_{jk} - \regCoef_{jk}) \pEgnfctnYest_k \otimes \pEgnfctnXest_j,\\  
\wh{A}_2 &= \sum_{k=1}^{\nFPCY}\sum_{j=1}^{\nFPCX} \regCoef_{jk} \left( \pEgnfctnYest_k \otimes \pEgnfctnXest_j - \egnfctnY_k \otimes \egnfctnX_j\right),\eal 
we observe that by \eqref{eq:measureOfError}, 
\bal\nn
\left\hsNormSq{\hsSpace{\FmeanX}{\FmeanY}}{\paraTransOp_{(\FmeanXest,\FmeanYest), (\FmeanX,\FmeanY)} \regOpEst - \regOp\right} 
&\le 3 \|\wh{A}_1\|_{\jointmean}^2 + 3\|\wh{A}_2\|_{\jointmean}^2 + 3\|A_3\|_{\jointmean}^2, \eal
with $A_3$ as per \eqref{eq:regOpOracle_decom}. 

In  analogy to the proof of Theorem~\ref{thm:rate_oracle}, we define events
\bgt\nn 
\wh\event_{1\nFPCX} = \wh\event_{1\nFPCX}(n) 
= \left\{ \egnvalX_{\nFPCX}\ge 2 \hsNorm{\hsSpaceAuto{\FmeanX}}{\pCovopXest - \covopX}\right\}.\egt
On $\wh\event_{1\nFPCX}$, Lemma~\ref{lem:cov_realistic} entails $\egnvalXest_j \ge \egnvalX_j/2$, for all $j=1,\ldots,\nFPCX$. 
Note that under the assumptions of Theorem~\ref{thm:rate_realistic}, $(\denRate_{\nDp}+n\inv) \egnvalX_{\nFPCX}^{-2} = o(1)$, as $n\ra\infty$.  
Thus, Lemma~\ref{lem:cov_realistic} entails that $\prob(\wh\event_{1\nFPCX})\ra 1$ as $n\ra \infty$. 
Following similar arguments to \eqref{eq:regCoefOracle1} and \eqref{eq:regCoefOracle1-2}, it holds on $\wh\event_{1\nFPCX}$ that 
\bal\label{eq:regCoefEst1}
\|\wh{A}_1\|_{\jointmean}^2 
&\le 8\sum_{k=1}^{\nFPCY}\sum_{j=1}^{\nFPCX} \egnvalX_j^{-2}(\crovalEst_{jk} - \croval_{jk})^2 + 8\sum_{k=1}^{\nFPCY}\sum_{j=1}^{\nFPCX} \egnvalX_j^{-2}(\egnvalXest_j - \egnvalX_j)^2 \regCoef_{jk}^2,\eal
where 
\bal
\sum_{k=1}^{\nFPCY}\sum_{j=1}^{\nFPCX} \egnvalX_j^{-2}(\egnvalXest_j - \egnvalX_j)^2 \regCoef_{jk}^2
&\le \const \hsNormSq{\hsSpaceAuto{\FmeanX}}{\pCovopXest-\covopX} \sum_{j=1}^{\nFPCX} j^{2\egnvalSpX-2\coefDecayX} 
= \op\left((\denRate_{\nDp}+n\inv)\nFPCX^{\egnvalSpX+1}\right). \label{eq:regCoefEst1-2}
\eal
We will show later that \note{if $\coefDecayX>\egnvalSpX+1$, $\coefDecayY>1$, w.o. $p$ series assumption of $\egnvalX_j$,}
\bal\label{eq:regCoefEst1-1}
&\sum_{k=1}^{\nFPCY}\sum_{j=1}^{\nFPCX}\egnvalX_j^{-2}(\crovalEst_{jk}-\croval_{jk})^2\\
&= \Op\left(n\inv\nFPCX^{\egnvalSpX+1}\right) +\Op\left(\denRate_{\nDp}\nFPCX^{2\egnvalSpX+1}\right) +\Op\left(\denRate_{\nDp}\nFPCX^{\egnvalSpX+1}\nFPCY\right) \\
&\quad+ \Op\left((\denRate_{\nDp}+n\inv) \sum_{k=1}^{\nFPCY}\left(k^{2\egnvalSpY-2\coefDecayY+2} + \sum_{k':\,k'\le k/2} {k'}^{2\egnvalSpY-2\coefDecayY}\right)\right)\\
&\quad +\Op\left((\denRate_{\nDp}^2+n^{-2})\nFPCX^{4\egnvalSpX+3} [1+(\denRate_{\nDp}+n\inv)\nFPCY^{2\egnvalSpY+3}]\right) +\Op\left((\denRate_{\nDp}^2+n^{-2})\nFPCX^{2\egnvalSpX+1}\nFPCY^{2\egnvalSpY+3}\right) \\
&\quad + 
\left\{\begin{array}{ll}
	\Op\left([1+(\denRate_{\nDp}+n\inv)\nFPCY^{2\egnvalSpY+3}](\denRate_{\nDp}^2+n^{-2}) \nFPCY^{4\egnvalSpY-2\coefDecayY+5}(\log \nFPCY)^2\right), &\text{if } \coefDecayY\le\egnvalSpY+1,\\
	\Op\left([1+(\denRate_{\nDp}+n\inv)\nFPCY^{2\egnvalSpY+3}](\denRate_{\nDp}^2+n^{-2}) \nFPCY^{2\egnvalSpY+3}\right), &\text{if }\coefDecayY>\egnvalSpY+1.
\end{array}\right.\eal

For $\wh{A}_2$, we note that \ref{ass:eigenval_spacingX} and Lemma~\ref{lem:cov_realistic} entail \note{(w.o. $p$ series assumption of $\egnvalX_j$)}
\bal\label{eq:egnfctn_est_rate}
\|\pEgnfctnXest_j-\egnfctnX_j\|_{\FmeanX}^2 
&\le \const j^{2\egnvalSpX+2} \hsNormSq{\hsSpaceAuto{\FmeanX}}{\pCovopXest-\covopX},\\
\|\pEgnfctnYest_k-\egnfctnY_k\|_{\FmeanY}^2 
&\le \const k^{2\egnvalSpY+2} \hsNormSq{\hsSpaceAuto{\FmeanY}}{\pCovopYest-\covopY},\eal
uniformly in $j,k$, which implies \note{(w.o. $p$ series assumption of $\egnvalX_j$)}
\bal\label{eq:regCoefEst2}
\|\wh{A}_2\|_{\jointmean}^2 
&=\Op\left((\denRate_{\nDp}+n\inv) \nFPCX \sum_{j=1}^{\nFPCX}j^{2\egnvalSpX-2\coefDecayX+2}+ (\denRate_{\nDp}+n\inv)\nFPCY \sum_{k=1}^{\nFPCY}k^{2\egnvalSpY-2\coefDecayY+2}\right).\eal
Combining \eqref{eq:regCoefEst1-2}, \eqref{eq:regCoefEst1-1}, \eqref{eq:regCoefEst2} and \eqref{eq:regCoefOracle3} yields \note{if $\coefDecayX>\egnvalSpX+1,\, \egnvalSpX\ge 1,\, \coefDecayY>1,\, (\denRate_{\nDp}+n\inv)\nFPCX^{2\egnvalSpX+2}=o(1),\, (\denRate_{\nDp}+n\inv)\nFPCY^{2\egnvalSpY+2}=o(1),$ (w.o. $p$ series assumption of $\egnvalX_j$)}
\bal\nn
&\left\hsNormSq{\hsSpace{\FmeanX}{\FmeanY}}{\paraTransOp_{(\FmeanXest,\FmeanYest), (\FmeanX,\FmeanY)} \regOpEst - \regOp\right} \\
&= \Op\left(n\inv\nFPCX^{\egnvalSpX+1}\right) +\Op\left(\denRate_{\nDp}\nFPCX^{2\egnvalSpX+1}\right) +\Op\left(\denRate_{\nDp}\nFPCX^{\egnvalSpX+1}\nFPCY\right) +O\left(\nFPCX^{-2\coefDecayX+1}+ \nFPCY^{-2\coefDecayY+1}\right) \\
&\quad +\Op\left((\denRate_{\nDp}^2+n^{-2})\nFPCX^{4\egnvalSpX+3}[1+(\denRate_{\nDp}+n\inv)\nFPCY^{2\egnvalSpY+3}]\right) +\Op\left((\denRate_{\nDp}^2+n^{-2})\nFPCX^{2\egnvalSpX+1}\nFPCY^{2\egnvalSpY+3}\right)\\
&\quad +\Op\left((\denRate_{\nDp}+n\inv)\nFPCY \sum_{k=1}^{\nFPCY}k^{2\egnvalSpY-2\coefDecayY+2}\right)\\
&\quad +\left\{\begin{array}{ll}
	\Op\left([1+(\denRate_{\nDp}+n\inv)\nFPCY^{2\egnvalSpY+3}](\denRate_{\nDp}^2+n^{-2}) \nFPCY^{4\egnvalSpY-2\coefDecayY+5}(\log \nFPCY)^2\right), &\text{if } \coefDecayY\le\egnvalSpY+1,\\
	\Op\left([1+(\denRate_{\nDp}+n\inv)\nFPCY^{2\egnvalSpY+3}](\denRate_{\nDp}^2+n^{-2}) \nFPCY^{2\egnvalSpY+3}\right), &\text{if }\coefDecayY>\egnvalSpY+1,
\end{array}\right.\eal
whence \eqref{eq:rate_realistic} follows with $\denRate_{\nDp}$ satisfying \ref{ass:nObsPerDen_rate}. 
\note{All the terms involving $n$ are the same as the results in the proof of Theorem~\ref{thm:rate_oracle}.}

Regarding \eqref{eq:regCoefEst1-1}, by the atomlessness of $\FmeanX$, $\FmeanY$, $\FmeanXest$ and $\FmeanYest$ and Proposition~\ref{prop:paraTrans}, 
\bal\nn
\crovalEst_{jk} - \croval_{jk}
&= \innerprod{\pCrosscovopXYest\pEgnfctnXest_j}{\pEgnfctnYest_k}_{\FmeanY} - \innerprod{\crosscovopXY\egnfctnX_j}{\egnfctnY_k}_{\FmeanY}\\
&= \innerprod{(\pCrosscovopXYest - \crosscovopXY)\egnfctnX_j}{\egnfctnY_k}_{\FmeanY} 
+\innerprod{\crosscovopXY(\pEgnfctnXest_j-\egnfctnX_j)}{\egnfctnY_k}_{\FmeanY} \\
&\quad+ \innerprod{\crosscovopXY\egnfctnX_j}{\pEgnfctnYest_k-\egnfctnY_k}_{\FmeanY} 
+\innerprod{(\pCrosscovopXYest - \crosscovopXY)\egnfctnX_j}{\pEgnfctnYest_k-\egnfctnY_k}_{\FmeanY} \\
&\quad+ \innerprod{(\pCrosscovopXYest - \crosscovopXY)(\pEgnfctnXest_j-\egnfctnX_j)}{\egnfctnY_k}_{\FmeanY} 
+ \innerprod{\crosscovopXY(\pEgnfctnXest_j-\egnfctnX_j)}{\pEgnfctnYest_k-\egnfctnY_k}_{\FmeanY} \\
&\quad+ \innerprod{(\pCrosscovopXYest - \crosscovopXY)(\pEgnfctnXest_j-\egnfctnX_j)}{\pEgnfctnYest_k-\egnfctnY_k}_{\FmeanY}\\
&\eqqcolon \wh{I}_{jk1} + \wh{I}_{jk2} + \wh{I}_{jk3} + \wh{I}_{jk4} + \wh{I}_{jk5} + \wh{I}_{jk6} + \wh{I}_{jk7},\eal
whence 
\bal\nn
\sum_{k=1}^{\nFPCY}\sum_{j=1}^{\nFPCX} \egnvalX_j^{-2}(\crovalEst_{jk} - \croval_{jk})^2
&\le 7\sum_{l=1}^7 \sum_{k=1}^{\nFPCY}\sum_{j=1}^{\nFPCX} \egnvalX_j^{-2} \wh{I}_{jkl}^2.\eal
For $\wh{I}_{jk4},\dots,\wh{I}_{jk7}$, applying Lemma~\ref{lem:cov_realistic} and \eqref{eq:egnfctn_est_rate} yields \note{(w.o. $p$ series assumption of $\egnvalX_j$)}
\bal\label{eq:regCoefEst1-1-4to7}
\sum_{k=1}^{\nFPCY}\sum_{j=1}^{\nFPCX} \egnvalX_j^{-2} \wh{I}_{jk4}^2
&\le \hsNormSq{\hsSpace{\FmeanX}{\FmeanY}}{\pCrosscovopXYest-\crosscovopXY} \sum_{j=1}^{\nFPCX}\egnvalX_j^{-2}  \sum_{k=1}^{\nFPCY}\|\pEgnfctnYest_k-\egnfctnY_k\|_{\FmeanY}^2\\
&= \Op\left((\denRate_{\nDp}^2+n^{-2}) \nFPCX^{2\egnvalSpX+1}\nFPCY^{2\egnvalSpY+3}\right), \\
\sum_{k=1}^{\nFPCY}\sum_{j=1}^{\nFPCX} \egnvalX_j^{-2} \wh{I}_{jk5}^2
&\le \hsNormSq{\hsSpace{\FmeanX}{\FmeanY}}{\pCrosscovopXYest-\crosscovopXY} \sum_{j=1}^{\nFPCX}\egnvalX_j^{-2}  \|\pEgnfctnXest_j-\egnfctnX_j\|_{\FmeanX}^2\\
&=\Op\left((\denRate_{\nDp}^2+n^{-2}) \nFPCX^{4\egnvalSpX+3}\right),\\
\sum_{k=1}^{\nFPCY}\sum_{j=1}^{\nFPCX} \egnvalX_j^{-2} \wh{I}_{jk6}^2
&\le \sum_{j=1}^{\nFPCX} \egnvalX_j^{-2}\|\crosscovopXY(\pEgnfctnXest_j-\egnfctnX_j)\|_{\FmeanY}^2 \sum_{k=1}^{\nFPCY}\|\pEgnfctnYest_k-\egnfctnY_k\|_{\FmeanY}^2 \\
&= \left(\sum_{k=1}^{\nFPCY}\sum_{j=1}^{\nFPCX} \egnvalX_j^{-2} \wh{I}_{jk3}^2\middle) \middle(\sum_{k=1}^{\nFPCY}\|\pEgnfctnYest_k-\egnfctnY_k\|_{\FmeanY}^2\right), \\
\sum_{k=1}^{\nFPCY}\sum_{j=1}^{\nFPCX} \egnvalX_j^{-2} \wh{I}_{jk7}^2
&\le\hsNormSq{\hsSpace{\FmeanX}{\FmeanY}}{\pCrosscovopXYest-\crosscovopXY} \sum_{k=1}^{\nFPCY} \|\pEgnfctnYest_k-\egnfctnY_k\|_{\FmeanY}^2\sum_{j=1}^{\nFPCX} \egnvalX_j^{-2}\|\pEgnfctnXest_j-\egnfctnX_j\|_{\FmeanX}^2\\
&=\Op\left((\denRate_{\nDp}^3+n^{-3}) \nFPCX^{4\egnvalSpX+3}\nFPCY^{2\egnvalSpY+3}\right).\eal

For $\wh{I}_{jk1}$, we observe that 
\bal\nn
\wh{I}_{jk1}
&= \innerprod{(\pCrosscovopXYest - \crosscovopXY)\egnfctnX_j}{\egnfctnY_k}_{\FmeanY}\\
&= \frac{1}{n}\sum_{i=1}^n \left(\innerprod{\Log_{\FmeanX}\XiEst}{\egnfctnX_j}_{\FmeanX}\innerprod{\Log_{\FmeanY}\YiEst}{\egnfctnY_k}_{\FmeanY} -  \innerprod{\Log_{\FmeanX}\Xi}{\egnfctnX_j}_{\FmeanX} \innerprod{\Log_{\FmeanY}\Yi}{\egnfctnY_k}_{\FmeanY}\right) \\ 
&\quad-\left(\innerprod{\Log_{\FmeanX}\FmeanXest}{\egnfctnX_j}_{\FmeanX}\innerprod{\Log_{\FmeanY}\FmeanYest}{\egnfctnY_k}_{\FmeanY} - \innerprod{\Log_{\FmeanX}\FmeanXoracle}{\egnfctnX_j}_{\FmeanX}\innerprod{\Log_{\FmeanY}\FmeanYoracle}{\egnfctnY_k}_{\FmeanY}\right)\\
&\quad+\frac{1}{n}\sum_{i=1}^n \innerprod{\Log_{\FmeanX}\Xi}{\egnfctnX_j}_{\FmeanX} \innerprod{\Log_{\FmeanY}\Yi}{\egnfctnY_k}_{\FmeanY} - \innerprod{\Log_{\FmeanX}\FmeanXoracle}{\egnfctnX_j}_{\FmeanX}\innerprod{\Log_{\FmeanY}\FmeanYoracle}{\egnfctnY_k}_{\FmeanY}\\
&\quad -\expect\left(\innerprod{\Log_{\FmeanX}\X}{\egnfctnX_j}_{\FmeanX} \innerprod{\Log_{\FmeanY}\Y}{\egnfctnY_k}_{\FmeanY}\right). \eal
By \ref{ass:distn_est_rate} and \ref{ass:variation},
\bal\nn
&\expect\left\{\left[\frac{1}{n}\sum_{i=1}^n \left(\innerprod{\Log_{\FmeanX}\XiEst}{\egnfctnX_j}_{\FmeanX}\innerprod{\Log_{\FmeanY}\YiEst}{\egnfctnY_k}_{\FmeanY} -  \innerprod{\Log_{\FmeanX}\Xi}{\egnfctnX_j}_{\FmeanX} \innerprod{\Log_{\FmeanY}\Yi}{\egnfctnY_k}_{\FmeanY}\right)\right]^2\right\}\\
&\le \frac{1}{n}\sum_{i=1}^n \expect\left\{\left(\innerprod{\Log_{\FmeanX}\XiEst}{\egnfctnX_j}_{\FmeanX}\innerprod{\Log_{\FmeanY}\YiEst}{\egnfctnY_k}_{\FmeanY} -  \innerprod{\Log_{\FmeanX}\Xi}{\egnfctnX_j}_{\FmeanX} \innerprod{\Log_{\FmeanY}\Yi}{\egnfctnY_k}_{\FmeanY}\right)^2\right\}\\
&\le \frac{3}{n}\sum_{i=1}^n \expect\left\{\innerprod{\Log_{\FmeanX}\XiEst-\Log_{\FmeanX}\Xi}{\egnfctnX_j}_{\FmeanX}^2 \innerprod{\Log_{\FmeanY}\Yi}{\egnfctnY_k}_{\FmeanY}^2\right\} \\
&\quad + \frac{3}{n}\sum_{i=1}^n \expect\left\{\innerprod{\Log_{\FmeanX}\Xi}{\egnfctnX_j}_{\FmeanX}^2 \innerprod{\Log_{\FmeanY}\YiEst-\Log_{\FmeanY}\Yi}{\egnfctnY_k}_{\FmeanY}^2\right\}\\
&\quad+ \frac{3}{n}\sum_{i=1}^n \expect\left\{\innerprod{\Log_{\FmeanX}\XiEst-\Log_{\FmeanX}\Xi}{\egnfctnX_j}_{\FmeanX}^2 \innerprod{\Log_{\FmeanY}\YiEst-\Log_{\FmeanY}\Yi}{\egnfctnY_k}_{\FmeanY}^2\right\}\\
&\le \const  \left(\egnvalY_k\frac{1}{n}\sum_{i=1}^n\left\{\expect\wdist^4(\XiEst,\Xi)\right\}\half 
+\egnvalX_j\frac{1}{n}\sum_{i=1}^n\left\{\expect\wdist^4(\YiEst,\Yi)\right\}\half \right)\\
&\quad
+\frac{3}{n}\sum_{i=1}^n \left\{\expect\wdist^4(\XiEst,\Xi)\right\}\half  \left\{\expect\wdist^4(\YiEst,\Yi)\right\}\half\\
&=O\left((\egnvalY_k + \egnvalX_j)\denRate_{\nDp} + \denRate_{\nDp}^2\right), \eal
and
\bal\nn
&\expect\left(\innerprod{\Log_{\FmeanX}\FmeanXest}{\egnfctnX_j}_{\FmeanX}\innerprod{\Log_{\FmeanY}\FmeanYest}{\egnfctnY_k}_{\FmeanY} - \innerprod{\Log_{\FmeanX}\FmeanXoracle}{\egnfctnX_j}_{\FmeanX}\innerprod{\Log_{\FmeanY}\FmeanYoracle}{\egnfctnY_k}_{\FmeanY}\right)^2\\
&\le 3\expect\left(\innerprod{\Log_{\FmeanX}\FmeanXest-\Log_{\FmeanX}\FmeanXoracle}{\egnfctnX_j}_{\FmeanX}^2 \innerprod{\Log_{\FmeanY}\FmeanYoracle}{\egnfctnY_k}_{\FmeanY}^2\right) \\
&\quad +3\expect\left(\innerprod{\Log_{\FmeanX}\FmeanXoracle}{\egnfctnX_j}_{\FmeanX}^2 \innerprod{\Log_{\FmeanY}\FmeanYest-\Log_{\FmeanY}\FmeanYoracle}{\egnfctnY_k}_{\FmeanY}^2\right) \\
&\quad +3\expect\left(\innerprod{\Log_{\FmeanX}\FmeanXest-\Log_{\FmeanX}\FmeanXoracle}{\egnfctnX_j}_{\FmeanX}^2 \innerprod{\Log_{\FmeanY}\FmeanYest-\Log_{\FmeanY}\FmeanYoracle}{\egnfctnY_k}_{\FmeanY}^2\right)\\
&\le 3\expect\left(\wdist^2(\FmeanXest,\FmeanXoracle) \frac{1}{n}\sum_{i=1}^n\innerprod{\Log_{\FmeanY}\Yi}{\egnfctnY_k}_{\FmeanY}^2\right) 
+3\expect\left(\frac{1}{n}\sum_{i=1}^n\innerprod{\Log_{\FmeanX}\Xi}{\egnfctnX_j}_{\FmeanX}^2 \wdist^2(\FmeanYest,\FmeanYoracle)\right) \\
&\quad +3\expect\left(\wdist^2(\FmeanXest,\FmeanXoracle) \wdist^2(\FmeanYest,\FmeanYoracle)\right)\\
&\le \frac{3}{n^2}\expect\left[\left(\sum_{i=1}^n\wdist^2(\XiEst,\Xi)\right) \left(\sum_{i=1}^n\innerprod{\Log_{\FmeanY}\Yi}{\egnfctnY_k}_{\FmeanY}^2\right)\right] \\
&\quad+\frac{3}{n^2}\expect\left[\left(\sum_{i=1}^n\innerprod{\Log_{\FmeanX}\Xi}{\egnfctnX_j}_{\FmeanX}^2\right) \left(\sum_{i=1}^n\wdist^2(\YiEst,\Yi)\right)\right] \\
&\quad +\frac{3}{n^2}\expect\left[\left(\sum_{i=1}^n\wdist^2(\XiEst,\Xi)\right) \left(\sum_{i=1}^n\wdist^2(\YiEst,\Yi)\right)\right]\\
&= O\left((\egnvalY_k + \egnvalX_j)\denRate_{\nDp} + \denRate_{\nDp}^2\right),\eal
uniformly in $j,k$. In conjunction with \eqref{eq:croval_oracle_decom} and \eqref{eq:croval_oracle_rate1}, it follows that 
\bal\label{eq:croval_est_rate1}
\expect(\wh{I}_{jk1}^2) = O\left((\egnvalY_k + \egnvalX_j)\denRate_{\nDp} + \denRate_{\nDp}^2 + n\inv\egnvalX_j\egnvalY_k\right),\eal
uniformly in $j,k$, which entails 
\bal\label{eq:regCoefEst1-1-1}
\sum_{k=1}^{\nFPCY}\sum_{j=1}^{\nFPCX} \egnvalX_j^{-2} \wh{I}_{jk1}^2 
&= \Op\left(\denRate_{\nDp}\nFPCX^{2\egnvalSpX+1} +\denRate_{\nDp}\nFPCX^{\egnvalSpX+1}\nFPCY +\denRate_{\nDp}^2\nFPCX^{2\egnvalSpX+1}\nFPCY +n\inv\nFPCX^{\egnvalSpX+1}\right). \eal 

For $\wh{I}_{jk2}$, analogous to the discussion of $I_{jk2}$ in the proof of Theorem~\ref{thm:rate_oracle}, we note that
\bal\nn
\wh{I}_{jk2}
&=\innerprod{\crosscovopXY(\pEgnfctnXest_j-\egnfctnX_j)}{\egnfctnY_k}_{\FmeanY}\\
&=\croval_{jk}\innerprod{\pEgnfctnXest_j-\egnfctnX_j}{\egnfctnX_j}_{\FmeanX} + \sum_{j':\,j'\ne j} \croval_{j'k} (\egnvalXest_{j'}-\egnvalX_j)\inv \innerprod{(\pCovopXest-\covopX)\pEgnfctnXest_j}{\egnfctnX_{j'}}_{\FmeanX}\\
&= \wh{I}_{jk2}^{(1)} + \wh{I}_{jk2}^{(2)} + \wh{I}_{jk2}^{(3)} + \wh{I}_{jk2}^{(4)},\eal
where 
\bal\nn
\wh{I}_{jk2}^{(1)} &\coloneqq \croval_{jk}\innerprod{\pEgnfctnXest_j-\egnfctnX_j}{\egnfctnX_j}_{\FmeanX},\\
\wh{I}_{jk2}^{(2)} &\coloneqq \sum_{j':\,j'\ne j} \croval_{j'k} (\egnvalX_{j'}-\egnvalX_j)\inv \innerprod{(\pCovopXest-\covopX)\egnfctnX_j}{\egnfctnX_{j'}}_{\FmeanX},\\
\wh{I}_{jk2}^{(3)} &\coloneqq \sum_{j':\,j'\ne j} \croval_{j'k} \left((\egnvalXest_{j'}-\egnvalX_j)\inv - (\egnvalX_{j'}-\egnvalX_j)\inv\right)\ \innerprod{(\pCovopXest-\covopX)\egnfctnX_j}{\egnfctnX_{j'}}_{\FmeanX},\\
\wh{I}_{jk2}^{(4)} &\coloneqq \sum_{j':\,j'\ne j} \croval_{j'k} (\egnvalXest_{j'}-\egnvalX_j)\inv \innerprod{(\pCovopXest-\covopX)(\pEgnfctnXest_j-\egnfctnX_j)}{\egnfctnX_{j'}}_{\FmeanX}.\eal
By \ref{ass:coef_decay} and \eqref{eq:egnfctn_est_rate}, \note{(w.o. $p$ series assumption of $\egnvalX_j$)}
\bal\label{eq:regCoefEst1-1-2-1}
\sum_{k=1}^{\nFPCY}\sum_{j=1}^{\nFPCX} \egnvalX_j^{-2} (\wh{I}_{jk2}^{(1)})^2
&\le \sum_{k=1}^{\nFPCY}\sum_{j=1}^{\nFPCX} \regCoef_{jk}^2 \|\pEgnfctnXest_j-\egnfctnX_j\|_{\FmeanX}^2
\le \const\sum_{j=1}^{\nFPCX} j^{-2\coefDecayX} \|\pEgnfctnXest_j-\egnfctnX_j\|_{\FmeanX}^2\\
&= \Op\left((\denRate_{\nDp}+n\inv) \sum_{j=1}^{\nFPCX} j^{2\egnvalSpX-2\coefDecayX+2}\right) 
= \op\left((\denRate_{\nDp}+n\inv) \nFPCX^{\egnvalSpX+1}\right). \note{\text{ if }\egnvalSpX+1<\coefDecayX}\eal
For $\wh{I}_{jk2}^{(2)}$, using similar arguments to the proof of \eqref{eq:croval_est_rate1}, it can be shown that 
\bal\nn
\expect\innerprod{(\pCovopXest-\covopX)\egnfctnX_j}{\egnfctnX_{j'}}_{\FmeanX}^2
=O\left((\egnvalX_j + \egnvalX_{j'})\denRate_{\nDp} + \denRate_{\nDp}^2 + n\inv\egnvalX_j\egnvalX_{j'}\right),\eal
uniformly in $j,j'$, and hence that
\bal\nn
\expect(\wh{I}_{jk2}^{(2)})^2
&= \expect\left[\sum_{j':\,j'\ne j} \egnvalX_{j'}\regCoef_{j'k} (\egnvalX_{j'}-\egnvalX_j)\inv \innerprod{(\pCovopXest-\covopX)\egnfctnX_j}{\egnfctnX_{j'}}_{\FmeanX} \right]^2\\ 
&\le \left[\sum_{j':\,j'\ne j} \egnvalX_{j'}\regCoef_{j'k}^2 (\egnvalX_{j'}-\egnvalX_j)^{-2}\right] \left[\sum_{j':\,j'\ne j} \egnvalX_{j'} \expect\innerprod{(\pCovopXest-\covopX)\egnfctnX_j}{\egnfctnX_{j'}}_{\FmeanX}^2 \right]\\
&\le \const k^{-2\coefDecayY} \left(\denRate_{\nDp} + (\denRate_{\nDp} + n\inv)\egnvalX_j\right) \left[\sum_{j':\,j'\ne j} \egnvalX_{j'}{j'}^{-2\coefDecayX} (\egnvalX_{j'}-\egnvalX_j)^{-2}\right]\\
&\le \const k^{-2\coefDecayY} \left(\denRate_{\nDp} + (\denRate_{\nDp} + n\inv)\egnvalX_j\right),\eal
where the last inequality follows from \eqref{eq:egnval_diff_rate} and similar arguments to \eqref{eq:sum-2-2-1}. This implies 
\bal\label{eq:regCoefEst1-1-2-2}
\sum_{k=1}^{\nFPCY}\sum_{j=1}^{\nFPCX}\egnvalX_j^{-2}(\wh{I}_{jk2}^{(2)})^2 
= \Op\left(\denRate_{\nDp}\nFPCX^{2\egnvalSpX+1} + n\inv\nFPCX^{\egnvalSpX+1} \right).\eal

\note{
	Define events 
	\bal\nn
	\wh\event_{2\nFPCX} = \wh\event_{2\nFPCX}(n)
	= \left\{(\egnvalXest_j-\egnvalX_{j'})^{-2} \le 2(\egnvalX_j-\egnvalX_{j'})^{-2},\text{ for all } j,j'=1,\dots,\nFPCX\text{ s.t. }j\ne j' \right\}.\eal
	Note that Lemma~\ref{lem:cov_realistic} and \eqref{eq:egnval_diff_rate} entail that 
	\bal\nn
	\frac{\left|(\egnvalXest_j-\egnvalX_{j'})^{-2} - (\egnvalX_j-\egnvalX_{j'})^{-2}\right|}{(\egnvalX_j-\egnvalX_{j'})^{-2}}
	&=\Op\left((\denRate_{\nDp}+n\inv)\half\nFPCX^{\egnvalSpX+1}\right),\eal
	uniformly in $j,j'$. 
	Hence, under the assumptions of Theorem~\ref{thm:rate_realistic}, $\prob(\wh\event_{2\nFPCX})\ra 1$ as $n\ra\infty$.
}For $\wh{I}_{jk2}^{(3)}$ and $\wh{I}_{jk2}^{(4)}$, following similar arguments as in  the proof of \eqref{eq:regCoefOracle1-1-2-3} and \eqref{eq:regCoefOracle1-1-2-4}, it can be shown that \note{on $\wh\event_{2\nFPCX}$,}
\bal\label{eq:regCoefEst1-1-2-3to4}
\sum_{k=1}^{\nFPCY}\sum_{j=1}^{\nFPCX} \egnvalX_j^{-2}(\wh{I}_{jk2}^{(3)})^2 
&= \Op\left((\denRate_{\nDp}^2+n^{-2}) \left(\nFPCX^{2\egnvalSpX+1} + \nFPCX^{6\egnvalSpX-2\coefDecayX+5} \right)\right),\\
\sum_{k=1}^{\nFPCY}\sum_{j=1}^{\nFPCX} \egnvalX_j^{-2}(\wh{I}_{jk2}^{(4)})^2 
&= \Op\left((\denRate_{\nDp}^2+n^{-2}) \nFPCX^{4\egnvalSpX+3}\right).\eal
Combining \eqref{eq:regCoefEst1-1-2-1}--\eqref{eq:regCoefEst1-1-2-3to4} yields 
\bal\label{eq:regCoefEst1-1-2}
\sum_{k=1}^{\nFPCY}\sum_{j=1}^{\nFPCX} \egnvalX_j^{-2}\wh{I}_{jk2}^2
= \Op\left(\denRate_{\nDp}\nFPCX^{2\egnvalSpX+1} + n\inv\nFPCX^{\egnvalSpX+1}\right) + \Op\left((\denRate_{\nDp}^2+n^{-2}) \nFPCX^{4\egnvalSpX+3}\right). \note{\text{ if }\egnvalSpX+1<\coefDecayX} \eal

For $\wh{I}_{jk3} =\innerprod{\crosscovopXY\egnfctnX_j}{\pEgnfctnYest_k-\egnfctnY_k}_{\FmeanY}
= \wh{I}_{jk3}^{(1)} + \wh{I}_{jk3}^{(2)} + \wh{I}_{jk3}^{(3)} + \wh{I}_{jk3}^{(4)}$, 
where 
\bal\nn
\wh{I}_{jk3}^{(1)} &\coloneqq \croval_{jk}\innerprod{\pEgnfctnYest_k-\egnfctnY_k}{\egnfctnY_k}_{\FmeanY},\\
\wh{I}_{jk3}^{(2)} &\coloneqq \sum_{k':\,k'\ne k} \croval_{jk'} (\egnvalY_{k'}-\egnvalY_k)\inv \innerprod{(\pCovopYest-\covopY)\egnfctnY_k}{\egnfctnY_{k'}}_{\FmeanY},\\
\wh{I}_{jk3}^{(3)} &\coloneqq \sum_{k':\,k'\ne k} \croval_{jk'} \left((\egnvalYest_{k'}-\egnvalY_k)\inv - (\egnvalY_{k'}-\egnvalY_k)\inv\right) \innerprod{(\pCovopYest-\covopY)\egnfctnY_k}{\egnfctnY_{k'}}_{\FmeanY},\\
\wh{I}_{jk3}^{(4)} &\coloneqq \sum_{k':\,k'\ne k} \croval_{jk'} (\egnvalYest_{k'}-\egnvalY_k)\inv \innerprod{(\pCovopYest-\covopY)(\pEgnfctnYest_k-\egnfctnY_k)}{\egnfctnY_{k'}}_{\FmeanY}.\eal
\note{Define events 
	\bal\nn
	\wh\event_{3\nFPCY} = \wh\event_{3\nFPCY}(n)
	= \left\{(\egnvalYest_k-\egnvalY_{k'})^{-2} \le 2(\egnvalY_k-\egnvalY_{k'})^{-2},\text{ for all } k,k'=1,\dots,\nFPCY\text{ s.t. }k\ne k' \right\}.\eal
}Using similar arguments as in  the proof of \eqref{eq:regCoefOracle1-1-3}, it can be shown that \note{under the assumptions of Theorem~\ref{thm:rate_realistic}, $\prob(\wh\event_{3\nFPCY})\ra 1$, as $n\ra\infty$\note{, since $(\denRate_{\nDp}+n\inv)\nFPCY^{2\egnvalSpY+2}\ra 0$} and that on $\wh\event_{3\nFPCY}$, \note{(w.o. $p$ series assumption of $\egnvalX_j$)}}
\bal\nn
\sum_{k=1}^{\nFPCY}\sum_{j=1}^{\nFPCX} \egnvalX_j^{-2} (\wh{I}_{jk3}^{(1)})^2
&\le \sum_{k=1}^{\nFPCY} \sum_{j=1}^{\nFPCX} \regCoef_{jk}^2 \|\pEgnfctnYest_k-\egnfctnY_k\|_{\FmeanY}^2 
=\Op\left((\denRate_{\nDp}+n\inv)\sum_{k=1}^{\nFPCY}k^{2\egnvalSpY-2\coefDecayY+2}\right),\\
\sum_{k=1}^{\nFPCY}\sum_{j=1}^{\nFPCX} \egnvalX_j^{-2} (\wh{I}_{jk3}^{(2)})^2
&= \Op\left((\denRate_{\nDp}+n\inv) \sum_{k=1}^{\nFPCY} \sum_{k':\,k'\ne k} {k'}^{-2\coefDecayY} (\egnvalY_{k'}-\egnvalY_k)^{-2} \right)\\
&=\Op\left((\denRate_{\nDp}+n\inv) \sum_{k=1}^{\nFPCY}\left(k^{2\egnvalSpY-2\coefDecayY+2} + \sum_{k':\,k'\le k/2} {k'}^{2\egnvalSpY-2\coefDecayY}\right)\right), \\
\sum_{k=1}^{\nFPCY}\sum_{j=1}^{\nFPCX} \egnvalX_j^{-2} (\wh{I}_{jk3}^{(3)})^2
&=\Op\left((\denRate_{\nDp}^2+n^{-2}) \sum_{k=1}^{\nFPCY} \sum_{k':\,k'\ne k} {k'}^{-2\coefDecayY}(\egnvalY_{k'}-\egnvalY_k)^{-4} \right)\\
&
=\Op\left((\denRate_{\nDp}^2+n^{-2})\nFPCY + (\denRate_{\nDp}^2+n^{-2})\nFPCY^{4\egnvalSpY-2\coefDecayY+5}\right),\\
\sum_{k=1}^{\nFPCY}\sum_{j=1}^{\nFPCX} \egnvalX_j^{-2} (\wh{I}_{jk3}^{(4)})^2
&=\Op\left((\denRate_{\nDp}^2+n^{-2}) \sum_{k=1}^{\nFPCY}  k^{2\egnvalSpY+2} \left(\sum_{k':\,k'\ne k} {k'}^{-\coefDecayY}|\egnvalY_{k'}-\egnvalY_k|\inv\right)^2\right)\\
&=\left\{\begin{array}{ll}
	\Op\left((\denRate_{\nDp}^2+n^{-2}) \nFPCY^{4\egnvalSpY-2\coefDecayY+5}(\log \nFPCY)^2\right), &\text{if }\egnvalSpY-\coefDecayY\ge -1,\\
	\Op\left((\denRate_{\nDp}^2+n^{-2}) \nFPCY^{2\egnvalSpY+3}\right), &\text{if }\egnvalSpY-\coefDecayY<-1.
\end{array}\right. \eal
Here, similar arguments to \eqref{eq:egnval_diff_rate} imply that  $\sum_{k':\,k'\ne k} {k'}^{-2\coefDecayY} (\egnvalY_{k'}-\egnvalY_k)^{-2}$ is bounded by a multiple of \note{(w.o. $p$ series assumption of $\egnvalX_j$)} 
\bal\nn
&k^{2\egnvalSpY}\left(\sum_{k':\,k'\ge 2k} {k'}^{-2\coefDecayY} \right) 
+ k^{2\egnvalSpY+2}\left(\sum_{\substack{k':\,k'\ne k,\, k/2<k'<2k}} |k'-k|^{-2} {k'}^{-2\coefDecayY} \right) + \left(\sum_{k':\,k'\le k/2} {k'}^{2\egnvalSpY-2\coefDecayY} \right)\\
&\le \const \left( k^{2\egnvalSpY-2\coefDecayY+2} + \sum_{k':\,k'\le k/2} {k'}^{2\egnvalSpY-2\coefDecayY} \right)\\
&\le\const \left\{\begin{array}{ll}
	k^{2\egnvalSpY-2\coefDecayY+2}, &\text{if }\egnvalSpY-\coefDecayY> -1/2,\\
	k, &\text{if }\egnvalSpY-\coefDecayY\in(-1,-1/2],\\
	1+\log k, &\text{if }\egnvalSpY-\coefDecayY=-1,\\
	1, &\text{if }\egnvalSpY-\coefDecayY<-1.
\end{array}\right.\eal
Therefore, \note{(w.o. $p$ series assumption of $\egnvalX_j$)} 
\bal\label{eq:regCoefEst1-1-3}
\sum_{k=1}^{\nFPCY}\sum_{j=1}^{\nFPCX}\egnvalX_j^{-2}\wh{I}_{jk3}^2 
&= \Op\left((\denRate_{\nDp}+n\inv) \sum_{k=1}^{\nFPCY}\left(k^{2\egnvalSpY-2\coefDecayY+2} + \sum_{k':\,k'\le k/2} {k'}^{2\egnvalSpY-2\coefDecayY}\right)\right)\\
&\quad +\left\{\begin{array}{ll}
	\Op\left((\denRate_{\nDp}^2+n^{-2}) \nFPCY^{4\egnvalSpY-2\coefDecayY+5}(\log \nFPCY)^2\right), &\text{if }\egnvalSpY-\coefDecayY\ge -1,\\
	\Op\left((\denRate_{\nDp}^2+n^{-2}) \nFPCY^{2\egnvalSpY+3}\right), &\text{if }\egnvalSpY-\coefDecayY<-1.
\end{array}\right.
\eal
Combining \eqref{eq:regCoefEst1-1-4to7}, \eqref{eq:regCoefEst1-1-1}, \eqref{eq:regCoefEst1-1-2} and \eqref{eq:regCoefEst1-1-3} yields \eqref{eq:regCoefEst1-1}, which completes the proof. 
\epf

We omit the proof of Corollary~\ref{cor:pred_realistic}, since it is analogous to that of Corollary~\ref{cor:pred_oracle}.

\bco
\bpf[Proof of Corollary~\ref{cor:pred_realistic}]
First note that for two distributions $\distnOne, \distnTwo \in\manifold$, if $\distnTwo$ is atomless, then 
$\wdist^2(\Exp_{\distnOne}\arbiFctn, \Exp_{\distnTwo}\paraTrans_{\distnOne,\distnTwo}\arbiFctn)= \wdist^2(\distnOne, \distnTwo)$. 
Thus, 
\begin{align*}
&\wdist^2\left(\YpredEst(\arbiDistnEst), \expect_\oplus(\Y|\X = \arbiDistn)\right)\\
&=\wdist^2\left(\Exp_{\FmeanYest} \int_{\dom} \regKernelEst(s,\cdot) \Log_{\FmeanXest} \arbiDistnEst (s)\diffop \FmeanXest(s),
\Exp_{\FmeanY}[\expect(\Log_{\FmeanY}\Y|\Log_{\FmeanX}\X=\Log_{\FmeanX}\arbiDistn)]\right)\\
&\le 2\wdist^2\left(\Exp_{\FmeanY} \paraTrans_{\FmeanYest,\FmeanY} \int_{\dom} \regKernelEst(s,\cdot) \Log_{\FmeanXest} \arbiDistnEst (s)\diffop \FmeanXest(s),\Exp_{\FmeanY}\int_{\dom}\regKernel(s,\cdot)\Log_{\FmeanX}\arbiDistn(s)\diffop\FmeanX(s)\right)\\ 
&\qquad + 2\wdist^2(\FmeanYest,\FmeanY)\\
&= 2\left\|\paraTrans_{\FmeanYest,\FmeanY} \int_{\dom} \regKernelEst(s,\cdot) \Log_{\FmeanXest} \arbiDistnEst (s)\diffop \FmeanXest(s) - \int_{\dom}\regKernel(s,\cdot)\Log_{\FmeanX}\arbiDistn(s)\diffop\FmeanX(s)\right\|_{\FmeanY}^2 
+ 2\wdist^2(\FmeanYest,\FmeanY).
\end{align*}
Note that by \ref{ass:atomlessFmeanEst} and Proposition~\ref{prop:paraTrans},
\be\nn \paraTrans_{\FmeanYest,\FmeanY} \int_{\dom} \regKernelEst(s,\cdot) \Log_{\FmeanXest} \arbiDistnEst (s)\diffop \FmeanXest(s) = \int_{\dom} \pRegKernelEst(s,\cdot) \pLogXest \arbiDistnEst (s)\diffop \FmeanX(s).\ee
Hence,
\begin{align*}
&\wdist^2\left(\YpredEst(\arbiDistnEst), \expect_\oplus(\Y|\X = \arbiDistn)\right)\\
&\le 2\left\|\int_{\dom} \pRegKernelEst(s,\cdot) \pLogXest \arbiDistnEst (s)\diffop \FmeanX(s) - \int_{\dom}\regKernel(s,\cdot)\Log_{\FmeanX}\arbiDistn(s)\diffop\FmeanX(s)\right\|_{\FmeanY}^2 
+ 2\wdist^2(\FmeanYest,\FmeanY)\\
&\le 4\left\|\int_{\dom} \left[\pRegKernelEst(s,\cdot) - \regKernel(s,\cdot)\right] \pLogXest \arbiDistnEst (s)\diffop \FmeanX(s)\right\|_{\FmeanY}^2\\
&\quad + 4\left\|\int_{\dom} \regKernel(s,\cdot) \left[ \pLogXest \arbiDistnEst(s)-\Log_{\FmeanX}\arbiDistn(s)\right] \diffop\FmeanX(s)\right\|_{\FmeanY}^2 
+ 2\wdist^2(\FmeanYest,\FmeanY)\\
&\le 4\|\pLogXest\arbiDistnEst\|_{\FmeanX}^2 \|\pRegKernelEst - \regKernel\|_{\jointmean}^2
+ 4\|\pLogXest\arbiDistnEst - \Log_{\FmeanX}\arbiDistn\|_{\FmeanX}^2 \|\regKernel\|_{\jointmean}^2 + 2\wdist^2(\FmeanYest,\FmeanY).
\end{align*}
Here, under \ref{ass:distn_est_rate}--\ref{ass:nObsPerDens} and \ref{ass:atomlessFmeanEst}, 
\begin{align*}
&\|\pLogXest\arbiDistnEst\|_{\FmeanX}^2 
\le 2\wdist^2(\FmeanX,\arbiDistn) +2\|\pLogXest\arbiDistnEst - \Log_{\FmeanX}\arbiDistn\|_{\FmeanX}^2,\\
&\|\pLogXest\arbiDistnEst - \Log_{\FmeanX}\arbiDistn\|_{\FmeanX}^2 = \Op(\denRate_{\nDp}) +\Op(n\inv).
\end{align*}
By \ref{ass:coef_decay}, $\|\regKernel\|_{\jointmean}^2 < \infty$. Lastly, $\wdist^2(\FmeanYest,\FmeanY) = \Op(\denRate_{\nDp})+ \Op(n\inv)$, which completes the proof by Theorem~\ref{thm:rate_realistic}.
\epf
\fi

\subsection{\update{Proofs for Section~\ref{sec:dts}}}\label{sec:proof_ts}
In this section, analogous to  the definitions in  Section~\ref{sec:proof_indep}, for random variables $X_n$ and a sequence of positive constants $c_n$, we will write $X_n=\Op(c_n)$ if
\bal\nn
\lim_{\largeConst\ra\infty} \limsup_{n\ra\infty} \sup_{\vjointdistn\in\vdistnfam} \prob_{\vjointdistn}(|X_n|>\largeConst c_n) =0,\eal
and $X_n =\op(c_n)$ if there exists $\largeConst_0>0$ such that
\bal\nn
\lim_{n\ra\infty} \sup_{\vjointdistn\in\vdistnfam} \prob_{\vjointdistn}(|X_n|>\largeConst_0 c_n) =0.\eal
For a sequence of deterministic quantities $a_n = a_n(\vjointdistn)$, we will write $a_n = O(c_n)$ if
\bal\nn
\sup_{n\ge 1} c_n\inv \sup_{\vjointdistn\in\vdistnfam} |a_n(\vjointdistn)| < \infty.\eal
We denote $\paraTrans_{\FmeanZoracle,\FmeanZ}\arbiFctn$ and $\paraTrans_{\FmeanZest,\FmeanZ}\arbiFctn$ by $\paraTrans\arbiFctn$, for $\arbiFctn$ in $\hilbert_{\FmeanZoracle}$ and $\hilbert_{\FmeanZest}$, respectively; 
we define 
$\pCovopZoracle\coloneqq \paraTransOp_{(\FmeanZoracle,\FmeanZoracle),(\FmeanZ,\FmeanZ)}\covopZlagZeroOracle$, 
$\pCovopZest\coloneqq \paraTransOp_{(\FmeanZest,\FmeanZest),(\FmeanZ,\FmeanZ)}\covopZlagZeroEst$, 
$\pCrosscovopZoracle\coloneqq \paraTransOp_{(\FmeanZoracle,\FmeanZoracle),(\FmeanZ,\FmeanZ)}\covopZlagOneOracle$,  $\pCrosscovopZest\coloneqq \paraTransOp_{(\FmeanZest,\FmeanZest),(\FmeanZ,\FmeanZ)}\covopZlagOneEst$, 
\bal\nn
\pRegKernelOracle \coloneqq \sum_{l=1}^{\nFPCZ} \sum_{j=1}^{\nFPCZ} \regCoefOracle_{jl} \pEgnfctnZoracle_l \otimes \pEgnfctnZoracle_j
\qaq \pRegKernelEst \coloneqq \sum_{l=1}^{\nFPCZ} \sum_{j=1}^{\nFPCZ} \regCoefEst_{jl} \pEgnfctnZest_l \otimes \pEgnfctnZest_j.\eal
By the third statement in Proposition~\ref{prop:paraTrans}, under \ref{ass:atomless_ts}, $\pRegKernelOracle$ is the kernel of $\paraTransOp_{(\FmeanZoracle,\FmeanZoracle), (\FmeanZ,\FmeanZ)}\regOpOracle$; under \ref{ass:atomless_ts} and \ref{ass:atomlessFmeanEst}, $\pRegKernelEst$ is the kernel of $\paraTransOp_{(\FmeanZest,\FmeanZest), (\FmeanZ,\FmeanZ)}\regOpEst$. 
Hence, 
\bal\label{eq:measureOfError_ts}
\hsNormSq{\hsSpaceAuto{\FmeanZ}}{\paraTransOp_{(\FmeanZoracle,\FmeanZoracle),(\FmeanZ,\FmeanZ)} \regOpOracle- \regOp}
&= \int_{\dom}\int_{\dom}\left[\pRegKernelOracle(s,t) - \regKernel(s,t)\right]^2 \diffop\FmeanZ(s) \diffop\FmeanZ(t)\\
\hsNormSq{\hsSpaceAuto{\FmeanZ}}{\paraTransOp_{(\FmeanZest,\FmeanZest),(\FmeanZ,\FmeanZ)} \regOpEst - \regOp}
&= \int_{\dom}\int_{\dom}\left[\pRegKernelEst(s,t) - \regKernel(s,t)\right]^2 \diffop\FmeanZ(s) \diffop\FmeanZ(t).\eal
Thus, for the proofs of Theorems~\ref{thm:rate_oracle_ts} and \ref{thm:rate_realistic_ts}, we will focus on the right hand sides in \eqref{eq:measureOfError_ts}.  To this end, we need to study the asymptotic properties of the estimators of the covariance operators, i.e., $\covopZlagZeroOracle$ and $\covopZlagOneOracle$ when the distributions $\Zi$ are fully observed, and $\covopZlagZeroEst$ and $\covopZlagOneEst$ when only samples of observations drawn from the $\Zi$ are available. 
We use the convention that $\innerprod{\pEgnfctnZoracle_j}{\egnfctnZ_j}_{\FmeanZ}\ge 0$ and $\innerprod{\pEgnfctnZest_j}{\egnfctnZ_j}_{\FmeanZ}\ge 0$ to determine the signs of the estimated eigenfunctions, $\egnfctnZoracle_j$ and $\egnfctnZest_j$ where choice of the signs may impact the validity of the results. 

We first focus on the case where $\Zi$ are fully observed. 

\blem\label{lem:cov_oracle_ts}
Assume \ref{ass:autoRegOpConv}--\ref{ass:variation_ts}. 
Furthermore, assume that the eigenvalues $\{\egnvalZ_j\}_{j=1}^\infty$ are distinct. Then
\bgt\nn
\hsNormSq{\hsSpaceAuto{\FmeanZ}}{\pCovopZoracle - \covopZlagZero} = \Op(n\inv) \qaq
\hsNormSq{\hsSpaceAuto{\FmeanZ}}{\pCrosscovopZoracle - \covopZlagOne} = \Op(n\inv).\egt
Furthermore, 
\bgt\nn
\sup_{j\ge 1} |\egnvalZoracle_j - \egnvalZ_j| \le \hsNorm{\hsSpaceAuto{\FmeanZ}}{\pCovopZoracle - \covopZlagZero}, \\
\|\pEgnfctnZoracle_j - \egnfctnZ_j\|_{\FmeanZ}
\le 2\sqrt{2} \hsNorm{\hsSpaceAuto{\FmeanZ}}{\pCovopZoracle - \covopZlagZero} / 
\min_{1\le j'\le j}\{\egnvalZ_{j'} - \egnvalZ_{j'+1}\},\text{ for all }j\ge 1. \egt
\elem
\bpf 
We note that due to stationarity, \ref{ass:autoRegOpConv} and \ref{ass:variation_ts}, 
\begin{gather}
\expect\left(\frac{1}{n}\sum_{i=1}^n\|\Log_{\FmeanZ}\Zi\|_{\FmeanZ}^2\right)
=\frac{1}{n}\sum_{i=1}^n \expect\left(\|\Log_{\FmeanZ}\Zi\|_{\FmeanZ}^2\right)
= \expect\left(\|\Log_{\FmeanZ}\Zone\|_{\FmeanZ}^2\right)
= \expect\left(\wdist^2(\Zone,\FmeanZ)\right) < \infty,\nn\\
\expect\left(\|\Log_{\FmeanZ}\FmeanZoracle\|_{\FmeanZ}^2\right)
=\expect\left(\left\|\frac{1}{n}\sum_{i=1}^n\Log_{\FmeanZ}\Zi\right\|_{\FmeanZ}^2\right) = O(n\inv), \label{eq:oracle_mean_rate_ts}\\
\expect\left(\left\hsNormSq{\hsSpaceAuto{\FmeanZ}}{\frac{1}{n}\sum_{i=1}^n\Log_{\FmeanZ}\Zi\otimes\Log_{\FmeanZ}\Zi-\covopZlagZero\right}\right)= O(n\inv), \label{eq:oracle_cov_lag0_rate}\\
\expect\left(\left\hsNormSq{\hsSpaceAuto{\FmeanZ}}{\frac{1}{n-1}\sum_{i=1}^{n-1}\Log_{\FmeanZ}\ZiPlusOne\otimes\Log_{\FmeanZ}\Zi-\covopZlagOne\right}\right)= O(n\inv), \label{eq:oracle_cov_lag1_rate}
\end{gather}
where \eqref{eq:oracle_mean_rate_ts}, \eqref{eq:oracle_cov_lag0_rate} and \eqref{eq:oracle_cov_lag1_rate} follow from Theorems~3.7, 4.1 and 4.7 of \citet{bosq:00}, respectively. Then the proof follows arguments similar to those in  Lemma~\ref{lem:cov_oracle}. 
\epf 

\bpf[Proof of Theorem~\ref{thm:rate_oracle_ts}]
The proof follows similar arguments as in the proof  of Theorem~\ref{thm:rate_oracle}. Here  we just discuss some of  the differences due to the serial dependence. 

Regarding the counterpart of \eqref{eq:regCoefOracle1-1-1},
\bal\nn
\sum_{l=1}^{\nFPCZ}\sum_{j=1}^{\nFPCZ} \egnvalX_j^{-2}\innerprod{(\pCrosscovopZoracle - \covopZlagOne)\egnfctnZ_j}{\egnfctnZ_l}_{\FmeanZ}^2
\le \sum_{j=1}^{\nFPCZ}\egnvalX_j^{-2} \|(\pCrosscovopZoracle - \covopZlagOne)\egnfctnZ_j\|_{\FmeanZ}^2 
= \Op\left(n\inv \nFPCZ^{2\egnvalSpZ+1}\right); \eal
regarding the counterpart of \eqref{eq:regCoefOracle1-1-2-2}, it can be shown similarly  to \eqref{eq:egnval_diff_rate} that 
\bal\nn
&\sum_{l=1}^{\nFPCZ}\sum_{j=1}^{\nFPCZ}\egnvalZ_j^{-2} \left(\sum_{j':\,j'\ne j} \crovalZlagOne_{j'l} (\egnvalZ_{j'}-\egnvalZ_j)\inv \innerprod{(\pCovopZoracle-\covopZlagZero)\egnfctnZ_j}{\egnfctnZ_{j'}}_{\FmeanZ}\right)^2\\
&\le \sum_{j=1}^{\nFPCZ}\egnvalZ_j^{-2} \|(\pCovopZoracle-\covopZlagZero)\egnfctnZ_j\|_{\FmeanZ}^2 \sum_{l=1}^{\nFPCZ}\sum_{j':\,j'\ne j} \crovalZlagOne_{j'l}^2 (\egnvalZ_{j'}-\egnvalZ_j)^{-2}\\
&\le \const \sum_{j=1}^{\nFPCZ}\egnvalZ_j^{-2} \|(\pCovopZoracle-\covopZlagZero)\egnfctnZ_j\|_{\FmeanZ}^2 \sum_{j':\,j'\ne j} \egnvalZ_{j'}^2 {j'}^{-2\coefDecayZx} (\egnvalZ_{j'}-\egnvalZ_j)^{-2}\\
&=\Op\left(n\inv\sum_{j=1}^{\nFPCZ}j^{4\egnvalSpZ-2\coefDecayZx+2}\right) 
=\op\left(n\inv\nFPCZ^{2\egnvalSpZ+1}\right) \note{\text{ if }\egnvalSpZ+1<\coefDecayZx};\eal
regarding the counterpart of the second term in \eqref{eq:regCoefOracle1-1-3-1to4},
\bal\nn
&\sum_{l=1}^{\nFPCZ}\sum_{j=1}^{\nFPCZ}\egnvalZ_j^{-2} \left(\sum_{l':\,l'\ne l} \crovalZlagOne_{jl'} (\egnvalZ_{l'}-\egnvalZ_l)\inv \innerprod{(\pCovopZoracle-\covopZlagZero)\egnfctnZ_l}{\egnfctnZ_{l'}}_{\FmeanZ}\right)^2\\
&\le \sum_{l=1}^{\nFPCZ}\sum_{j=1}^{\nFPCZ} \|(\pCovopZoracle-\covopZlagZero)\egnfctnZ_l\|_{\FmeanZ}^2 \sum_{l':\,l'\ne l} \regCoef_{jl'}^2 (\egnvalZ_{l'}-\egnvalZ_l)^{-2}\\
&\le\const  \sum_{l=1}^{\nFPCZ} \|(\pCovopZoracle-\covopZlagZero)\egnfctnZ_l\|_{\FmeanZ}^2 \sum_{l':\,l'\ne l} {l'}^{-2\coefDecayZy}(\egnvalZ_{l'}-\egnvalZ_l)^{-2}\\
&\le \const \left(n\inv\sum_{l=1}^{\nFPCZ}l^{2\egnvalSpZ-2\coefDecayZy+2}\right)
=\op\left(n\inv\nFPCZ^{2\egnvalSpZ+1}\right). \note{\text{ if }\egnvalSpZ+1<\coefDecayZx,\, \coefDecayZy>1} \eal
Hence, under the assumptions of Theorem~\ref{thm:rate_oracle_ts}, \note{if $n\inv\nFPCZ^{2\egnvalSpZ+2} = o(1)$, $\egnvalSpZ+1<\coefDecayZx$, $\coefDecayZy>1$} 
\bal\nn
\hsNormSq{\hsSpace{\FmeanZ}{\FmeanZ}}{\paraTransOp_{(\FmeanZoracle,\FmeanZoracle), (\FmeanZ,\FmeanZ)} \regOpOracle - \regOp}
&=\Op\left(n\inv\nFPCZ^{2\egnvalSpZ+1}\right) +\Op\left(n\inv\nFPCZ\sum_{j=1}^{\nFPCZ}j^{2\egnvalSpZ-2\coefDecayZy+2}\right)  +O\left(\nFPCZ^{1-2\coefDecayZx}+ \nFPCZ^{1-2\coefDecayZy}\right)\\
&\quad +\Op\left(n^{-2}\nFPCZ^{4\egnvalSpZ+4}\right) +\Op\left(n^{-3}\nFPCZ^{6\egnvalSpZ+6}\right). \eal
\epf

Next, we move on to the case where the distributions $\Zi$ are not fully observed and hence need to be estimated from the corresponding samples of measurements drawn from $\Zi$. Again, we first obtain the asymptotic properties of the estimates of the autocovariance operators, $\covopZlagZeroEst$ and $\covopZlagOneEst$. 

\blem\label{lem:cov_realistic_ts}
Assume \ref{ass:distn_est_rate}, 
\ref{ass:atomlessFmeanEst}, \ref{ass:autoRegOpConv}, \ref{ass:atomless_ts}, \ref{ass:variation_ts} and \ref{ass:nObsPerDens_ts}. 
Furthermore, assume that the eigenvalues $\{\egnvalZ_j\}_{j=1}^\infty$ are distinct. Then
\bgt\nn
\hsNormSq{\hsSpaceAuto{\FmeanZ}}{\pCovopZest - \covopZlagZero} = \Op(\denRate_{\nDp}+n\inv)\qaq
\hsNormSq{\hsSpaceAuto{\FmeanZ}}{\pCrosscovopZest - \covopZlagOne} = \Op(\denRate_{\nDp}+n\inv).\egt
Furthermore, 
\begin{gather*}
\sup_{j\ge 1} |\egnvalZest_j - \egnvalZ_j| \le \hsNorm{\hsSpaceAuto{\FmeanZ}}{\pCovopZest - \covopZlagZero}, \\
\|\pEgnfctnZest_j - \egnfctnZ_j\|_{\FmeanZ}
\le 2\sqrt{2} \hsNorm{\hsSpaceAuto{\FmeanZ}}{\pCovopZest - \covopZlagZero} / 
\min_{1\le j'\le j}\{\egnvalZ_{j'} - \egnvalZ_{j'+1}\},\text{ for all }j\ge 1.
\end{gather*}
\elem
The proof of Lemma~\ref{lem:cov_realistic_ts} is analogous to the proof of Lemma~\ref{lem:cov_realistic} and the proof of Theorem~\ref{thm:rate_realistic_ts} is similar to the proofs of Theorems~\ref{thm:rate_realistic} and \ref{thm:rate_oracle_ts} and therefore there is no need for  further details. The same applies to  the proofs of Corollaries~\ref{cor:pred_oracle_ts} and \ref{cor:pred_realistic_ts}, which are analogous to the proof of Corollary~\ref{cor:pred_oracle}.

\section{Estimation for Distribution-to-Scalar Regression}\label{sec:d2sEst}
For the distribution-to-scalar regression as per \eqref{eq:d2sreg} in Section~\ref{sec:cov_indep}, suppose $\{(\Xi,\sYi)\}_{i=1}^n$ are $n$ independent realizations of $(\X,\sY)$. When $\Xi$ are fully observed, the regression coefficient function $\regKernelDtoS$ can be estimated by 
\bal\nn
\regKernelDtoSoracle = \sum_{j=1}^{\nFPCX}\egnvalXoracle_j\inv n\inv\sum_{i=1}^n\innerprod{\sYi\Log_{\FmeanXoracle}\Xi}{\egnfctnXoracle_j}_{\FmeanXoracle}\egnfctnXoracle_j, \eal
where $\egnvalXoracle_j$ and $\egnfctnXoracle_j$ are the eigenvalues and eigenfunctions of $\covopXoracle$ as defined in Section~\ref{sec:coef_est_indep}. 
Similar to the distribution-to-distribution regression case, given a constant $\factor> 1$, we assume
\ben[label = (A\arabic*'), series = ass_d2s]
\item\label{ass:atomless_d2s} 
With probability equal to 1, the random distributions $\X$ are atomless.
\item\label{ass:variation_d2s}
$\expect(\|\Log_{\FmeanX}\X\|_{\FmeanX}^4)<\infty$, and $\expect(\innerprod{\Log_{\FmeanX}\X}{\egnfctnX_j}_{\FmeanX}^4) \le \factor \egnvalX_j^2$, for all $j\ge 1$.
\item\label{ass:eigenval_spacingX_d2s}
For $j\ge 1$, $\egnvalX_j - \egnvalX_{j+1} \ge \factor\inv j^{-\egnvalSpX - 1}$, where $\egnvalSpX\ge 1$ is a constant.
\item\label{ass:coef_decay_d2s}
For $j\ge 1$, $|\egnvalX_j\inv \innerprod{\expect(\sY\Log_{\FmeanX}\X)}{\egnfctnX_j}_{\FmeanX}|\le \factor j^{-\coefDecayX}$, where $\coefDecayX > \egnvalSpX+1$ is a constant.
\item\label{ass:nFPC_rate_d2s} 
$n\inv\nFPCX^{2\egnvalSpX+2}\ra 0$, as $n\ra\infty$.
\een
Let $\distnfamDtoS = \distnfamDtoS(\factor,\egnvalSpX,\coefDecayX)$ denote the set of distributions $\jointdistn$ of $(\X,\sY)$ that satisfy \ref{ass:atomless_d2s}--\ref{ass:coef_decay_d2s}.
\bthm\label{thm:rate_oracle_d2s}
If \ref{ass:atomless_d2s}--\ref{ass:nFPC_rate_d2s} hold, then \note{($\coefDecayX>\egnvalSpX+1,\, \egnvalSpX\ge 1,\, n\inv\nFPCX^{2\egnvalSpX+2}=o(1)$)}
\bal\nn
\lim_{\largeConst\ra\infty} \limsup_{n\ra\infty} \sup_{\jointdistnDtoS\in\distnfamDtoS} \prob_{\jointdistnDtoS}\left(\left\|\paraTrans_{\FmeanXoracle,\FmeanX}\regKernelDtoSoracle-\regKernelDtoS\right\|_{\FmeanX}^2 > \largeConst \max\left\{n\inv\nFPCX^{\egnvalSpX+1}, n^{-2}\nFPCX^{4\egnvalSpX+3}, \nFPCX^{-2\coefDecayX+1}\right\}\right) = 0.\eal
Choosing $\nFPCX\sim n^{1/(\egnvalSpX+2\coefDecayX)}$, we have 
\bal\label{eq:rate_oracle_d2s_n}
\lim_{\largeConst\ra\infty} \limsup_{n\ra\infty} \sup_{\jointdistnDtoS\in\distnfamDtoS} \prob_{\jointdistnDtoS}\left(\left\|\paraTrans_{\FmeanXoracle,\FmeanX}\regKernelDtoSoracle-\regKernelDtoS\right\|_{\FmeanX}^2 > \largeConst n^{-(2\coefDecayX-1)/(\egnvalSpX+2\coefDecayX)}\right) = 0.\eal
\ethm
As per \eqref{eq:rate_oracle_d2s_n}, the rate of convergence of $\regKernelDtoSoracle$ matches the minimax rate of the function-to-scalar linear regression based on fully observed functions developed by \citet{hall:07:1}. 

When $\Xi$ are not fully observed but rather only samples of measurements $\{\dpX_{il}\}_{l=1}^{\nDpXi}$ drawn from $\Xi$ are available, an estimate of the regression coefficient function $\regKernelDtoS$ is given by 
\bal\label{eq:regKernelDtoSest}
\regKernelDtoSest = \sum_{j=1}^{\nFPCX}\egnvalXest_j\inv n\inv\sum_{i=1}^n\innerprod{\sYi\Log_{\FmeanXest}\Xi}{\egnfctnXest_j}_{\FmeanXest}\egnfctnXest_j,\eal 
where $\egnvalXest_j$ and $\egnfctnXest_j$ are the eigenvalues and eigenfunctions of $\covopXest$ as defined in Section~\ref{sec:coef_est_indep}. Furthermore, we assume 
\ben[label = (A\arabic*'), resume = ass_d2s]
\item \label{ass:nObsPerDens_d2s}
There exists a sequence $\nDp = \nDp(n)$ such that $\min\{\nDpXi: i=1,2,\ldots,n\} \ge \nDp$ and that $\nDp\ra\infty$ as $n\ra \infty$.
\item \label{ass:nObsPerDen_rate_d2s}
For $\denRate_{\nDp}$ in \ref{ass:distn_est_rate}, $\denRate_{\nDp} \le \factor n\inv\nFPCX^{-\egnvalSpX}$, for all $n$.
\een
Then the data-based estimator $\regKernelEst$ is found to achieve the same rate as the estimator $\regKernelOracle$ based on fully observed distributions as shown in Theorem~\ref{thm:rate_oracle_d2s}. 
\bthm\label{thm:rate_realistic_d2s}
If \ref{ass:distn_est_rate}, 
\ref{ass:atomlessFmeanEst} and \ref{ass:atomless_d2s}--\ref{ass:nObsPerDen_rate_d2s} hold, then \note{($\coefDecayX>\egnvalSpX+1,\, \egnvalSpX\ge 1,\, n\inv\nFPCX^{2\egnvalSpX+2}=o(1)$)}
\bal\nn
\lim_{\largeConst\ra\infty} \limsup_{n\ra\infty} \sup_{\jointdistnDtoS\in\distnfamDtoS} \prob_{\jointdistnDtoS}\left(\left\|\paraTrans_{\FmeanXest,\FmeanX}\regKernelDtoSest-\regKernelDtoS\right\|_{\FmeanX}^2 > \largeConst \max\left\{n\inv\nFPCX^{\egnvalSpX+1}, n^{-2}\nFPCX^{4\egnvalSpX+3}, \nFPCX^{-2\coefDecayX+1}\right\}\right) = 0.\eal
Choosing $\nFPCX\sim n^{1/(\egnvalSpX+2\coefDecayX)}$, we have 
\bal\nn
\lim_{\largeConst\ra\infty} \limsup_{n\ra\infty} \sup_{\jointdistnDtoS\in\distnfamDtoS} \prob_{\jointdistnDtoS}\left(\left\|\paraTrans_{\FmeanXest,\FmeanX}\regKernelDtoSest-\regKernelDtoS\right\|_{\FmeanX}^2 > \largeConst n^{-(2\coefDecayX-1)/(\egnvalSpX+2\coefDecayX)}\right) = 0.\eal
\ethm

Proofs of Theorems~\ref{thm:rate_oracle_d2s} and \ref{thm:rate_realistic_d2s} are analogous to those of Theorem~\ref{thm:rate_oracle} and \ref{thm:rate_realistic_d2s} and are therefore omitted. 

\section{\update{An Example for Explicit Construction of the Autoregressive Model}}\label{sec:eg-WAR}

We consider $\dom=[0,1]$ and $\FmeanZ$ with bounded density $\maxDenFmeanZ \coloneqq \sup_{x\in\dom}\denFmeanZ(x) < \infty$, where we denote the density and cdf of $\FmeanZ$ by $\denFmeanZ$ and $\cdfFmeanZ$, respectively. 
We set 
\bal\nn
\regKernel = \sum_{l=1}^{\infty}\sum_{j=1}^{\infty}\regCoef_{jl}(\basisL_l\circ\cdfFmeanZ)\otimes(\basisL_j\circ\cdfFmeanZ),\qaq
\noise_i = \sum_{l=1}^{\infty}\errCoef_{il}\basisL_l\circ\cdfFmeanZ.\eal
Here, $\basisL_j$ are basis functions as per \eqref{eq:egbases}, 
\bal\nn
&\regCoef_{kl} = \sgnFractn\frac{j^{-\epntCfX+1}l^{-\epntCfY-1}}{\rzeta(\epntCfY)},\\
&\errCoef_{il}\sim \mathrm{Unif}(-\maxErrCoef_l,\maxErrCoef_l),\quad
\text{with } \maxErrCoef_l = (1-\sgnFractn) \frac{l^{-\epntCfY-1}\rzeta(\epntCfX+\epntCfY)}{2\sqrt{2}\pi\maxDenFmeanZ \rzeta(\epntCfY)^2},\eal
where $\epntCfX>3/2$ and $\epntCfY>1$ are constants, and $\rzeta(s) = \sum_{j=1}^{\infty}j^{-s}$. 
Define 
\bal\nn \Log_{\FmeanZ}\Zi = \sum_{r=0}^{\infty}\regOp^r(\noise_{i-r}),\quad i\in\intgSet.\eal
We will show in the following that $\{\Zi\}_{i\in\intgSet}$ is a stationary process taking values in $\manifold$ which satisfies the proposed model in \eqref{eq:autoreg}.

For this, it suffices to show that $\sum_{r=0}^{\infty}\regOp^r(\noise_{i-r})\in\Log_{\FmeanZ}\manifold$ with probability 1 and that \ref{ass:autoRegOpConv} holds. 
Observe that 
\bal\nn
\regOp(\noise_{i-1})
&= \sum_{l=1}^{\infty}\left(\sum_{j_1=1}^{\infty}\regCoef_{j_1 l}\errCoef_{i-1,\,j_1}\right)\basisL_l\circ\cdfFmeanZ,\\
\regOp^2(\noise_{i-2})
&= \sum_{l=1}^{\infty}\left(\sum_{j_2=1}^{\infty}\regCoef_{j_2 l} \sum_{j_1=1}^{\infty}\regCoef_{j_1 j_2}\errCoef_{i-2,\,j_1}\right)\basisL_l\circ\cdfFmeanZ,\\
&\vdots\\
\regOp^r(\noise_{i-r})
&= \sum_{l=1}^{\infty}\left(\sum_{j_r=1}^{\infty}\regCoef_{j_r l} \sum_{j_{r-1}=1}^{\infty}\regCoef_{j_{r-1} j_r} \cdots \sum_{j_1=1}^{\infty}\regCoef_{j_1 j_2} \errCoef_{i-r,\,j_1}\right)\basisL_l\circ\cdfFmeanZ,\eal
whence 
\bal\nn
\sum_{r=0}^{\infty}\regOp^r(\noise_{i-r}) =\sum_{l=1}^{\infty}\left[\errCoef_{il}+\sum_{r=1}^{\infty}\left(\sum_{j_r=1}^{\infty}\regCoef_{j_r l} \sum_{j_{r-1}=1}^{\infty}\regCoef_{j_{r-1} j_r} \cdots \sum_{j_1=1}^{\infty}\regCoef_{j_1 j_2} \errCoef_{i-r,\,j_1}\right)\right] \basisL_l\circ\cdfFmeanZ.\eal
To show $\sum_{r=0}^{\infty}\regOp^r(\noise_{i-r})\in\Log_{\FmeanZ}\manifold$ with probability 1, we will show that 
\bal\label{eq:convex_ts}
\sup_{t\in\dom} \left|\frac{\diffop}{\diffop t}\sum_{r=0}^{\infty}\regOp^r(\noise_{i-r})(t)\right|\le 1.\eal

Since $\sup_{t\in [0,1]}|\basisL_l'(t)| = 2\sqrt{2}\pi l$, we have
\bal\nn
&\sup_{t\in\dom}\sum_{r=0}^{\infty}|(\regOp^r(\noise_{i-r}))'(t)|\\
&\le \sum_{l=1}^{\infty}\left[\maxErrCoef_l+\sum_{r=1}^{\infty}\left(\sum_{j_r=1}^{\infty}\regCoef_{j_r l} \sum_{j_{r-1}=1}^{\infty}\regCoef_{j_{r-1} j_r} \cdots \sum_{j_1=1}^{\infty}\regCoef_{j_1 j_2} \maxErrCoef_{j_1}\right)\right] 2\sqrt{2}\pi l \maxDenFmeanZ\\
&= (1-\sgnFractn) \sum_{l=1}^{\infty}\left[ \frac{l^{-\epntCfY}\rzeta(\epntCfX+\epntCfY)}{\rzeta(\epntCfY)^2} +  \sum_{r=1}^{\infty}\left(\sum_{j_r=1}^{\infty}\regCoef_{j_r l} \sum_{j_{r-1}=1}^{\infty}\regCoef_{j_{r-1} j_r} \cdots \sum_{j_1=1}^{\infty}\regCoef_{j_1 j_2}  \frac{j_1^{-\epntCfY-1}\rzeta(\epntCfX+\epntCfY)}{\rzeta(\epntCfY)^2}l\right)\right]\\
&= (1-\sgnFractn)\left[ \frac{\rzeta(\epntCfX+\epntCfY)}{\rzeta(\epntCfY)} + \frac{\rzeta(\epntCfX+\epntCfY)}{\rzeta(\epntCfY)^2}\sum_{r=1}^{\infty}\sum_{l=1}^{\infty}l\left(\sum_{j_r=1}^{\infty}\regCoef_{j_r l} \sum_{j_{r-1}=1}^{\infty}\regCoef_{j_{r-1} j_r} \cdots \sum_{j_1=1}^{\infty}\regCoef_{j_1 j_2} j_1^{-\epntCfY-1} \right)\right].\eal
We observe that
\bal\nn
&\sum_{j_1=1}^{\infty}\regCoef_{j_1 j_2}  j_1^{-\epntCfY-1}
=\sum_{j_1=1}^{\infty}\sgnFractn\frac{j_1^{-\epntCfX-\epntCfY}j_2^{-\epntCfY-1}}{\rzeta(\epntCfY)}
=j_2^{-\epntCfY-1} \frac{\sgnFractn\rzeta(\epntCfX+\epntCfY)}{\rzeta(\epntCfY)},\\
&\sum_{j_2=1}^{\infty}\regCoef_{j_2 j_3} \sum_{j_1=1}^{\infty}\regCoef_{j_1 j_2} j_1^{-\epntCfY-1}
=\sum_{j_2=1}^{\infty}\regCoef_{j_2 j_3}j_2^{-\epntCfY-1} \frac{\sgnFractn\rzeta(\epntCfX+\epntCfY)}{\rzeta(\epntCfY)}
= j_3^{-\epntCfY-1} \frac{\sgnFractn^2\rzeta(\epntCfX+\epntCfY)^2}{\rzeta(\epntCfY)^2},\\
&\quad\vdots\\
&\sum_{j_r=1}^{\infty}\regCoef_{j_r l} \sum_{j_{r-1}=1}^{\infty}\regCoef_{j_{r-1} j_r} \cdots \sum_{j_1=1}^{\infty}\regCoef_{j_1 j_2} j_1^{-\epntCfY-1}
=l^{-\epntCfY-1} \frac{\sgnFractn^r\rzeta(\epntCfX+\epntCfY)^r}{\rzeta(\epntCfY)^r}. \eal
Therefore, 
\bal\nn
&(1-\sgnFractn)\left[ \frac{\rzeta(\epntCfX+\epntCfY)}{\rzeta(\epntCfY)} + \frac{\rzeta(\epntCfX+\epntCfY)}{\rzeta(\epntCfY)^2}\sum_{r=1}^{\infty}\sum_{l=1}^{\infty}l\left(\sum_{j_r=1}^{\infty}\regCoef_{j_r l} \sum_{j_{r-1}=1}^{\infty}\regCoef_{j_{r-1} j_r} \cdots \sum_{j_1=1}^{\infty}\regCoef_{j_1 j_2} j_1^{-\epntCfY-1} \right)\right]\\
&=(1-\sgnFractn)\left[ \frac{\rzeta(\epntCfX+\epntCfY)}{\rzeta(\epntCfY)} + \frac{\rzeta(\epntCfX+\epntCfY)}{\rzeta(\epntCfY)^2}\sum_{r=1}^{\infty} \sum_{l=1}^{\infty}l^{-\epntCfY} \frac{\sgnFractn^r\rzeta(\epntCfX+\epntCfY)^r}{\rzeta(\epntCfY)^r}\right]\\
&=(1-\sgnFractn)\left[ \frac{\rzeta(\epntCfX+\epntCfY)}{\rzeta(\epntCfY)} + \frac{\rzeta(\epntCfX+\epntCfY)}{\rzeta(\epntCfY)^2}\sum_{r=1}^{\infty} \rzeta(\epntCfY)  \frac{\sgnFractn^r\rzeta(\epntCfX+\epntCfY)^r}{\rzeta(\epntCfY)^r}\right]\\
&=(1-\sgnFractn)\sum_{r=1}^{\infty}   \frac{\sgnFractn^{r-1}\rzeta(\epntCfX+\epntCfY)^r}{\rzeta(\epntCfY)^r}
=(1-\sgnFractn) \frac{\frac{\rzeta(\epntCfX+\epntCfY)}{\rzeta(\epntCfY)}}{1-\frac{\sgnFractn\rzeta(\epntCfX+\epntCfY)}{\rzeta(\epntCfY)}}
=\frac{(1-\sgnFractn)\rzeta(\epntCfX+\epntCfY)}{\rzeta(\epntCfY)-\sgnFractn\rzeta(\epntCfX+\epntCfY)}.\eal
Thus, $\sum_{r=0}^{\infty}(\regOp^r(\noise_{i-r}))'$ uniformly converges and hence
\bal\nn
\sup_{t\in\dom} \left|\frac{\diffop}{\diffop t}\sum_{r=0}^{\infty}\regOp^r(\noise_{i-r})(t)\right| 
= \sup_{t\in\dom}\sum_{r=0}^{\infty}|(\regOp^r(\noise_{i-r}))'(t)|
\le \frac{(1-\sgnFractn)\rzeta(\epntCfX+\epntCfY)}{\rzeta(\epntCfY)-\sgnFractn\rzeta(\epntCfX+\epntCfY)}.\eal
Note that
\bal\nn
\frac{(1-\sgnFractn)\rzeta(\epntCfX+\epntCfY)}{\rzeta(\epntCfY)-\sgnFractn\rzeta(\epntCfX+\epntCfY)}\le 1 
&\Leftrightarrow (1-\sgnFractn)\rzeta(\epntCfX+\epntCfY) \le\rzeta(\epntCfY)-\sgnFractn\rzeta(\epntCfX+\epntCfY)\\
&\Leftrightarrow \rzeta(\epntCfX+\epntCfY) \le\rzeta(\epntCfY),\eal
whence \eqref{eq:convex_ts} follows. 
Similarly, it can be shown that $\sum_{r=0}^{\infty}|\regOp^r(\noise_{i-r})|$ uniformly converges. Applying the dominated convergence theorem yields 
\bal\nn
\sum_{r=0}^{\infty}\regOp^r(\noise_{i+1-r})
= \regOp\left(\sum_{r=1}^{\infty}\regOp^{r-1}(\noise_{i+1-r})\right) + \noise_{i+1}
= \regOp\left(\sum_{r=0}^{\infty}\regOp^r(\noise_{i-r})\right) + \noise_{i+1},\eal
whence model \eqref{eq:autoreg} holds. 

We also note that the stationarity condition \ref{ass:autoRegOpConv} holds.
In fact, for any $\arbiFctn\in\tangentspace{\FmeanZ}$ such that $\|\arbiFctn\|_{\FmeanZ}>0$, 
\bal\nn
\regOp\arbiFctn
&= \sum_{l=1}^{\infty}\sum_{j_1=1}^{\infty}\regCoef_{j_1 l}\innerprod{\arbiFctn}{\basisL_{j_1}\circ\cdfFmeanZ}_{\FmeanZ}\basisL_l\circ\cdfFmeanZ,\\
\regOp^2\arbiFctn
&= \sum_{l=1}^{\infty}\sum_{j_2=1}^{\infty}\regCoef_{j_2 l}\sum_{j_1=1}^{\infty}\regCoef_{j_1j_2}\innerprod{\arbiFctn}{\basisL_{j_1}\circ\cdfFmeanZ}_{\FmeanZ}\basisL_l\circ\cdfFmeanZ,\\
&\vdots\\
\regOp^r\arbiFctn
&= \sum_{l=1}^{\infty}\sum_{j_r=1}^{\infty}\regCoef_{j_r l}\sum_{j_{r-1}=1}^{\infty}\regCoef_{j_{r-1}j_r}\cdots\sum_{j_1=1}^{\infty}\regCoef_{j_1j_2}\innerprod{\arbiFctn}{\basisL_{j_1}\circ\cdfFmeanZ}_{\FmeanZ}\basisL_l\circ\cdfFmeanZ. \eal
Hence, 
\bal\nn
&\|\regOp^r\arbiFctn\|_{\FmeanZ}^2\\
&= \sum_{l=1}^{\infty}\left(\sum_{j_r=1}^{\infty}\regCoef_{j_r l}\sum_{j_{r-1}=1}^{\infty}\regCoef_{j_{r-1}j_r}\cdots\sum_{j_1=1}^{\infty}\regCoef_{j_1j_2}\innerprod{\arbiFctn}{\basisL_{j_1}\circ\cdfFmeanZ}_{\FmeanZ}\right)^2\\
&= \sum_{l=1}^{\infty}\left(\frac{\sgnFractn^r}{\rzeta(\epntCfY)^r} \sum_{j_r=1}^{\infty}j_r^{-\epntCfX+1}l^{-\epntCfY-1} \sum_{j_{r-1}=1}^{\infty}j_{r-1}^{-\epntCfX+1}j_r^{-\epntCfY-1} \cdots\sum_{j_1=1}^{\infty}j_1^{-\epntCfX+1}j_2^{-\epntCfY-1}\innerprod{\arbiFctn}{\basisL_{j_1}\circ\cdfFmeanZ}_{\FmeanZ}\right)^2\\
&= \sum_{l=1}^{\infty}\left(\frac{\sgnFractn^r}{\rzeta(\epntCfY)^r}l^{-\epntCfY-1} \sum_{j_r=1}^{\infty}j_r^{-\epntCfX-\epntCfY} \sum_{j_{r-1}=1}^{\infty}j_{r-1}^{-\epntCfX-\epntCfY} \cdots \sum_{j_2=1}^{\infty}j_2^{-\epntCfX-\epntCfY} \sum_{j_1=1}^{\infty}j_1^{-\epntCfX+1}\innerprod{\arbiFctn}{\basisL_{j_1}\circ\cdfFmeanZ}_{\FmeanZ}\right)^2\\
&= \sum_{l=1}^{\infty}\left(\frac{\sgnFractn^r\rzeta(\epntCfX+\epntCfY)^{r-1}}{\rzeta(\epntCfY)^r}l^{-\epntCfY-1} \sum_{j_1=1}^{\infty}j_1^{-\epntCfX+1}\innerprod{\arbiFctn}{\basisL_{j_1}\circ\cdfFmeanZ}_{\FmeanZ} \right)^2\\
&= \frac{\sgnFractn^{2r}\rzeta(\epntCfX+\epntCfY)^{2r-2}}{\rzeta(\epntCfY)^{2r}} \rzeta(2\epntCfY+2) \left( \sum_{j_1=1}^{\infty}j_1^{-\epntCfX+1}\innerprod{\arbiFctn}{\basisL_{j_1}\circ\cdfFmeanZ}_{\FmeanZ} \right)^2\\
&\le \frac{\sgnFractn^{2r}\rzeta(\epntCfX+\epntCfY)^{2r-2}}{\rzeta(\epntCfY)^{2r}} \rzeta(2\epntCfY+2)  \sum_{j_1=1}^{\infty}j_1^{-2\epntCfX+2} \sum_{j_1=1}^{\infty}\innerprod{\arbiFctn}{\basisL_{j_1}\circ\cdfFmeanZ}_{\FmeanZ}^2\\
&= \frac{\sgnFractn^{2r}\rzeta(\epntCfX+\epntCfY)^{2r-2}}{\rzeta(\epntCfY)^{2r}} \rzeta(2\epntCfY+2)\rzeta(2\epntCfX-2) \|\arbiFctn\|_{\FmeanZ}^2.\eal
Therefore, $\|\regOp^r\|_{\hilbert_{\FmeanZ}}<1$ if 
\bal\nn
\frac{\sgnFractn^{2r}\rzeta(\epntCfX+\epntCfY)^{2r-2}}{\rzeta(\epntCfY)^{2r}} \rzeta(2\epntCfY+2)\rzeta(2\epntCfX-2) < 1, \eal
i.e.,
\bal\label{eq:autoRegOpConv-special}
\sgnFractn\frac{\rzeta(\epntCfX+\epntCfY)}{\rzeta(\epntCfY)} <  \left[\frac{\rzeta(\epntCfX+\epntCfY)^2}{\rzeta(2\epntCfY+2)\rzeta(2\epntCfX-2)}\right]^{1/(2r)}.\eal
Note that $\sgnFractn\rzeta(\epntCfX+\epntCfY)/\rzeta(\epntCfY)<1$ and that by Cauchy--Schwarz inquality, $\rzeta(\epntCfX+\epntCfY)^2\le\rzeta(2\epntCfY+2)\rzeta(2\epntCfX-2)$, whence the right hand side of \eqref{eq:autoRegOpConv-special} $\uparrow 1$, as $r\ra\infty$. Therefore, there always exists $r\in\pint$ such that  \eqref{eq:autoRegOpConv-special} holds.

\section{Implementation}\label{sec:imple}

\subsection{Tuning Parameter Selection}\label{sec:tuning}
\subsubsection{Choosing $\nFPCX$ and $\nFPCY$ for the Independent Case} The following data-based method for  choosing  the numbers of predictor and response FPCs $\nFPCX$ and $\nFPCY$, respectively, as defined in \eqref{eq:regOpEst}, when one has an i.i.d. sample of distributions $\{\Xi,\Yi\}_{i=1}^n$, was found to be adequate in  practical applications.  
We first choose the number of FPCs $\nFPCY$ for the logarithms of the response distributions $\Log_{\FmeanYest}\Yi$, by applying either leave-one-curve-out cross-validation or thresholding according to cumulative fraction of variance explained (FVE). For cross-validation, the objective function to be minimized is the discrepancy between predicted  and observed trajectories, i.e.,
\be\nn
\nFPCY = \argmin_{\nFPCY'} \s1n \left\|\Log_{\FmeanYest} \YiEst - \sum_{k=1}^{\nFPCY'} \innerprod{\Log_{\FmeanYest} \YiEst}{\egnfctnYest_{k,-i}}_{\FmeanYest} \egnfctnYest_{k,-i} \right\|_{\FmeanYest}^2,
\ee
where $\{\egnfctnYest_{k,-i}\}_{k\ge 1}$ are  the estimated eigenfunctions from the functional principal component analysis (FPCA) of the $i$th-curve-left-out sample  $\{\Log_{\FmeanYest}\YlEst\}_{i'\ne i}$. 
For the choice by FVE, $\nFPCY$ is chosen such that at least $100(1-\alpha)\%$ of the variance is explained by the first $\nFPCY$ FPCs,  where users need to specify $\alpha$, with the common  choice  $\alpha=0.05$.  

Now turning to the choice of $\nFPCX$, we use leave-one-curve-out cross-validation by minimizing the difference between the curves constructed with FPCs predicted by linear regression and the observed trajectories: 
\be\nn \nFPCX = \argmin_{\nFPCX'} \s1n \left\|\Log_{\FmeanYest} \YiEst - \sum_{k=1}^{\nFPCY} \sum_{j=1}^{\nFPCX'} \regCoefEst_{jk,-i}\innerprod{\Log_{\FmeanXest}\XiEst}{\egnfctnXest_{j,-i}}_{\FmeanXest}\egnfctnYest_{k,-i} \right\|_{\FmeanYest}^2,\ee
where $\regCoefEst_{jk,-i} = \egnvalXest_{j,-i}\inv\crovalEst_{jk,-i}$ with 
$\crovalEst_{jk,-i} = (n-1)\inv\sum_{i'\ne i} \innerprod{\Log_{\FmeanXest}\XiEstVar} {\egnfctnXest_{j,-i}} _{\FmeanXest}$  $\cdot\innerprod{\Log_{\FmeanYest}\YiEstVar} {\egnfctnYest_{k,-i}} _{\FmeanYest}$, 
and $\{\egnvalXest_{j,-i}\}_{j\ge 1}$ and $\{\egnfctnXest_{j,-i}\}_{j\ge 1}$ are the estimated eigenvalues and eigenfunctions from the FPCA of the $i$th-curve-left-out sample $\{\Log_{\FmeanXest}\XlEst\}_{i'\ne i}$, respectively. 

We mention that in practical implementations we replace leave-one-curve-out cross validation by 5-fold cross-validation when 
$n>30$. 

\subsubsection{Choosing $\nFPCX$ for Distribution-Valued Time Series}
Here, we use a cross-validation approach proposed by \citet{berg:18} to select the number of FPCs $\nFPCZ$ as defined in \eqref{eq:autoRegOpEst} for the  time series case. Note that there is no second truncation parameter $\nFPCY$. 
Following \citet{berg:18}, we first divide the observed time series into a training set and a testing set, and then apply $k$-fold cross validation on the training set to choose the number of FPCs $\nFPCX$; secondly, the regression coefficient function $\regKernel$ will be estimated on the whole training set with the optimal $\nFPCX$; lastly, the estimate of $\regKernel$ will be applied on the testing set to evaluate the performance of out-of-sample prediction.

\subsection{Boundary Projection}\label{sec:projection}
We discuss the independent scenario only; the time series case is analogous. 
If \eqref{eq:out-of-set-issue} happens, we update the fit by a projection onto the boundary of $\Log_{\FmeanYest}\manifold$ along the line segment connecting the origin $0$ and the original fit $\regOpEst(\Log_{\FmeanXest} \XiEst)$. Specifically, we multiply the original estimate $\regOpEst(\Log_{\FmeanXest} \XiEst)$ by a constant $\projconst_i$ such that 
\be\label{eq:projection}
\projconst_i = \max \{\projconst\in[0,1]\colon \projconst\regOpEst(\Log_{\FmeanXest} \XiEst) + \id \text{ is non-decreasing}\}.
\ee
Note that $\projconst_i = 1$ when $\regOpEst(\Log_{\FmeanXest} \XiEst)\in \Log_{\FmeanYest}\manifold$. In our implementation, the fitted logarithmic response is then given by $\projconst_i\regOpEst(\Log_{\FmeanXest} \XiEst)$ for all $i=1,\ldots,n$.

\section{Additional Simulations}\label{sec:simu_supp}

\subsection{\update{Illustration of Asymptotic Results in Theorem~\ref{thm:rate_oracle}}}\label{sec:simu_asymp}

Regardless of the distribution estimation method, we consider the case where distributions are fully observed. 
To illustrate the asymptotic results in Theorem~\ref{thm:rate_oracle}, we generate data as follows such that we can derive the decay rates in  \ref{ass:eigenval_spacingX}--\ref{ass:coef_decay} and hence obtain the value of $\indepRate(n)$ in Theorem~\ref{thm:rate_oracle}. 
With $\maxDbasisL_j$, $\maxDenFmeanX$ and $\maxDenFmeanY$ defined as below \eqref{eq:distort} and $\usedFPCX = \usedFPCY = 20$, for $j=1,\dots,\usedFPCX$ and $k=1,\dots,\usedFPCY$, we set $\regCoef_{jk} = \sgnFractn j^{-\epntCfX} k^{-\epntCfY}\maxDbasisL_j \maxDenFmeanX (\maxDbasisL_k\maxDenFmeanY \sum_{l=1}^\infty l^{-\epntCfY} )\inv = \sgnFractn j^{-\epntCfX+1} k^{-\epntCfY-1}\maxDenFmeanX (\maxDenFmeanY\sum_{l=1}^\infty l^{-\epntCfY})\inv$, 
where $\sgnFractn\in(0,1)$, $\epntCfX>3/2$, and $\epntCfY>1$ are constants. 
Taking $\convSeries_{1j} = j^{-\epntX}$ with $\epntX > 1$, data were generated as follows: 
\ben[label = Step \arabic*:, itemindent=1.1em]
\item Generate $\frcoef_{ij} \sim \mathrm{Unif}(-\convSeries_{1j}(\maxDbasisL_j\maxDenFmeanX\sum_{l=1}^\infty \convSeries_{1l})\inv, \convSeries_{1j}(\maxDbasisL_j\maxDenFmeanX\sum_{l=1}^\infty \convSeries_{1l})\inv)$ independently for $i=1,\dots,n$ and $j=1,\dots,\usedFPCX$, 
whence $\Log_{\FmeanX}\Xi = \sum_{j=1}^{\usedFPCX} \frcoef_{ij} \basisL_j\circ\cdfFmeanX$, with $\basisL_j$ as per \eqref{eq:egbases},  $\regOp(\Log_{\FmeanX}\Xi) = \sum_{k=1}^{\usedFPCY} \sum_{j=1}^{\usedFPCX} \regCoef_{jk}\frcoef_{ij}\basisL_k\circ\cdfFmeanY$, and $\Xi 
= \Exp_{\FmeanX}(\sum_{j=1}^{\usedFPCX} \frcoef_{ij} \basisL_j\circ\cdfFmeanX)$. 
\item Sample $\errCoef_{ik}\sim \mathrm{Unif}(-\maxErrCoef_k,\maxErrCoef_k)$, independently for $i=1,\dots,n$ and $k=1,\dots,\usedFPCY$, 
where $\maxErrCoef_k 
= (1-\sgnFractn) k^{-\epntCfY}(\maxDbasisL_k\maxDenFmeanY \sum_{l=1}^\infty l^{-\epntCfY})\inv $.  
Let $\Yi = \Exp_{\FmeanY}(\regOp(\Log_{\FmeanX}\Xi) + \sum_{k=1}^{\usedFPCY} \errCoef_{ik}  \basisL_k\circ\cdfFmeanY)$.
\een
We considered the two cases with different choices of the Fr\'echet mean distributions $\FmeanX$ and $\FmeanY$ as in Section~\ref{sec:simu}. 
Taking $\sgnFractn=0.9$, $\epntX = 1.5$, $\epntCfX= 8.5$ and $\epntCfY = 5$, 
we simulated 500 runs for each $n\in\{20,100,500\}$. 

\update{From \eqref{eq:measureOfError} one finds that among the terms that determine  the convergence rate of $\regOpOracle$ in Theorem~\ref{thm:rate_oracle}, the terms $\nFPCX^{1-2\coefDecayX}$ and $\nFPCY^{1-2\coefDecayY}$ correspond to the bias, i.e., $\|\regKernel - \sum_{k=1}^{\nFPCY}\sum_{j=1}^{\nFPCX}\regCoef_{jk}\egnfctnY_k\otimes\egnfctnX_j\|_{\jointmean}^2$, and the other terms correspond to the variance, i.e., $\|\paraTransKernel\regKernelOracle - \sum_{k=1}^{\nFPCY}\sum_{j=1}^{\nFPCX}\regCoef_{jk}\egnfctnY_k\otimes\egnfctnX_j\|_{\jointmean}^2$, where $\paraTransKernel\regKernelOracle$ is defined above \eqref{eq:measureOfError}.
	Since in simulations only a finite number of basis functions can be included in the generation of the regression coefficient function $\regKernel=\sum_{k=1}^{\usedFPCY}\sum_{j=1}^{\usedFPCX}\regCoef_{jk}(\basisL_k\circ\cdfFmeanY)\otimes(\basisL_j\circ\cdfFmeanX)$, the number of included eigenfunctions in $\regOpOracle$ is necessarily bounded ($\nFPCX\le\usedFPCX$, $\nFPCY\le\usedFPCY$) and does not increase as $n$ increases. Therefore, we focus here exclusively on  the variance part,  which converges with a rate of  $\Op(n^{-1})$ according to Theorem~\ref{thm:rate_oracle}.}

\begin{figure}[hbt!]
	\centering
	\begin{subfigure}[c]{.4\textwidth}
		\includegraphics[width=.9\linewidth]{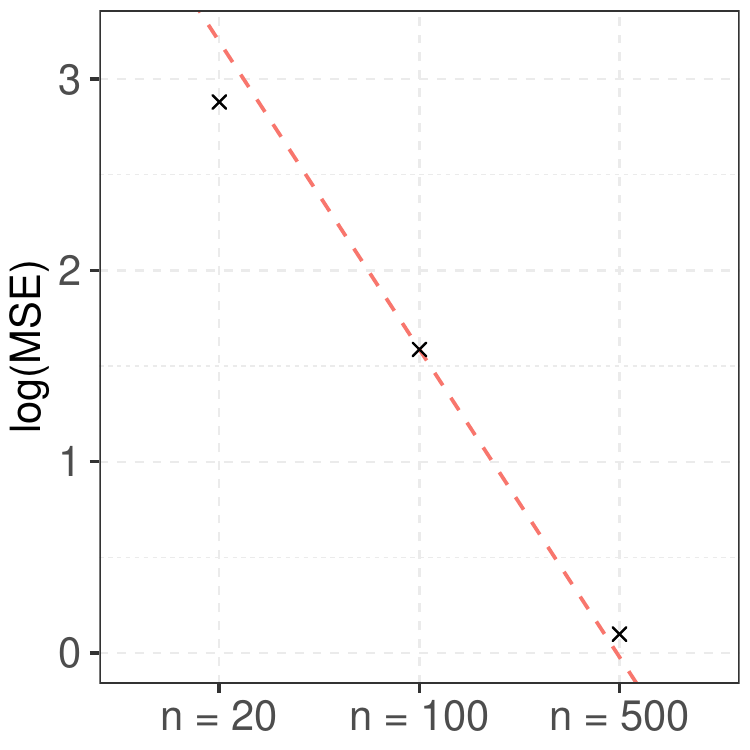}
		\caption{Case~\ref{case:d2d_tgaus}.}
	\end{subfigure}
	\begin{subfigure}[c]{.4\textwidth}
		\includegraphics[width=.9\linewidth]{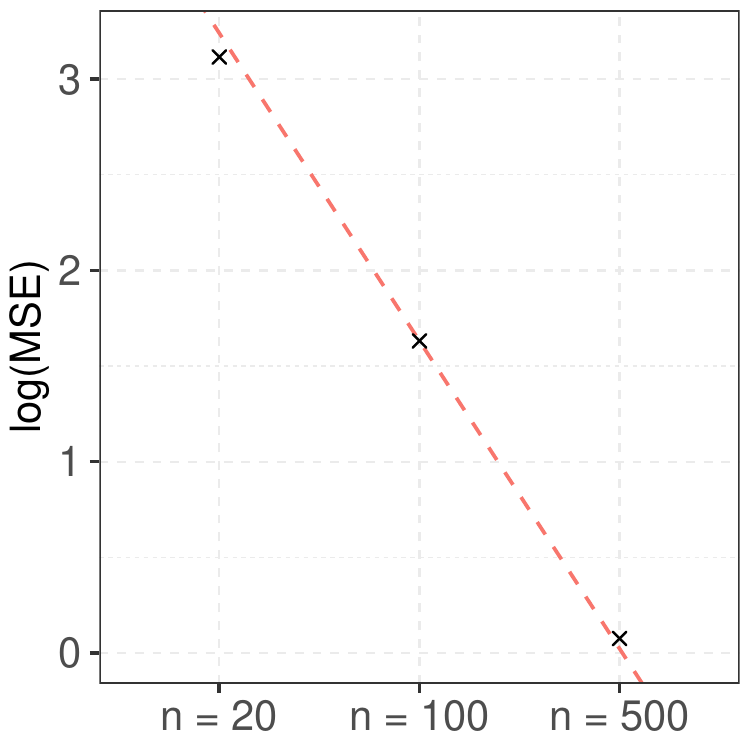}
		\caption{Case~\ref{case:d2d_beta}.}
	\end{subfigure}
	\caption{Illustration of Theorem~\ref{thm:rate_oracle} using simulations. The $\mathrm{MSE}_n$ across 500 runs on a log scale,  represented by the crosses ``$\times$'', are shown for $n\in\{20,100,500\}$ in the two cases. 
		The dashed lines represent the theoretical $\Op$ rate passing through the point at which $n=100$, i.e., $\log(\mathrm{MSE}_n) = -1\cdot[\log(n)-\log(100)] + \log(\mathrm{MSE}_{100})$.
	} \label{fig:simu_asympCheck_box}
\end{figure}

\update{For each run $l=1,\dots,500$, we computed the variance part, $\mathrm{SE}_{n,l} =\|\paraTransKernel\regKernelOracle_l - \sum_{k=1}^{\nFPCY}\sum_{j=1}^{\nFPCX}\regCoef_{jk}\egnfctnY_k\otimes\egnfctnX_j\|_{\jointmean}^2$, where $\regKernelOracle_l$ is the estimate obtained in the $l$th run. 
	We show in Figure~\ref{fig:simu_asympCheck_box} the mean of $\mathrm{SE}_{n,l}$ across MC runs, $\mathrm{MSE}_n = 500\inv\sum_{l=1}^{500}\mathrm{SE}_{n,l}$, on a log scale for each $n\in\{20,100,500\}$ as well as the theoretical rate represented by the dashed line passing through the point at which $n=100$, i.e., $\log(\mathrm{MSE}_n) = -1\cdot[\log(n)-\log(100)] + \log(\mathrm{MSE}_{100})$. This suggests that indeed  $\expect\|\paraTransKernel\regKernelOracle - \sum_{k=1}^{\nFPCY}\sum_{j=1}^{\nFPCX}\regCoef_{jk}\egnfctnY_k\otimes\egnfctnX_j\|_{\jointmean}^2$ converges with a rate close to $n^{-1}$, which aligns with the results in Theorem~\ref{thm:rate_oracle}.
}

\subsection{\update{Robustness of the Proposed Method in the Regression between Gaussian Distributions}}\label{sec:simu_robust_gaus}
\begin{figure}[hbt!]
	\centering
	\includegraphics[width=0.45\textwidth]{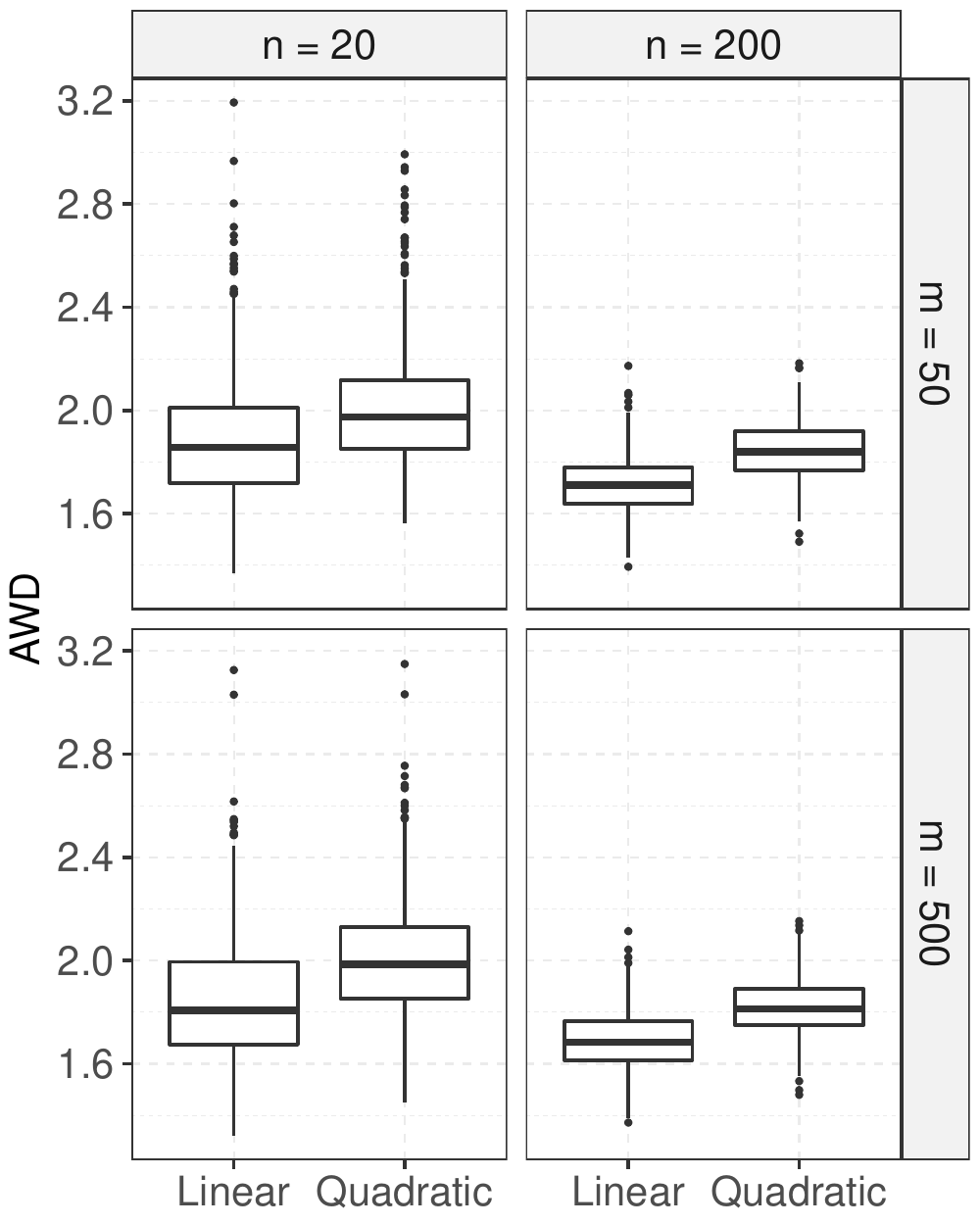}
	\caption{Boxplots of the out-of-sample AWDs as per \eqref{eq:AWD} for the linear and quadratic cases, each with $(n,\nDp)\in\{20,200\}\times \{50,500\}$.}
	\label{fig:simu_robust}
\end{figure}

Suppose that the predictor and response distributions are both Gaussian, $\X = N(\Xmean,\Xsd^2)$ and $\Y=N(\Ymean,\Ysd^2)$, where $(\Xmean,\Xsd,\Ymean,\Ysd)$ is a random vector with a joint distribution on $\real\times\real_+\times\real\times\real_+$ and $\EXsd,\EYsd>0$. 
For simulations, we independently sample $\Ximean\sim N(0,1^2)$, $\errInYmean_i\sim N(0,0.5^2)$, $\Xisd\sim \mathrm{Gamma}(0.5,0.5)$, and $\Yisd\sim\mathrm{Gamma}(1,0.5)$, independently for $i=1,\dots,n$. 
We consider different cases of generating  $\Yimean$: (1) Linear case: $\Yimean = 1 + \Ximean + \errInYmean_i$; (2) Quadratic case: $\Yimean = 1 + \Ximean -0.5 \Ximean^2 + \errInYmean_i$. 
Then i.i.d. samples of size $\nDp$ are drawn from each of $\{\Xi\}_{i=1}^n$ and $\{\Yi\}_{i=1}^n$. 
We note that $(\Xi,\Yi)$ generated as per the linear case satisfy the proposed model in \eqref{eq:d2dreg}, which does not hold for the quadratic case. 
Four scenarios were considered with $n\in\{20,200\}$ and $\nDp\in\{50,500\}$, and 500 runs were executed for each $(n,\nDp)$ pair and each case. 
The out-of-sample AWDs for 200 new predictors as per \eqref{eq:AWD} were computed for each run. The results for both linear and quadratic cases are summarized in Figure~\ref{fig:simu_robust}, where it can be seen that the performance of the proposed method does not worsen too much when the model becomes invalid in the quadratic case as compared to the linear case.

\subsection{\update{Distribution-to-Scalar Regression: Comparison with the Gaussian Process Regression Method}}\label{sec:simu_d2s}

\begin{figure}[hbt!]
	\centering
	\begin{subfigure}[c]{.4\textwidth}
		\includegraphics[width=.9\linewidth]{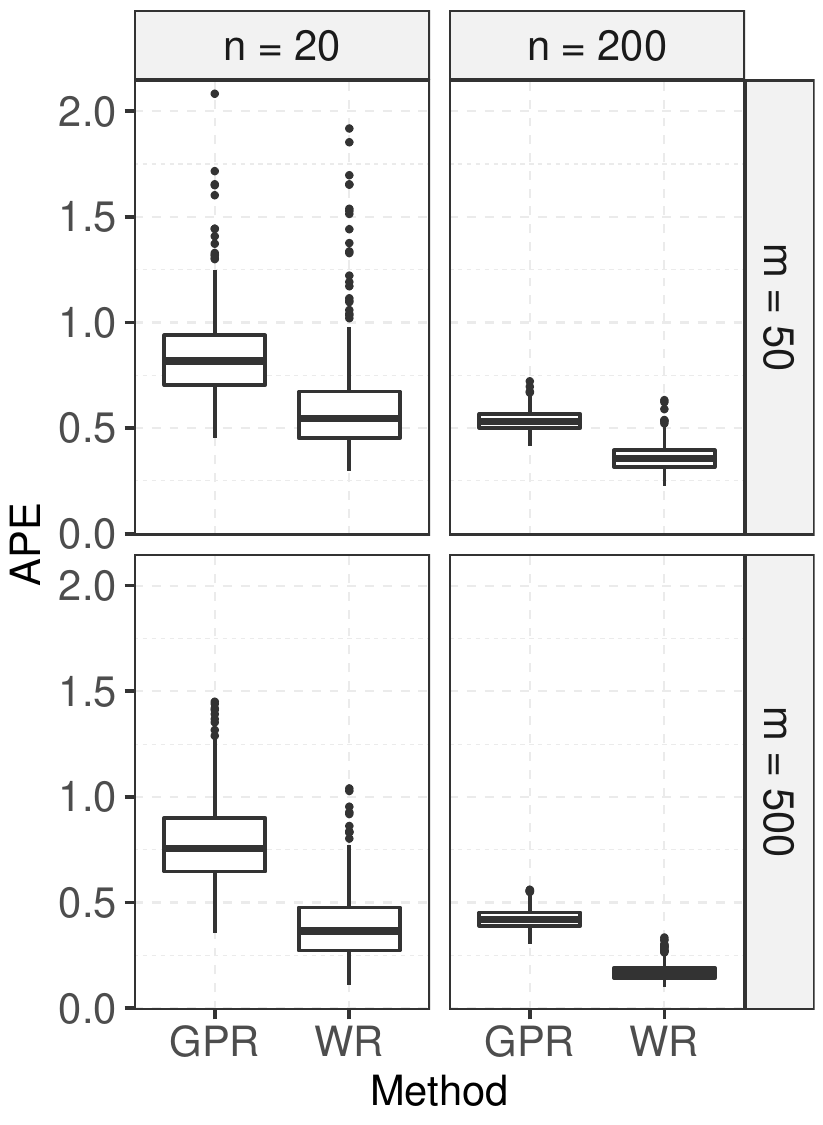}
		\vspace{-0.5em}
		\caption{Case~\ref{case:d2s_gaus_gausErr}.}
	\end{subfigure}
	\begin{subfigure}[c]{.4\textwidth}
		\includegraphics[width=.9\linewidth]{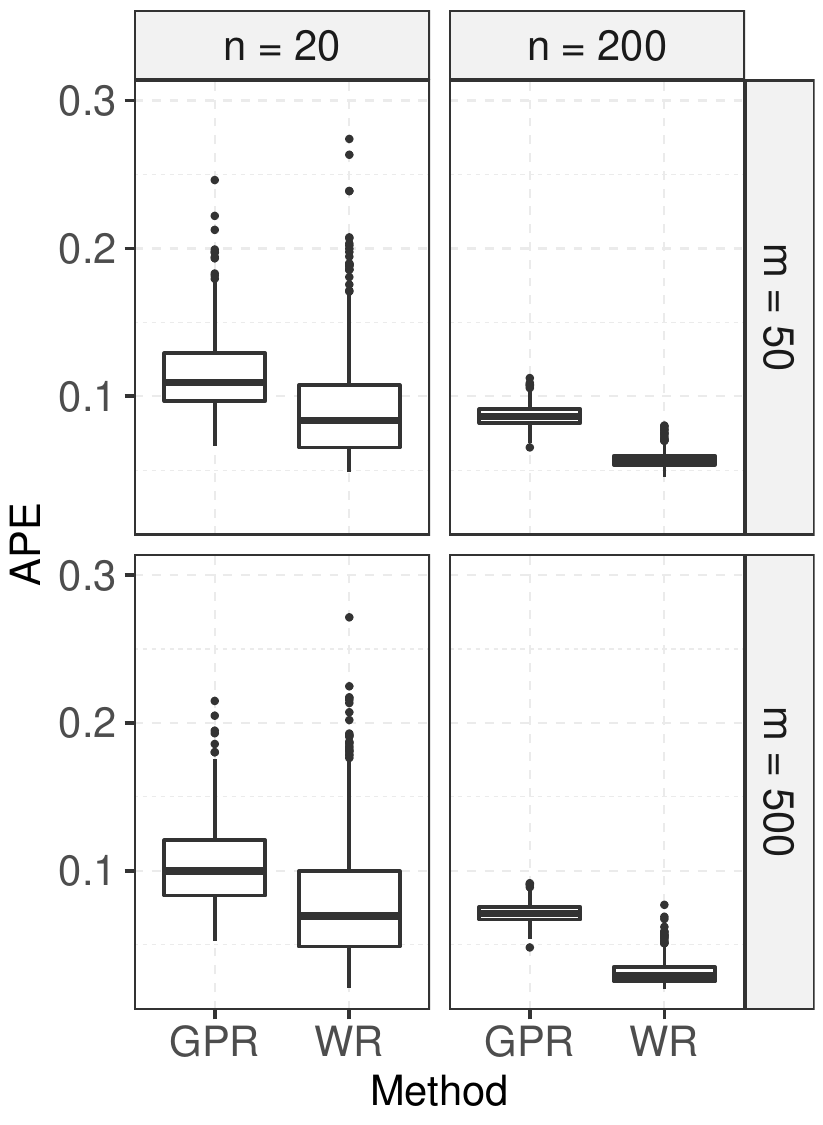}
		\vspace{-0.5em}
		\caption{Case~\ref{case:d2s_beta_gausErr}.}
	\end{subfigure}
	\caption{Boxplots of the out-of-sample APEs as per \eqref{eq:ape} for the GPR and proposed WR methods for distribution-to-scalar regression. One outlier is omitted for Case~\ref{case:d2s_gaus_gausErr} for better visual comparison.}
	\label{fig:simu_d2s}
\end{figure}

In this section, we compare the proposed distribution-to-scalar Wasserstein regression as per \eqref{eq:d2sreg} in Section~\ref{sec:cov_indep} with a Gaussian process regression (GPR) method proposed by \citet{bach:17}. The data is generated by
$\sYi = \expect(\sY) +  \innerprod{\regKernelDtoS}{\Log_{\FmeanX}\Xi}_{\FmeanX} + \errInsY_i$, where $\regKernelDtoS(s) = 3 - 2\cdfFmeanX(s) + \cdfFmeanX(s)^2$, for $s\in\dom$, 
and the distributional predictors $\Xi$ and random noise $\errInsY_i$ are generated in two cases as follows. 
\renewcommand{\thecase}{2.\arabic{case}}
\begin{case} \label{case:d2s_gaus_gausErr}
	$\Xi = \distortmap_{A_i}\# N(\Ximean,\Xisd^2)$, where $\Ximean\sim N(0,1^2)$, $\Xisd\sim\mathrm{Gamma}(0.5,0.5)$, $A_i\sim\mathrm{Unif}\{\pm\pi,$\newline$\pm 2\pi,\pm 3\pi\}$, and $\errInsY_i\sim N(0,1^2)$, independently for $i=1,\dots,n$.
\end{case}
\begin{case}\label{case:d2s_beta_gausErr}
	$\Xi = \distortmap_{A_i}\#\mathrm{Beta}(\betashape_{1i},\betashape_{2i})$, where $\betashape_{1i}\sim\mathrm{Unif}(1,5)$, $\betashape_{2i}\sim\mathrm{Unif}(1,5)$, $A_i\sim\mathrm{Unif}\{\pm\pi,\pm 2\pi,$\newline$\pm 3\pi\}$, and $\errInsY_i\sim N(0,0.2^2)$, independently for $i=1,\dots,n$.
\end{case}
Here, $\distortmap_a$ is defined as per \eqref{eq:distort}. 
Then, i.i.d. samples of size $\nDp$ are drawn from each of $\{\Xi\}_{i=1}^n$. 
Five hundred runs were executed for each $(n,\nDp)$ pair with $n\in\{20,200\}$ and $\nDp\in\{50,500\}$ and each case considered. The out-of-sample average prediction errors (APEs) for 200 new predictors were computed for each run. Specifically, \bgt\label{eq:ape}
\mathrm{APE}(n,\nDp) = \frac{1}{200}\sum_{i=n+1}^{n+200} \left|\expect(\sYi\mid\Log_{\FmeanX}\Xi) - \sYiEst\right|,\egt 
where $\sYiEst$ denotes the predicted value for $\sYi$; for the proposed method, $\sYiEst = \overline{\sY} + \innerprod{\regKernelDtoSest}{\Log_{\FmeanXest}\XiEst}_{\FmeanXest}$, $\overline{\sY} = n\inv\sum_{i=1}^n\sYi$, $\regKernelDtoSest$ is as per \eqref{eq:regKernelDtoSest}, and $\XiEst$ and $\FmeanXest$ are the estimates for $\Xi$ and $\FmeanX$ as described in Section~\ref{sec:coef_est_indep}. 
The results of the GPR method and the proposed Wasserstein regression (WR) method are summarized in the boxplots of Figure~\ref{fig:simu_d2s}. The proposed method is found to outperform the GPR method when model \eqref{eq:d2sreg} is true.

\section{Autoregressive Modeling of the Mortality Data for Sweden}\label{sec:dts_sweden}

We chose Sweden as an example because demographic data are available for a longer time span and are of very high quality. 
Here, the model was trained on the time series between 1961 and 2001, and the out-of-sample prediction was evaluated for the following 15 years up to 2016, where the yearly distribution was predicted from the fitted distribution time series, using the predicted distributions from previous years. 

\begin{figure}[hbt!]
	\centering
	\includegraphics[width=.65\textwidth]{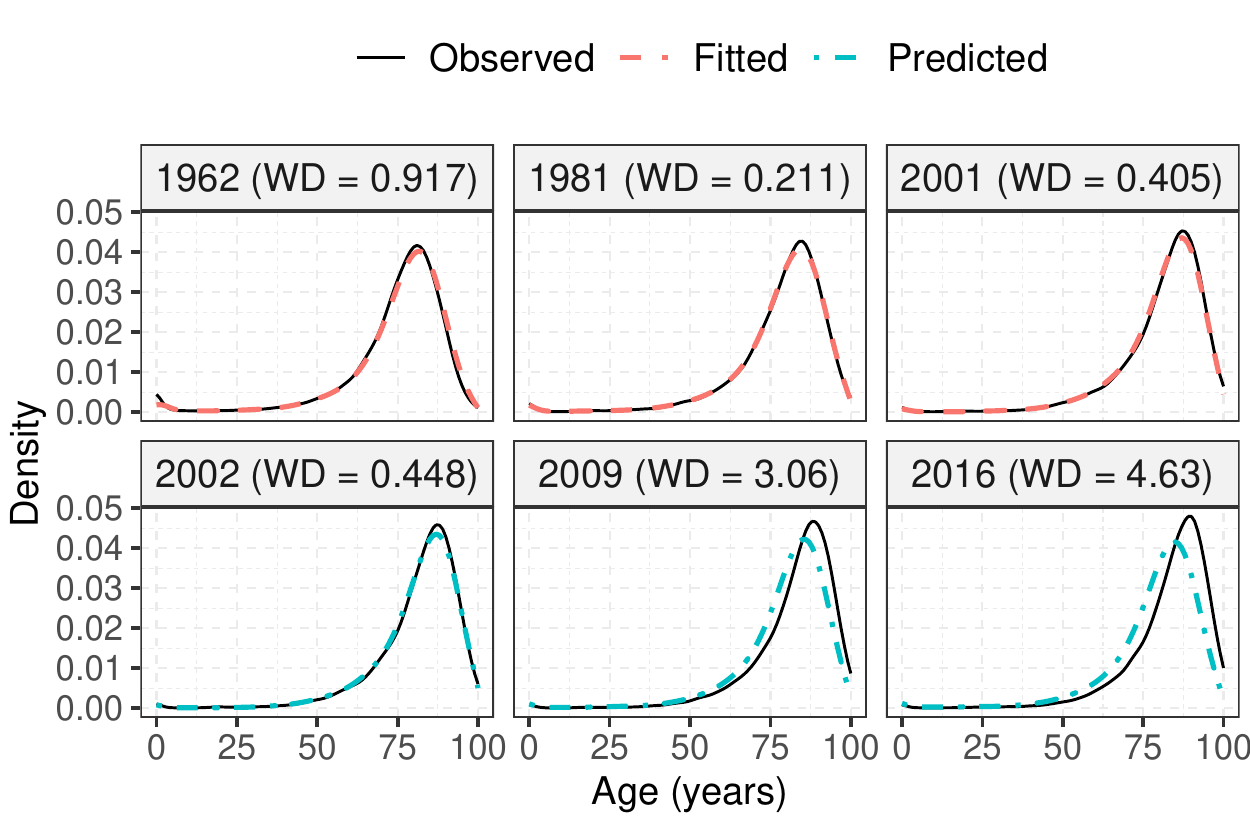}
	\caption{Implementation of the autoregressive distribution time series model for age-at-death distributions of females in Sweden.  Training period: 1961--2001 (with fitting results for three years shown in the top row). Prediction period: 2002--2016 (with prediction results for three years shown in the bottom row). The fitting/prediction Wasserstein discrepancies (WDs) are as shown for each panel.}\label{fig:dts_sweden}
\end{figure}

As can been seen from  the fitted distributions for the training period as shown in the first row of Figure~\ref{fig:dts_sweden}, they are all close to the observed distributions. 
For the prediction period, the predicted densities displayed in the second row of the figure increasingly deviate from the observed distributions going from 2002 to 2016, where the observed densities have a mode increasingly shifting to the right compared to the prediction. This means that the rightward mortality distribution shift outpaces the model expectation and longevity extension is accelerating for Sweden. 

\section{\update{Multivariate Extension}}\label{sec:disc}
Consider the Wasserstein space of probability measures on $\real^p$ with finite second moments, $\manifold(\real^p)$, for $p>1$. 
For two given measures $\refDistn,\arbiDistn\in\manifold(\real^p)$, any map $\trans\colon\real^p\ra\real^p$ that minimizes Monge's transport problem $\optTrans{\refDistn}{\arbiDistn} = \argmin_{\trans \#\refDistn = \arbiDistn} \int_{\real^p} \|\trans(x) - x\|^2 \diffop\refDistn(x)$ is called an optimal transport map. Such optimal transport maps uniquely exist if $\refDistn$ is absolutely continuous with respect to Lebesgue measure on $\real^p$ (referred to as ``a.c.'' hereafter) \citep[Theorem 6.2.4,][]{ambr:08}. Their construction is however computationally demanding and difficult to implement in practice.
We note that for $p=1$, $\optTrans{\refDistn}{\arbiDistn} = \quantileArbiDistn\circ\cdfRefDistn$, where $\cdfRefDistn$ and $\quantileArbiDistn$ are the cdf of $\refDistn$ and the quantile function of $\arbiDistn$, respectively. 
The notions of tangent spaces, exponential maps and log maps at $\refDistn$ can be analogously defined as for $\manifold(\real)$: 
$\tangentspace{\refDistn} = {\overline{\{t(\optTrans{\refDistn}{\arbiDistn} - \id):\, \arbiDistn\in\manifold,\, t >0 \}}}^{\hilbert_{\refDistn}}$; 
$\Exp_{\refDistn}\arbiFctn = (\arbiFctn + \id)\#\refDistn$ for functions of the form $\arbiFctn=t(\optTrans{\refDistn}{\arbiDistn}-\id)$; 
$\Log_{\refDistn}\arbiDistn = \optTrans{\refDistn}{\arbiDistn} - \id$, for $\arbiDistn\in\manifold(\real^p)$. Also, the tangent space $\tangentspace{\refDistn}$ is a subspace of $\hilbert_{\refDistn}$. 

Let $(\X,\Y)$ be a pair of random elements with a joint distribution on $\manifold(\real^p)\times \manifold(\real^p)$, 
assumed to be square integrable in the sense that $\expect\wdist^2(\arbiDistn,\X)<\infty$ and $\expect\wdist^2(\arbiDistn,\Y)<\infty$ for some (and thus for all) $\arbiDistn\in\manifold(\real^p)$. 
While $\manifold(\real^p)$ is not a Hadamard space for $p>1$, there exists an element $\arbiDistn\in\manifold(\real^p)$ that minimizes $\arbiDistn\mapsto\expect\wdist^2(\arbiDistn,\X)$; such minimizers are unique and referred to as the Fr\'echet mean of $\X$, $\FmeanX$, if $\X$ is a.c. (with positive probability) \citep{bigo:12}. 
We assume
\ben[label = (C\arabic*)]
\item \label{ass:acRp}With probability 1, both $\X$ and $\Y$ are a.c. and the corresponding densities, $f_1$ and $f_2$, are bounded, i.e., there exists a constant $\denBound>0$ such that $\sup_{\argu\in\real^p}f_1(\argu)<\denBound$ and $\sup_{\argu\in\real^p}f_2(\argu)<\denBound$.
\een
Under \ref{ass:acRp}, $\FmeanX$ and $\FmeanY$ are also a.c. with bounded densities \citep[Theorem~5.5.2,][]{pana:20}, whence it follows that the optimal transport maps from $\FmeanX$ and $\FmeanY$ to any $\arbiDistn\in\manifold(\real^p)$ uniquely exist, denoted by $\optTrans{\FmeanX}{\arbiDistn}$ and $\optTrans{\FmeanX}{\arbiDistn}$, respectively. The proposed distribution-to-distribution regression model  in \eqref{eq:d2dreg} for $\manifold(\dom)$ with $\dom\subseteq\real$ can hence be generalized to $\manifold(\real^p)$ with $p>1$. 

Considering $n$ independent realizations of $(\X,\Y)$, $\{(\Xi,\Yi)\}_{i=1}^n$, the empirical Fr\'echet means, $\FmeanXoracle=\argmin_{\arbiDistn\in\manifold(\real^p)}\sum_{i=1}^n\wdist^2(\arbiDistn,\Xi)$ and $\FmeanYoracle=\argmin_{\arbiDistn\in\manifold(\real^p)}\sum_{i=1}^n\wdist^2(\arbiDistn,\Yi)$, uniquely exist \citep{ague:11,alva:11} and are a.c. with bounded densities under \ref{ass:acRp} \citep{ague:11}. Similar statements hold for the empirical Fr\'echet means based on estimates of $\Xi$ and $\Yi$, $\XiEst$ and $\YiEst$, when $\Xi$ and $\Yi$ are not fully observed and only samples of measurements drawn from them are available if employing a distribution estimation method that guarantees the absolute continuity of $\XiEst$ and $\YiEst$. Hence, the proposed estimation method in Section~\ref{sec:d2dEst} can be extended to this case. 
Regarding the implementations, the empirical Fr\'echet means can be computed by the steepest descent algorithm \citep{zeme:19}, and the optimal transport maps can be obtained by the computation of a power diagram \citep{aure:87,aure:98,meri:11,levy:18} or approximate algorithms that are computationally efficient \citep{cutu:13,gene:16}. 

\unappendix
\references

\end{document}

%% file: macros_Wreg24_arXiv.tex
\usepackage{geometry,float}
\usepackage{amssymb}
\usepackage{amstext}
\usepackage{bm}
\usepackage{booktabs}
\usepackage{graphics}
\usepackage{graphicx}
\usepackage{amsthm}
\usepackage{amsmath}
\usepackage{latexsym}
\usepackage{color}
\usepackage{epsfig,epsf,psfrag}
\usepackage{sectsty}
\usepackage{graphics}
\usepackage{epstopdf}
\usepackage{subcaption}
\usepackage{multirow}
\usepackage{graphicx}
\usepackage{sectsty}
\usepackage{natbib}
\usepackage{rotate}
\usepackage{lscape}
\usepackage{enumerate}
\usepackage{bbding}
\usepackage{xspace}

\def\update{}
\def\revtwo{}

\def\note#1{\iffalse #1 \fi}

\newtheorem{prop}{Proposition}

\sectionfont{\centering\large}
\subsectionfont{\bf\normalsize}
\subsubsectionfont{\it\normalsize}

\let\counterwithin\relax
\usepackage{chngcntr}

\usepackage{titlesec}

\newcommand\unappendix{
	\titleformat{\section}{\centering\bf\large}{}{0em}{}
}

\definecolor{DarkRed}{rgb}{.7,0,.4}

\renewcommand{\baselinestretch}{1.6} 
\newcommand{\single}{\renewcommand{\baselinestretch}{1.2}\normalsize}

\renewenvironment{thebibliography}[1]{
	\begin{oldthebibliography}{#1}
		\setlength{\parskip}{0ex}
		\setlength{\itemsep}{0ex}
	}
	{
	\end{oldthebibliography}
}

\newcommand{\bea}{\begin{eqnarray*}}
	\newcommand{\eea}{\end{eqnarray*}}
\newcommand{\be}{\begin{eqnarray}}
\newcommand{\ee}{\end{eqnarray}}
\newcommand{\beq}{\begin{equation}}
\newcommand{\eeq}{\end{equation}}
\newcommand{\bal}{\begin{equation}\aligned}
\newcommand{\eal}{\endaligned\end{equation}}
\newcommand{\bgt}{\begin{equation}\begin{gathered}}
\newcommand{\egt}{\end{gathered}\end{equation}}
\newcommand{\ed}{\end{document}}

\newcommand{\btab}{\begin{tabular}}
\newcommand{\etab}{\end{tabular}}

\newcommand{\bi}{\begin{itemize}}
\newcommand{\ei}{\end{itemize}}
\newcommand{\bfi}{\begin{figure}}
\newcommand{\efi}{\end{figure}}
\newcommand{\ben}{\begin{enumerate}}
\newcommand{\een}{\end{enumerate}}
\newcommand{\bay}{\begin{array}}
\newcommand{\eay}{\end{array}}

\newcommand{\nn}{\nonumber}

\def\bco{\iffalse}

\def\var{{\rm var}}

\newcommand{\bc}{\begin{center}}
\newcommand{\ec}{\end{center}}

\def\s2{\sigma^2}
\def\hs2{\hat{\sigma}^2}

\def\ra{\rightarrow}

\DeclareMathOperator*{\argmin}{argmin}

\def\inv{^{-1}}

\bibpunct{(}{)}{;}{a}{}{,}

\usepackage{float}
\usepackage{color}

\definecolor{DarkBlue}{rgb}{0,.08,.45}
\definecolor{DarkRed}{rgb}{.7,0,.4}

\newcommand{\bpmt}{\begin{pmatrix}}
	\newcommand{\epmt}{\end{pmatrix}}

\newcommand{\bsp}{\begin{split}}
	\newcommand{\esp}{\end{split}}

\newcommand{\bdes}{\begin{description}}
	\newcommand{\edes}{\end{description}}

\newtheorem{thm}{Theorem}
\newtheorem{lem}{Lemma}
\newtheorem{cor}{Corollary}

\newtheorem*{lem*}{Lemma}
{
	\theoremstyle{remark}
	\newtheorem{assumption}{Assumption}

	\newtheorem{case}{Case} 
}
\counterwithin{case}{subsection} 

\newcommand{\bass}{\begin{assumption}}
	\newcommand{\eass}{\end{assumption}}
\newcommand{\bthm}{\begin{thm}}
	\newcommand{\ethm}{\end{thm}}
\newcommand{\blem}{\begin{lem}}
	\newcommand{\elem}{\end{lem}}
\newcommand{\bcor}{\begin{cor}}
	\newcommand{\ecor}{\end{cor}}
\newcommand{\bpf}{\begin{proof}}
	\newcommand{\epf}{\end{proof}}

\def\bco{\iffalse}

\def\var{{\rm var}}

\def\id{{\rm id}}

\def\hs{\hspace{0.2cm}}

\def\inv{^{-1}}

\def\i1{_{i1}}

\def\half{^{1/2}}

\def \mbb {\mathbb}
\def \mbf {\mathbf}

\def\wh{\widehat}
\def\wt{\widetilde}

\def\ra{\rightarrow}

\def\half{^{1/2}}

\def\s1n{\sum_{i=1}^n}
\def\p1n{\prod_{i=1}^n}
\def\i01{\int_0^1}

\usepackage[T1]{fontenc}
\usepackage[latin9]{inputenc}
\usepackage{mathtools}
\usepackage[shortlabels]{enumitem}
\usepackage{setspace}
\usepackage{url}
\usepackage{xargs}[2008/03/08]
\makeatletter

\@ifundefined{date}{}{\date{}}
\usepackage{ifpdf} 
\ifpdf 

 \IfFileExists{lmodern.sty}{\usepackage{lmodern}}{}

\fi 

\usepackage{mathrsfs}
\usepackage{ifthen}

\usepackage{xcolor}
\usepackage{fancyhdr}
\usepackage[hidelinks]{hyperref}
\hypersetup{
	colorlinks,
	linkcolor={red!50!black},
	citecolor={blue!50!black},
	urlcolor={blue!80!black}
}

\newcommand\Author{Chen, Lin \& M\"uller}
\newcommand\Title{Wasserstein Regression}

\usepackage{babel}

\def\expect{\mathbb{E}}
\def\prob{\mathbb{P}}
\def\real{\mathbb{R}}
\def\pint{\mathbb{N}_+}
\def\Op{O_p}
\def\manifold{\mathcal{W}}
\def\denBound{R}
\def\manifoldLQD{\manifold_{\denBound}^{\text{ac}}}
\def\op{o_p}
\def\hilbert{\mathcal L^2}
\def\hsSpace#1#2{\mathcal H_{#1,#2}}
\def\hsSpaceAuto#1{\mathcal H_{#1}}
\def\hsNorm#1#2{\|#2\|_{#1}}
\def\hsNormSq#1#2{\|#2\|_{#1}^2}
\def\hsOp{\mathcal A}
\def\hsOpVar{\mathcal A'}
\def\diffop{\mathrm{d}}
\newcommandx\tangentspace[1]{T_{#1}}
\def\wdist{d_W}
\def\innerprod#1#2{\langle#1,#2\rangle}
\def\Log{\mathrm{Log}}
\def\Exp{\mathrm{Exp}}

\def\borel{\mathscr{B}}

\def\dom{D}
\def\quantile{F^{-1}}
\def\cdf{F}
\def\den{f}
\def\refDistn{\mu_{*}}
\def\arbiDistn{\mu}
\def\arbiDistnEst{\widehat\arbiDistn}
\def\cdfRefDistn{\cdf_{*}}
\def\cdfArbiDistn{\cdf}
\def\cdfArbiDistnEst{\widehat\cdfArbiDistn}
\def\quantileArbiDistn{\quantile}
\def\quantileArbiDistnEst{\widehat F^{-1}}

\def\distnOne{\mu_1}
\def\distnTwo{\mu_2}

\def\cdfDistnTwo{\cdf_2}
\def\quantileDistnOne{\quantile_1}
\def\quantileDistnTwo{\quantile_2}

\def\arbiDistnVar{\mu'}
\def\distnOneVar{\mu'_1}
\def\distnTwoVar{\mu'_2}

\def\FmeanX{\nu_{1\oplus}}
\def\cdfFmeanX{\cdf_{1\oplus}}
\def\quantileFmeanX{\quantile_{1\oplus}}

\def\frcoef{\chi}

\def\denFmeanX{\den_{1\oplus}}
\def\denFmeanY{\den_{2\oplus}}

\def\FmeanXoracle{\widetilde\nu_{1\oplus}}

\def\quantileFmeanXoracle{\widetilde F^{-1}_{1\oplus}}

\def\FmeanXest{\widehat\nu_{1\oplus}}
\def\cdfFmeanXest{\widehat\cdf_{1\oplus}}
\def\quantileFmeanXest{\widehat F^{-1}_{1\oplus}}

\def\FmeanY{\nu_{2\oplus}}
\def\cdfFmeanY{\cdf_{2\oplus}}
\def\quantileFmeanY{\quantile_{2\oplus}}

\def\FmeanYoracle{\widetilde\nu_{2\oplus}}

\def\quantileFmeanYoracle{\widetilde F^{-1}_{2\oplus}}

\def\FmeanYest{\widehat\nu_{2\oplus}}
\def\cdfFmeanYest{\widehat\cdf_{2\oplus}}
\def\quantileFmeanYest{\widehat F^{-1}_{2\oplus}}

\def\jointmean{\FmeanX\times\FmeanY}

\def\X{\nu_1}

\def\quantileX{\quantile_1}

\def\Xi{\nu_{1i}}

\def\quantileXi{\quantile_{1i}}

\def\XiEst{\widehat\nu_{1i}}
\def\XiEstVar{\widehat\nu_{1i'}}

\def\quantileXiEst{\widehat F^{-1}_{1i}}

\def\Y{\nu_2}

\def\quantileY{\quantile_2}

\def\YpredOracle{\widetilde\nu_2} 
\def\YpredEst{\widehat\nu_2} 

\def\Yi{\nu_{2i}}

\def\quantileYi{\quantile_{2i}}

\def\YiEst{\widehat\nu_{2i}}
\def\YiEstVar{\widehat\nu_{2i'}}

\def\quantileYiEst{\widehat F^{-1}_{2i}}

\def\XlEst{\widehat \nu_{1i'}}
\def\YlEst{\widehat \nu_{2i'}}

\def\nFPCX{J}
\def\nFPCY{K}
\def\covop{\mathcal{C}}
\def\covopX{\covop_{\X}}
\def\covopY{\covop_{\Y}}
\def\covopXoracle{\widetilde\covop_{\X}}
\def\covopYoracle{\widetilde\covop_{\Y}}
\def\covopXest{\widehat\covop_{\X}}
\def\covopYest{\widehat\covop_{\Y}}
\def\crosscovopXY{\covop_{\X\Y}}
\def\crosscovopXYoracle{\widetilde\covop_{\X\Y}} 
\def\crosscovopXYest{\widehat\covop_{\X\Y}} 

\def\egnfctnX{\phi}
\def\egnfctnY{\psi}
\def\egnfctnXoracle{\widetilde\phi}
\def\egnfctnYoracle{\widetilde\psi}
\def\egnfctnXest{\widehat\phi}
\def\egnfctnYest{\widehat\psi}
\def\egnvalX{\lambda}
\def\egnvalY{\varsigma}
\def\egnvalXoracle{\widetilde\lambda}
\def\egnvalYoracle{\widetilde\varsigma}
\def\egnvalXest{\widehat\lambda}
\def\egnvalYest{\widehat\varsigma}
\def\croval{\xi}
\def\crovalOracle{\widetilde\xi}
\def\crovalEst{\widehat\xi}
\def\egnvalSpX{\theta}
\def\egnvalSpY{\vartheta}
\def\coefDecayX{\rho}
\def\coefDecayY{\varrho}
\def\autoBound{q}

\def\noise{\varepsilon}

\def\arbiFctn{g}
\def\arbiFctnTwo{h}

\def\argu{r}

\def\const{{\rm const.}}
\def\denRate{\tau}
\def\indepRate{\alpha}
\def\tsRate{\zeta}

\def\geodesic{\gamma}
\def\paraTrans{\mathrm{P}}
\def\paraTransOp{\mathcal{P}}
\def\paraTransKernel{\mathcal{Q}}

\def\regOp{\Gamma}
\def\regOpOracle{\widetilde\Gamma}
\def\regOpEst{\widehat\Gamma}
\def\regKernel{\beta}
\def\regKernelDtoS{\beta_1}
\def\regKernelOracle{\widetilde\beta}
\def\regKernelEst{\widehat\beta}
\def\regCoef{b}

\def\regCoefOracle{\widetilde b}
\def\regCoefEst{\widehat b}

\def\linRegOp{\Gamma_{\mathrm{E}}}
\def\linRegCoef{\beta}
\def\geoRegOp{\Gamma_{\mathrm{W}}}

\def\dpX{X}
\def\dpY{Y}
\def\nDp{m}
\def\nDpXi{\nDp_{\Xi}}
\def\nDpYi{\nDp_{\Yi}}
\def\projconst{\eta}

\def\lag{r}
\def\Zone{\mu_1}

\def\Ztwo{\mu_2}

\def\Zi{\mu_i}
\def\ZiEst{\widehat\mu_i}

\def\Zn{\mu_n}

\def\ZnPlusOne{\mu_{n+1}}

\def\ZiPlusOne{\mu_{i+1}}

\def\ZiPlusOneEst{\widehat\mu_{i+1}}
\def\ZiPluslag{\mu_{i+\lag}}

\def\FmeanZ{\mu_\oplus}
\def\FmeanZoracle{\widetilde\mu_\oplus}
\def\FmeanZest{\widehat\mu_\oplus}

\def\nDpZi{\nDp_{\Zi}}

\def\covopZlagGen{\covop_\lag} 
\def\covopZlagZero{\covop_0}
\def\covopZlagOne{\covop_1}
\def\covopZlagZeroOracle{\widetilde\covop_0}
\def\covopZlagOneOracle{\widetilde\covop_1}
\def\covopZlagZeroEst{\widehat\covop_0}
\def\covopZlagOneEst{\widehat\covop_1}

\def\egnvalZ{\egnvalX}
\def\egnvalZoracle{\egnvalXoracle}
\def\egnvalZest{\egnvalXest}
\def\crovalZlagOne{\croval}
\def\crovalZlagOneOracle{\widetilde\crovalZlagOne}
\def\crovalZlagOneEst{\widehat\crovalZlagOne}

\def\egnfctnZ{\egnfctnX}
\def\egnfctnZoracle{\egnfctnXoracle}
\def\egnfctnZest{\egnfctnXest}
\def\nFPCZ{\nFPCX}

\def\egnvalSpZ{\egnvalSpX}
\def\coefDecayZx{\coefDecayX}
\def\coefDecayZy{\coefDecayY}

\def\event{\mathcal{E}}

\def\redundant#1{\iffalse#1\fi}

\def\qaq{\quad\text{and}\quad}
\def\usedFPCX{\nFPCX^*}
\def\usedFPCY{\nFPCY^*}
\def\enFPCY{\varkappa}
\def\distnfam{\mathscr{F}}
\def\distnfamDtoS{\mathscr{F}'}
\def\vdistnfam{\mathscr{G}}
\def\jointdistn{\mathcal{F}}
\def\jointdistnDtoS{\mathcal{F}'}
\def\vjointdistn{\mathcal{G}}
\def\factor{C}

\def\pEgnfctnXoracle{\paraTrans\egnfctnXoracle}
\def\pEgnfctnXest{\paraTrans\egnfctnXest}
\def\pEgnfctnYoracle{\paraTrans\egnfctnYoracle}
\def\pEgnfctnYest{\paraTrans\egnfctnYest}

\def\pLogXoracle{\paraTrans\Log_{\FmeanXoracle}}
\def\pLogXest{\paraTrans\Log_{\FmeanXest}}

\def\pEgnfctnZoracle{\paraTrans\egnfctnZoracle}
\def\pEgnfctnZest{\paraTrans\egnfctnZest}

\def\pCovopXoracle{\paraTransOp\covopXoracle}
\def\pCovopXest{\paraTransOp\covopXest}
\def\pCovopYoracle{\paraTransOp\covopYoracle}
\def\pCovopYest{\paraTransOp\covopYest}
\def\pCrosscovopXYoracle{\paraTransOp\crosscovopXYoracle}
\def\pCrosscovopXYest{\paraTransOp\crosscovopXYest}

\def\pCovopZoracle{\paraTransOp\covopZlagZeroOracle}
\def\pCovopZest{\paraTransOp\covopZlagZeroEst}
\def\pCrosscovopZoracle{\paraTransOp\covopZlagOneOracle}
\def\pCrosscovopZest{\paraTransOp\covopZlagOneEst}

\def\pRegKernelOracle{\paraTransKernel\regKernelOracle}
\def\pRegKernelEst{\paraTransKernel\regKernelEst}

\def\largeConst{M}

\def\linCoef{\regCoef^*}
\def\convSeries{\upsilon}
\def\basisL{\varphi}
\def\intgSet{\mbb{Z}}
\def\distortmap{g}
\def\maxDbasisL{\kappa}
\def\maxDenFmeanX{\denBound_{1\oplus}}
\def\maxDenFmeanY{\denBound_{2\oplus}}
\def\sgnFractn{p}
\def\epntX{a_0}
\def\epntCfX{a_1}
\def\epntCfY{a_2}
\def\errCoef{E}
\def\maxErrCoef{e}

\def\sY{Y} 
\def\sYi{Y_i} 
\def\sYiEst{\wh{Y}_i}
\def\regKernelDtoSoracle{\wt\beta_1}
\def\regKernelDtoSest{\wh\beta_1}
\def\betashape{S}
\def\errInsY{\epsilon} 

\def\Xmean{u_1}
\def\Xsd{\sigma_1}
\def\Ymean{u_2}
\def\Ysd{\sigma_2}

\def\EXsd{\expect\Xsd}

\def\EYsd{\expect\Ysd}
 
\def\Ximean{u_{i1}}
\def\Xisd{\sigma_{i1}}
\def\Yimean{u_{2i}}
\def\Yisd{\sigma_{2i}}
\def\errInYmean{\epsilon} 

\def\cdfFmeanZ{F_{\mu_\oplus}}
\def\denFmeanZ{f_{\mu_\oplus}}
\def\maxDenFmeanZ{\denBound_{\mu_\oplus}}
\def\rzeta{\varpi} 

\def\trans{\mathbf{t}}
\def\optTrans#1#2{\trans_{#1,#2}}